\documentstyle[new_cite, graphicx, times]{mn}

\newif\ifAMStwofonts
\AMStwofontstrue

\def\einstein{{\it Einstein~\/}}

\def\et{{et al.\ }}

\def\rosat{{\it ROSAT~\/}}

\def\gsim{\mathrel{\hbox{\rlap{\hbox{\lower4pt\hbox{$\sim$}}}{\raise2pt\hbox{$>$}}}}}
\def\vvm{{\hbox{$\langle V/V_{max}  \rangle$}}}
\newcommand{\ls}{\mathrel{\hbox{\rlap{\hbox{\lower4pt\hbox{$\sim$}}}\hbox{$<$}}}}
\newcommand{\gs}{\mathrel{\hbox{\rlap{\hbox{\lower4pt\hbox{$\sim$}}}\hbox{$>$}}}}
\def\la{\mathrel{\hbox{\rlap{\hbox{\lower4pt\hbox{$\sim$}}}{\raise2pt\hbox{$<$}}}}}
\def\ga{\mathrel{\hbox{\rlap{\hbox{\lower4pt\hbox{$\sim$}}}{\raise2pt\hbox{$>$}}}}}

\def\arcs{{\hbox{$^{\prime\prime}$}}}

\def\dg{^{\circ}}
\def\H0{{\rm ~km~s^{-1}~Mpc^{-1}}}

\def\ga{{\rm\thinspace gauss}}

\def\et{{\it et al.~\/}}

\def\feii{{Fe~\textsc{ii}}}
\def\hb{{H$\beta$}}
\def\oiii{{[O~\textsc{iii}]}}

\title[The 1/4 keV Seyfert galaxy sample]
	{A complete sample of Seyfert galaxies selected at 1/4 keV} 
\author[S. Vaughan \et]
	{S.~Vaughan$^{1,2}$, 
        R.~Edelson$^{1,3}$, R. S.~Warwick$^1$, M.~A.~Malkan$^3$, M.~R.~Goad$^1$\\
$^1$X-Ray Astronomy Group; Department of Physics and Astronomy; Leicester 
	University; Leicester LE1 7RH; U.K.\\
$^2$Institute of Astronomy; Madingley Road; Cambridge CB3
0HA; U.K.\\
$^3$Department of Physics and Astronomy; University of California, 
	Los Angeles; Los Angeles, CA 90095-1562; U.S.A.
}
\date{Accepted 18/6/2001; submitted 11/4/2001; in original form 7/3/2001}
\pagerange{\pageref{firstpage}--\pageref{lastpage}}
\pubyear{2001}
\begin{document}
\maketitle
\label{firstpage}

\begin{abstract}
We have used the \rosat\ Bright Source Catalogue to extract a complete
sample of sources selected in the band from 0.1--0.4~keV. This  1/4
keV-selected sample is comprised of 54 Seyfert galaxies, 25 BL
Lacertae objects,  4 clusters and 27 Galactic stars or
binaries. Seyfert-type galaxies with  ``ultrasoft'' X-ray spectra  can
very often be classed optically as  Narrow-line Seyfert 1s (NLS1s).
Such objects are readily detected in  1/4~keV surveys; the sample
reported here contains 20 NLS1s, corresponding  to a 40\% fraction of
the Seyferts. Optical spectra of the Seyfert galaxies were gathered
for correlative analysis, which confirmed the well-known relations
between X-ray slope and optical spectral properties (e.g., \oiii/\hb\
ratio; \feii\ strength, \hb\ width). The various intercorrelations are
most likely driven, fundamentally, by the shape of the photoionising
continuum in Seyfert nuclei. We argue that  a steep X-ray spectrum is
a better indicator of an ``extreme''  set of physical properties in
Seyfert galaxies than is the narrowness of the  optical \hb\ line.

The correlation studies were also used to isolate a number of Seyfert galaxies
with apparently ``anomalous'' properties. Of
particular interest are the six objects with relatively weak permitted
line emission (\hb\ and \feii) and weak optical continua.  Such
objects are rare in most surveys, but two of these (IC~3599 and
NGC~5905) are known to be transient active galactic nuclei in which the
X-ray flux has faded by factors $\sim$100.  If the other four objects also 
turn out to be transient, this would demonstrate that 1/4 keV surveys
provide an efficient way of finding an interesting class of object.  

Finally, the luminosity function of the 1/4 keV-selected Seyfert galaxies 
was determined and broken down into subsamples to investigate the relative 
space densities of Seyferts when separated on the basis of either X-ray slope 
or \hb\ line width.
\end{abstract}


\section{Introduction}
\label{sect:intro}

Although Active Galactic Nuclei (AGN) exhibit an extremely wide variety of 
properties it has, nevertheless, been posited that such objects can be
 characterised in terms of a relatively small number of underlying 
physical parameters. Multi-wavelength observations of well-defined source 
samples derived from large-area surveys, such as the ultraviolet-excess 
Palomar Bright Quasar Survey (BQS;  \ncite{SG83}),
have proved a particularly effective means of identifying the key physical 
drivers in AGN.  For instance, a principle component analysis of 87 BQS 
objects indicated strong correlations between optical \feii, 
\oiii~$\lambda$5007 line strengths and the velocity width and asymmetry of 
\hb\ (\ncite{BOR92}; henceforth BG92).  This approach was extended into the 
X-ray regime by \scite{LAO97} and others, who found correlations between 
these optical properties and the slope of the soft X-ray continuum 
($\alpha_X$).  Most notably, Seyfert galaxies and quasars with very steep 
soft X-ray spectra -- `ultrasoft' Seyferts, defined here by
$\alpha_{X}>1.7$ -- tend to have narrow \hb, strong  \feii\  and weak
\oiii\ lines.  Many of these ultrasoft sources are, in fact,
Narrow-Line Seyfert 1 (NLS1) galaxies (\ncite{BOL96}),  a sub-class of AGN   
originally identified on the basis of  their strong and
narrow permitted line emission (e.g., FWHM \hb~$\leq 2000$~km~s$^{-1}$
and \oiii/\hb$\leq 3$; \ncite{OP85}).  The remarkable properties of NLS1
galaxies have often been interpreted as due to an extremal value 
of some underlying physical parameter, most frequently the mass accretion rate
relative to the Eddington limit (BG92; \ncite{POU95}).

Of course, the gross properties of an AGN sample depend critically on the 
waveband of the original survey. For instance, hard X-ray selection
introduces a bias against the detection of objects with a steep X-ray 
spectrum and therefore against NLS1s. An illustration of this effect 
is provided by the HEAO-1 A2 hard X-ray survey (\ncite{PIC82})
which failed to find any ultrasoft or NLS1s. In contrast ultrasoft 
Seyferts were detected in both the \rosat\ X-ray (0.1-2.4~keV) and 
extreme-ultraviolet (EUV; 60--210~eV)  all-sky surveys (e.g., \ncite{PUC92}; 
\ncite{GRU98}; \ncite{POU93}; \ncite{EDE99}; \ncite{GRU01}), although the well-studied
samples derived from these surveys are either rather small (e.g. \ncite{EDE99}),
incomplete or biased in favour of sources with a particular spectral form. 

In this paper we report on the first large, complete sample of Seyfert 
galaxies selected in the softest accessible (0.1-0.4 keV) X-ray band
(which hereafter is referred to as the 1/4 keV band). The paper is organized as
follows.  The method used to select the full list of sources from the \rosat\ 
survey data  is described in \S~2 together with the identification of 
a complete 1/4~keV-selected Seyfert galaxy sample. The statistical properties 
of the Seyfert galaxy sample are then investigated in \S~\ref{sect:analysis}. 
This section includes a detailed study of the correlations which exist 
between parameters describing the optical to soft X-ray continuum and optical 
emission line properties of the Seyfert galaxies. A derivation of
the luminosity function of the 1/4~keV-selected Seyfert galaxy sample is 
also presented. \S~\ref{sect:discussion} goes on to discuss the implications 
of our results, followed by a brief summary of the main conclusions of
the paper. Two appendices are also included. The first provides
source lists and further information relating to the two remaining 
source subsamples, namely other extragalactic sources (i.e., excluding Seyfert 
galaxies) and sources with Galactic counterparts.  Appendix B
gives details of an optical programme carried out at the 4.2m William Herschel
Telescope (WHT) on La Palma and the 3.0m Shane Telescope at
the Lick Observatory which, together with published data, provides nearly 
complete optical spectroscopy for the Seyfert galaxy sample. 
Throughout this work values of $ H_0 = 50 $~km~s$^{-1}$~Mpc$^{-1}$ and 
$ q_0 = 0.5 $ are assumed and all measured parameters are quoted for the 
rest frame of the source.

\section{Sample selection}
\label{sect:sample}

During 1990 \rosat\ performed an all-sky survey in both the soft
X-ray and EUV bands.  The former employed the \rosat\ X-ray telescope
and position sensitive detector (PSPC; \ncite{TRU91}) and resulted in
the \rosat\ Bright Source Catalogue (RBSC; \ncite{VOG99}) of 18,811
soft X-ray sources.  A high fraction of these X-ray sources are AGN
(e.g., \ncite{THO98}; \ncite{KRA00}). We utilize the RBSC in  deriving a new sample
based on a count rate threshold in the 0.1-0.4 keV (i.e. 1/4 keV)
\rosat\ band.

The three selection criteria used to construct a preliminary source list were 
as follows:

\begin{enumerate}

\item The 0.1--0.4~keV count rate $C \geq 0.25$~ct~s$^{-1}$.\\

\item Galactic column $N_{H} \leq 1.5 \times 10^{20}$~cm$^{-2}$ 
(using $N_{H}$ measurements from \ncite{DIC90}).\\

\item Dec~$\delta \geq 0 \dg$.

\end{enumerate}

The  PSPC C- and H-band (0.1-0.4 keV and 0.5--2.0~keV respectively)
count  rates, $C$ and $H$, were  derived in  an  approximate  way from  the
RBSC     hardness     ratio     ($HR1=(H-C)/(H+C)$) and full-band
(0.1--2.4 keV) count rate $T$:

\[ C \approx \frac{(1 - HR1)}{2}T \]

\[ H \approx \frac{(1 + HR1)}{2}T \]

In practice this  method recovered the correct  values to  within
0.01~ct/s compared to the values published in \scite{SCH00}.
Criteria (ii) and (iii) delimit the survey area to regions around the Lockman
Hole (\ncite{LOC86}) totalling $\sim$0.6~sr ($\sim 5$\% of the sky).
Selection in terms of line-of-sight column is crucial; the $N_H$ limit
of $1.5 \times 10^{20}$~cm$^{-2}$ corresponds to a foreground optical
depth $ \tau \approx 2 $ in the 0.1--0.4~keV band.  The  total number of 
X-ray sources satisfying the above  criteria is 110.

\begin{table*}
 \caption{The Seyfert galaxy sample. $N_{H}$ is in units of $10^{20}$ cm$^{-2}$.
\label{table:sample}
}
\small
 \begin{tabular}{@{}llrrrrrrrrrrrr@{}}
\hline
 	&	 		&	&		&	R.A.			&	Dec.			&	$T$	 		&	 	 	&		&			&	$C$	&	$H$	&		&		\\
No	&	Name		&Type	&	$z$	&	(J2000)			&	(J2000)			&	(ct/s)			&	$HR1$		&$\alpha_{X}$	& 	$N_{H}$		&	(ct/s)	&	(ct/s)	&	$m_{V}$	&	Ref	\\
(1)	&	(2)		&(3)	&	(4)	&	(5)			&	(6)			&	(7)			&	(8)		&	(9)	&	(10)		&	(11)	&	(12)	&	(13)	&	(14)	\\
\hline
1	&	1E 0919+515	&NLS1	&	0.161	&	09	22	46	&	51	20	46	&	0.42$\pm$0.03	&	-0.69$\pm$0.05	&	2.2	&	1.43		&	0.354	&	0.065	&	17.9	&	L, St	\\
2	&	Mrk 110		&BLS1	&	0.035	&	09	25	12	&       52	17	16	&	1.69$\pm$0.06	&	-0.19$\pm$0.03	&	1.2	&	1.47		&	1.006	&	0.685	&	16.0	&	BG92	\\
3	&	US 0656 	&S1	&	0.160	&	09	30	17	&	47	07	25	&	0.46$\pm$0.03	&	-0.24$\pm$0.07	&	1.4	&	1.47		&	0.283	&	0.173	&	16.5	&	--	\\
4	&	PG 0953+414	&BLS1	&	0.234	&	09	56	52	&	41	15	24	&	0.95$\pm$0.06	&	-0.55$\pm$0.04	&	1.6	&	0.79		&	0.735	&	0.213	&	14.5	&	BG92	\\
5	&	IRAS 10026	&BLS1	&	0.178	&	10	05	42	&	43	32	44	&	0.66$\pm$0.04	&	-0.61$\pm$0.04	&	1.8	&	1.08		&	0.534	&	0.129	&	16.5	&	G	\\
\hline
6	&	RX J1008+46	&BLS1	&	0.388	&	10	08	30	&	46	29	57	&	0.42$\pm$0.03	&	-0.33$\pm$0.07	&	1.2	&	0.91		&	0.278	&	0.140	&	18.9	&	W	\\
7	&	Ton 1187	&BLS1	&	0.07	&	10	13	03	&	35	51	31	&	1.38$\pm$0.06	&	-0.44$\pm$0.03	&	1.7	&	1.38		&	0.992	&	0.386	&	17.4	&	G	\\
8	&	RX J1019+37	&NLS1	&	0.133	&	10	19	00	&	37	52	49	&	0.78$\pm$0.04	&	-0.07$\pm$0.05	&	1.0	&	1.38		&	0.415	&	0.360	&	15.0	&	W	\\
9	&	Mrk 141		&BLS1	&	0.042	&	10	19	12	&	63	58	02	&	0.51$\pm$0.04	&	-0.44$\pm$0.06	&	1.6	&	1.24		&	0.364	&	0.142	&	15.4	&	G	\\
10	&	Mrk 142 	&NLS1	&	0.045	&	10	25	31	&	51	40	39	&	1.75$\pm$0.06	&	-0.61$\pm$0.02	&	1.9	&	1.14		&	1.406	&	0.341	&	16.1	&	G, L	\\
\hline
11	&	RX J1026+55	&BLS1	&	0.119	&	10	26	52	&	55	09	13	&	0.37$\pm$0.03	&	-0.63$\pm$0.05	&	2.0	&	0.80		&	0.304	&	0.069	&	17.7	&	W	\\
12	&	RE J1034+396	&NLS1	&	0.424	&	10	34	38	&	39	38	34	&	2.66$\pm$0.09	&	-0.74$\pm$0.02	&	2.2	&	1.02		&	2.312	&	0.346	&	15.6	&	P	\\
13	&	RX J1046+52	&BLS1	&	0.499	&	10	46	14	&	52	56	00	&	0.32$\pm$0.03	&	-0.82$\pm$0.04	&	2.6	&	1.08		&	0.293	&	0.029	&	17.5	&	W	\\
14	&	RX J1050+55	&BLS1	&	0.331	&	10	50	55	&	55	27	31	&	0.37$\pm$0.03	&	-0.69$\pm$0.05	&	2.0	&	0.77     	&	0.316	&	0.058	&	17.0	&	W, G   \\
15	&	RX J1054+48	&BLS1	&	0.266	&	10	54	44	&	48	31	45	&	0.45$\pm$0.03	&	-0.36$\pm$0.06	&	1.5	&	1.23		&	0.309	&	0.145	&	15.7	&	W	\\
\hline
16	&	EXO 1055+60	&BLS1	&	0.150	&	10	58	30	&	60	16	02	&	0.39$\pm$0.03	&	-0.76$\pm$0.04	&	2.0	&	0.61		&	0.342	&	0.047	&	17.0	&	G	\\
17	&	RX J1117+65	&BLS1	&	0.147	&	11	17	10	&	65	22	10	&	0.55$\pm$0.03	&	-0.72$\pm$0.03	&	2.0     &	1.00		&	0.469	&	0.076	&	16.7	&	G	\\
18	&	PG 1116+21	&BLS1	&	0.177	&	11	19	08	&	21	19	14	&	1.03$\pm$0.06	&	-0.48$\pm$0.05	&	1.7	&	1.28		&	0.759	&	0.267	&	15.2	&	G	\\
19	&	EXO 1128+691	&NLS1	&	0.043	&	11	31	05	&	68	51	55	&	1.58$\pm$0.05	&	-0.43$\pm$0.02	&	1.6	&	1.32		&	1.129	&	0.450	&	16.5	&	Be	\\
20	&	RX J1138+57	&BLS1	&	0.116	&	11	38	49	&	57	42	45	&	0.58$\pm$0.04	&	-0.48$\pm$0.06	&	1.6	&	1.09		&	0.428	&	0.150	&	16.5	&	W	\\
\hline
21	&	NGC 4051	&NLS1	&	0.002	&	12	03	08	&	44	31	54	&	3.92$\pm$0.11	&	-0.45$\pm$0.02	&	1.7	&	1.32		&	2.841	&	1.077	&	10.8	&	L	\\
22	&	RX J1209+32	&NLS1	&	0.145	&	12	09	46	&	32	17	02	&	0.57$\pm$0.06	&	-0.62$\pm$0.09	&	1.9	&	1.06		&	0.458	&	0.107	&	17.7	&	W	\\
23	&	RX J1226+32	&BLS1	&	0.243	&	12	26	23	&	32	44	30	&	0.37$\pm$0.03	&	-0.38$\pm$0.07	&	1.6	&	1.41		&	0.255	&	0.114	&	17.2	&	W	\\
24	&	RX J1232+49	&NLS1	&	0.262	&	12	32	20	&	49	57	31	&	0.34$\pm$0.03	&	-0.60$\pm$0.07	&	2.0	&	1.31		&	0.272	&	0.068	&	17.0	&	W	\\
25	&	Ton 83		&NLS1	&	0.29	&	12	33	41	&	31	01	03	&	0.50$\pm$0.03	&	-0.52$\pm$0.05	&	1.8	&	1.35		&	0.378	&	0.119	&	16.2	&	S	\\
\hline
26	&	MCG +08-23-067	&NLS1	&	0.03	&	12	36	51	&	45	39	06	&	0.53$\pm$0.04	&	-0.27$\pm$0.06	&	1.3	&	1.37		&	0.334	&	0.192	&	16.0	&	W	\\
27	&	IC 3599		&NLS1/S2&	0.022	&	12	37	41	&	26	42	29	&	5.10$\pm$0.11	&	-0.63$\pm$0.01	&	2.0	&	1.29		&	4.152	&	0.943	&	15.6	&	S, G	\\
28	&	Was 61		&BLS1	&	0.044	&	12	42	11	&	33	17	03	&	0.86$\pm$0.04	&	-0.27$\pm$0.04	&	1.1	&	1.33		&	0.545	&	0.313	&	15.4	&	S, G	\\
29	&	RX J1244+58	&NLS1	&	0.198	&	12	44	41	&	58	56	29	&	0.33$\pm$0.04	&	-0.78$\pm$0.05	&	2.1	&	1.21		&	0.293	&	0.036	&	18.0	&	W	\\
30	&	RX J1258+23	&BLS1	&	0.075	&	12	58	51	&	23	55	32	&	0.43$\pm$0.03	&	-0.38$\pm$0.06	&	1.5	&	1.30		&	0.297	&	0.133	&	17.0	&	W	\\
\hline
31	&	RX J1312+26	&BLS1	&	0.06	&	13	12	59	&	26	28	25	&	0.42$\pm$0.04	&	-0.41$\pm$0.07	&	1.4	&	1.05		&	0.295	&	0.123	&	16.8	&	S, G	\\
32	&	Ton 1571	&NLS1	&	0.075	&	13	14	22	&	34	29	40	&	0.65$\pm$0.04	&	-0.63$\pm$0.04	&	1.9	&	0.99		&	0.530	&	0.120	&	16.3	&	S	\\
33	&	RX J1319+52	&NLS1	&	0.092	&	13	19	57	&	52	35	33	&	0.66$\pm$0.04	&	-0.52$\pm$0.05	&	1.7	&	1.19		&	0.504	&	0.159	&	17.3	&	W	\\
34	&	RX J1328+24	&BLS1	&	0.223	&	13	28	20	&	24	09	27	&	0.32$\pm$0.03	&	-0.66$\pm$0.07	&	2.0	&	1.16		&	0.261	&	0.053	&	17.7	&	W	\\
35	&	IRAS 13349+243	&BLS1	&	0.108	&	13	37	18	&	24	23	06	&	2.53$\pm$0.09	&	-0.65$\pm$0.02	&	2.0	&	1.16		&	2.085	&	0.442	&	15.0	&	S	\\
\hline
36	&	RX J1339+40	&NLS1	&	0.118	&	13	39	28	&	40	32	29	&	0.35$\pm$0.03	&	-0.58$\pm$0.06	&	1.7	&	0.87		&	0.276	&	0.073	&	17.7	&	W	\\
37	&	RX J1342+38	&BLS1	&	0.176	&	13	42	31	&	38	29	08	&	0.37$\pm$0.03	&	-0.37$\pm$0.06	&	1.5	&	0.91		&	0.250	&	0.115	&	18.0	&	W	\\
38	&	PG 1341+25	&BLS1	&	0.087	&	13	43	56	&	25	38	45	&	0.50$\pm$0.04	&	-0.56$\pm$0.06	&	1.8	&	1.09		&	0.393	&	0.111	&	16.6	&	W	\\
39	&	Mrk 663 	&S2	&	0.026	&	13	54	20	&	32	55	47	&	0.84$\pm$0.05	&	0.36$\pm$0.05	&	0.4	&	1.24		&	0.267	&	0.568	&	14.6	&	S	\\
40	&	RX J1355+56	&BLS1	&	0.122	&	13	55	15	&	56	12	44	&	0.55$\pm$0.06	&	-0.55$\pm$0.08	&	1.8	&	1.13		&	0.423	&	0.123	&	17.1	&	S, G	\\
\hline
41	&	PG 1402+261	&BLS1	&	0.164	&	14	05	16	&	25	55	36	&	0.65$\pm$0.04	&	-0.54$\pm$0.05	&	1.9	&	1.47		&	0.500	&	0.149	&	15.6	&	S, L	\\
42	&	PG 1415+451	&BLS1	&	0.114	&	14	17	00	&	44	55	56	&	0.50$\pm$0.03	&	-0.66$\pm$0.03	&	2.0	&	1.22		&	0.413	&	0.085	&	15.7	&	S, L	\\
43	&	RX J1426+39	&S1	&	0.081	&	14	26	30	&	39	03	48	&	0.36$\pm$0.03	&	-0.54$\pm$0.05	&	1.6	&	0.93		&	0.279	&	0.083	&	16.0	&	--	\\
44	&	Mrk 684		&NLS1	&	0.046	&	14	31	04	&	28	17	16	&	0.58$\pm$0.04	&	-0.23$\pm$0.06	&	1.4	&	1.48		&	0.355	&	0.222	&	15.2	&	S	\\
45	&	Mrk 478		&NLS1	&	0.079	&	14	42	07	&	35	26	32	&	5.78$\pm$0.10	&	-0.70$\pm$0.01	&	2.1	&	1.03		&	4.916	&	0.867	&	15.0	&	G, L	\\
\hline
46	&	PG 1444+407	&BLS1	&	0.267	&	14	46	45	&	40	35	10	&	0.34$\pm$0.03	&	-0.57$\pm$0.05	&	1.8	&	1.25		&	0.269	&	0.074	&	16.0	&	S	\\
47	& 	RX J1448+35	&BLS1	&	0.113	&	14	48	25	&	35	59	55	&	0.37$\pm$0.03	&	-0.71$\pm$0.06	&	2.1	&	1.07		&	0.312	&	0.053	&	16.4	&	W	\\
48	&	NGC 5905 	&H~\textsc{ii}&	0.011	&	15	15	23	&	55	30	57	&	0.31$\pm$0.02	&	-0.99$\pm$0.01	&	3.6	&	1.44		&	0.308	&	0.002	&	12.5	&	S	\\
49	&	RX J1529+56	&BLS1	&	0.099	&	15	29	07	&	56	16	04	&	0.74$\pm$0.03	&	-0.43$\pm$0.03	&	1.7	&	1.29		&	0.531	&	0.212	&	15.8	&	W	\\
50	&	MCG +06-36-003	&BLS1	&	0.070	&	16	13	01	&	37	16	56	&	0.42$\pm$0.03	&	-0.31$\pm$0.06	&	1.4	&	1.24		&	0.272	&	0.143	&	15.5	&	W	\\
\hline
51	&	RX J1618+36	&NLS1	&	0.034	&	16	18	09	&	36	19	50	&	0.85$\pm$0.03	&	-0.43$\pm$0.03	&	1.6	&	1.28		&	0.608	&	0.242	&	16.9	&	S, G	\\
52	&	RX J1619+40	&BLS1	&	0.038	&	16	19	51	&	40	58	34	&	0.51$\pm$0.03	&	-0.54$\pm$0.04	&	1.6	&	0.93		&	0.393	&	0.117	&	16.0	&	S, W	\\
53	&	RX J1629+40	&BLS1	&	0.272	&	16	29	01	&	40	07	53	&	0.79$\pm$0.03	&	-0.79$\pm$0.02	&	2.3	&	0.85		&	0.702	&	0.082	&	19.0	&	S	\\
54	&	RX J1646+39	&NLS1	&	0.10	&	16	46	25	&	39	29	21	&	0.40$\pm$0.03	&	-0.42$\pm$0.06	&	1.6	&	1.31		&	0.281	&	0.115	&	17.1	&	S, G	\\
\hline		
\end{tabular}
\end{table*}

The initial classification  of the  X-ray sources in our source list
was  based on  the   optical  identifications   of \scite{SCH00} 
together with information obtained from the
{\tt SIMBAD}\footnote{{\tt http://simbad.u-strasbg.fr/Simbad}}    and
{\tt   NED}\footnote{{\tt http://nedwww.ipac.caltech.edu/}} on-line
catalogues.  This analysis  showed that the 110 X-ray sources comprise
27 Galactic objects (3 cataclysmic variable stars,  6 white dwarfs,
17 active coronal stars and one super-soft source) and 83 
extragalactic objects (54 Seyfert galaxies or quasars, 25 BL Lac  objects, 4
clusters  of galaxies).  Table~\ref{table:sample} shows the Seyfert
subsample, which is the focus of the present work,  but for
completeness the full lists of non-Seyfert sources are presented in
Appendix~\ref{append:sample}.

In Table~\ref{table:sample} the columns provide  the following information: 
(2) The source name; (3) optical type (see \S~\ref{sect:properties}); (4)
the redshift;  (5)  and  (6) the  right ascension and declination as
tabulated  in the RBSC; (7) the full-band count rate;  (8) hardness
ratio; (9) an X-ray slope (derived as in \ncite{EDE99}); (10) Galactic
column density; (11)  and (12) derived 
C- and H-band count rates; (13) optical
V-band magnitude; (14) reference for the optical information (W
indicates WHT data - see Appendix~\ref{append:optical_data} , S indicates 
Shane data - see Appendix~\ref{append:optical_data}, Be indicates data from
\scite{BED88}, G from \scite{GRU99}, L from  \scite{LIP93}, P from
\scite{PUC95}, St from \scite{STE89}).  The tabulated information is
derived from the published RBSC data (e.g., \ncite{VOG99} and
\ncite{SCH00}) except for the Galactic $N_{H}$ (\ncite{DIC90}).

\subsection{Sample completeness}
\label{sect:complete}

The completeness of the sample was checked using  the  \vvm\  test  of
\scite{SCH68}.    For  the  Seyfert 
sample   \vvm~$=  0.52 \pm 0.04  $ indicating no
evidence for either  strong evolution or  incompleteness. The  BL Lac
sample (Appendix~\ref{append:sample}) shows evidence for either
incompleteness or negative 
evolution, with \vvm~$= 0.37 \pm  0.06 $. This
effect has been seen in previous samples  of X-ray  selected BL Lacs,
first in  the \einstein Medium Sensitivity Survey (\ncite{MAC84}) and
also in \rosat\ selected samples (\ncite{BAD98}),  and may be due to
cosmological evolution of some subset of the BL Lac population.

The  Galactic sample  has  \vvm~$=  0.38 \pm
0.06 $.  This apparent  incompleteness is  most  likely due  to the
source populations  falling  below  the  Euclidean  prediction  at
distances comparable to the scale  height of the Galactic disc.  In
order to check this, distance estimates for  17  of the sources  were
taken from the literature  (e.g., \ncite{STR93}; \ncite{VEN97}) or derived 
from the distance moduli of the stars with well-known
spectral type. The mean distance to  the ten main sequence stars  with
known distances is $\sim 90$~pc, and the mean distance to the older stars
(white dwarfs, RS CVn systems) is $\sim 180$~pc. These are comparable
to the scale heights of the {\it young  thin disc} of the  Galaxy
($\sim 100$~pc) and  the {\it old thin  disc}  ($\sim  300$~pc),
respectively (e.g.,  \ncite{HEY97}; \ncite{VALL00}  and references
therein).

\section{Analysis}
\label{sect:analysis}

In order  to measure  the optical properties  of the  {\it  complete}
sample   of  Seyfert  galaxies, new optical  spectra have  been acquired
for  38 of  the  54 objects listed in Table~\ref{table:sample} 
(labelled as W or S  in column 14 of the Table).  These  data, when combined  
with  the previously  published  optical  work and non-simultaneous \rosat\
survey data, provide  detailed information about the  properties of
the Seyfert sample.  Details of the new optical observations and
the optical data reduction are given in Appendix~\ref{append:optical_data}.

The full optical dataset was used to define the global properties of the 
sample, as described in the next subsection, and to search for correlations 
between the various  observed properties as discussed in
Section~\ref{sect:correlation}.  Section~\ref{sect:outlyers} considers
the objects which lie at extreme ends of these correlations. 
Finally the luminosity functions of the complete sample of Seyfert galaxies
and of various subsamples of objects are derived in 
section~\ref{sect:lum_function}.

\subsection{General sample properties}
\label{sect:properties}

The Seyfert galaxy sample comprises roughly equal numbers of
``ultrasoft'' $\alpha_{X}>1.7$, sources  (28) and ``normal'' spectrum
objects (26) with an overall  mean $\langle \alpha_{X} \rangle
=1.73$. The sample can also be divided on the basis of \hb\ line
width. Twenty objects conform to the standard definition of an NLS1
leaving 31 broad-line Seyfert 1s (BLS1s) or Seyfert 2s (including one
H~\textsc{ii} galaxy) and two objects for which \hb\ 
measurements are not available.   (Two objects appear to be transient
in nature, and hence their exact identification is non-trivial, 
but have been left in the Seyfert sample. These are
discussed further in \S~\ref{sect:outlyers}.) 
The mean redshift of the Seyfert sample is  $\langle z \rangle =
0.134$ with a standard deviation of 0.103.

\subsection{Correlations between parameters}
\label{sect:correlation}

\begin{table*}
\small
 \caption{Results of non-parametric correlation tests. Each entry
  shows the Spearman rank-order correlation coefficient $R_{S}$ and
  the number of objects included in the test. 
  Correlations significant at the $>99$\% level are shown in bold. 
  \label{table:correlation}}
 \begin{tabular}{@{}lrrrrrrrrrrrr@{}}
\hline
		&	\hb	&	\oiii	&	EW	&H$\alpha$/	&	peak	&	\feii/	&	\oiii/	&		&		&		\\
		&	FWHM	&	FWHM	&	\feii	&	\hb	&\oiii/\hb	&	\hb	&	\hb	&$\alpha_{opt}$	&$\alpha_{OX}$	&$\alpha_{X}$	\\
(1)		&	(2)	&	(3)	&	(4)	&	(5)	&	(6)	&	(7)	&	(8)	&	(9)	&	(10)	&	(11)	\\
\hline																		 
$\log(\nu L_{1/4})$&	0.26	&     0.33 	&	0.13	&	0.16	&	-0.23	&	-0.14	&{\bf	-0.43}	&{\bf -0.63}	&	-0.18	&	0.24	\\
		&	(51)	&	(42)	&	(48)	&	(39)	&	(35)	&	(47)	&	(43)	&	(42)	&	(54)	&	(54)	\\
\hline																		 
\hb		&		&	-0.14	&	-0.13	&	 0.10	&	0.40	&	-0.16	&	-0.06	&	-0.03	&	0.21	&     	-0.31 	\\
FWHM		&		&	(40)	&	(47)	&	(39)	&	(35)	&	(47)	&	(43)	&	(41)	&	(51)	&	(51)	\\
\hline																		 
\oiii		&		&		&	0.13	&	0.03	&	-0.40	&	0.04	&	-0.19	&	-0.36	&	0.02	&	0.07	\\
FWHM		&		&		&	(40)	&	(36)	&	(32)	&	(39)	&	(40)	&	(38)	&	(41)	&	(42)	\\
\hline																		 
EW		&		&		&		&	-0.32	& -0.37		&{\bf 0.84}	&{\bf -0.48}	&{\bf -0.50}	&	0.27	&{\bf 	0.45}	\\
\feii		&		&		&		&	(38)	&	(35)	&	(47)	&	(42)	&	(42)	&	(48)	&	(48)	\\
\hline																		 
H$\alpha$/	&		&		&		&		&	0.10	&	-0.16	&	0.36	&	0.25	&	-0.07	&	0.08	\\
\hb		&		&		&		&		&	(31)	&	(39)	&	(36)	&	(37)	&	(39)	&	(39)	\\
\hline																		
peak		&		&		&		&		&		&	-0.14	&{\bf 0.72}	&	 0.29	&      	-0.35 	&{\bf 	-0.53}	\\
\oiii/\hb	&		&		&		&		&		&	(35)	&	(35)	&	(31)	&	(35)	&	(35)	\\
\hline																		 
\feii/		&		&		&		&		&		&		&	-0.30	&	-0.30	&	0.16	&	-0.28	\\
\hb		&		&		&		&		&		&		&	(42)	&	(41)	&	(47)	&	(47)	\\
\hline																		 
\oiii/		&		&		&		&		&		&		&		& 	-0.20	&     	-0.33 	&	-0.20	\\
\hb		&		&		&		&		&		&		&		&	(42)	&	(43)	&	(43)	\\
\hline																		 
$\alpha_{opt}$	&		&		&		&		&		&		&		&		&	-0.02	&     	-0.33 	\\
		&		&		&		&		&		&		&		&		&	(42)	&	(42)	\\
\hline																		 
$\alpha_{OX}$	&		&		&		&		&		&		&		&		&		&	-0.08	\\
		&		&		&		&		&		&		&		&		&		&	(54)	\\
\hline
\end{tabular}
\end{table*}

\begin{figure*}
\centering
\hbox{
\includegraphics[width=4.8 cm, height=5.5 cm, angle=270]{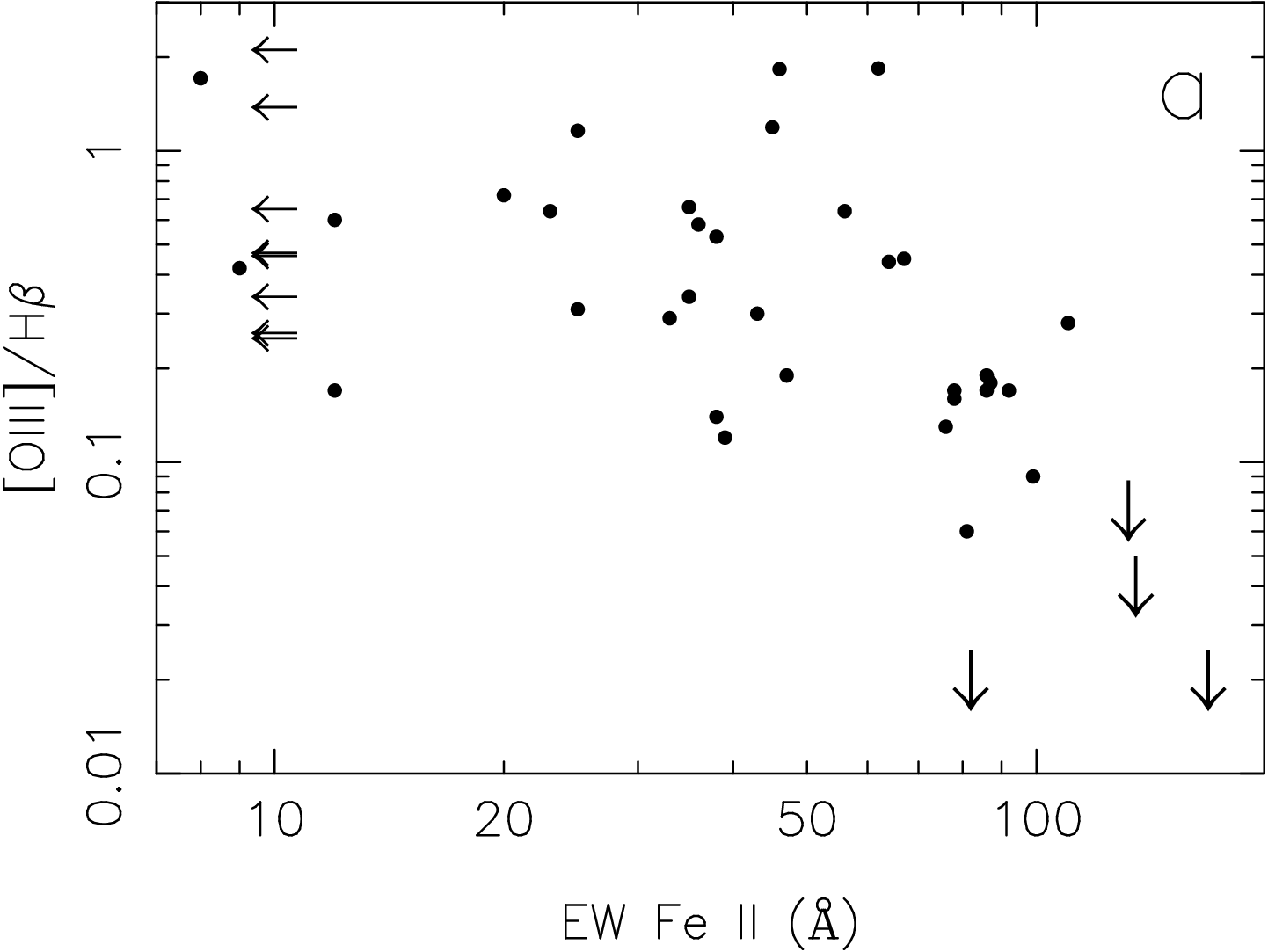}
\hspace{0.1 cm}
\includegraphics[width=4.8 cm, height=5.5 cm, angle=270]{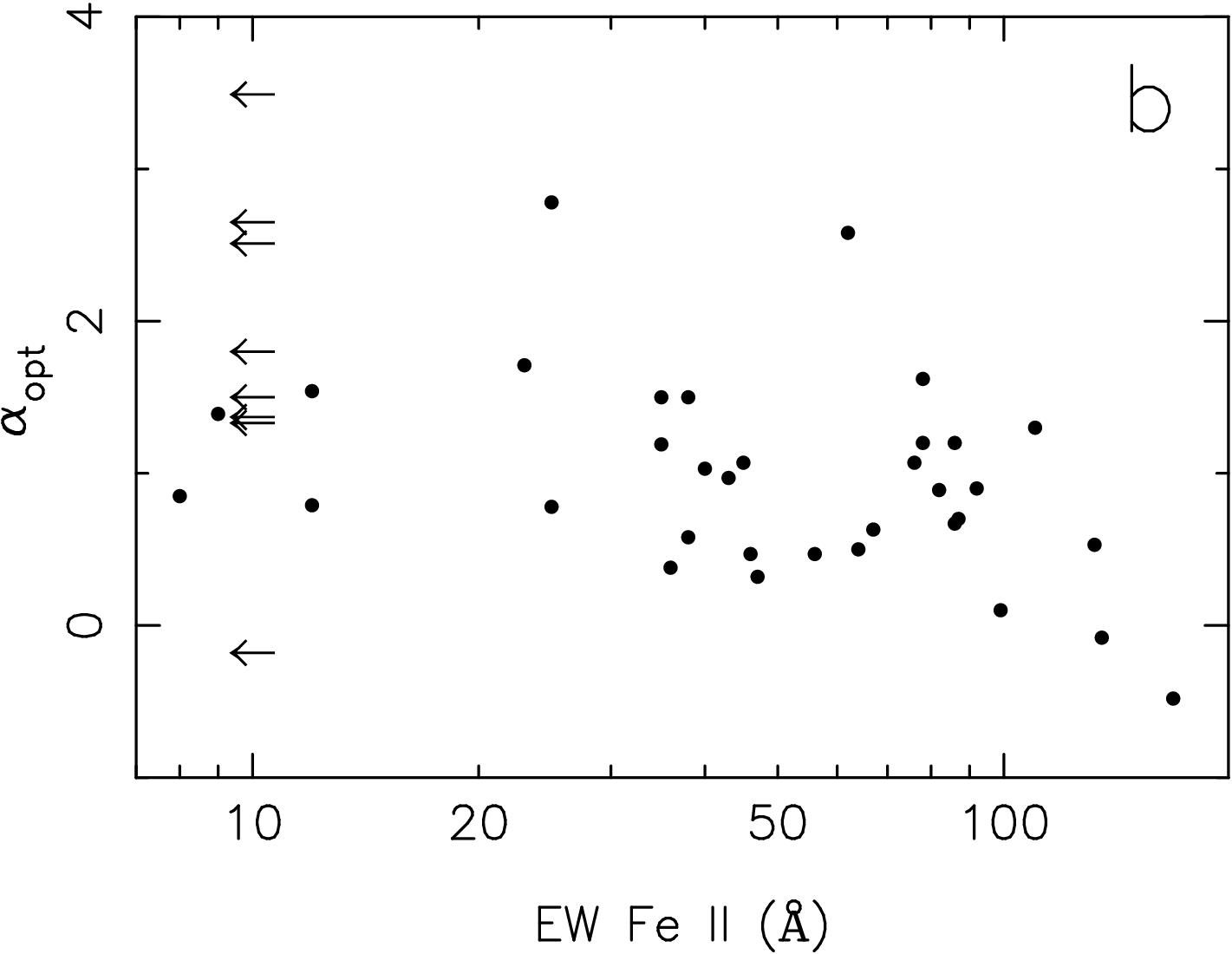}
\hspace{0.1 cm}
\includegraphics[width=4.8 cm, height=5.5 cm, angle=270]{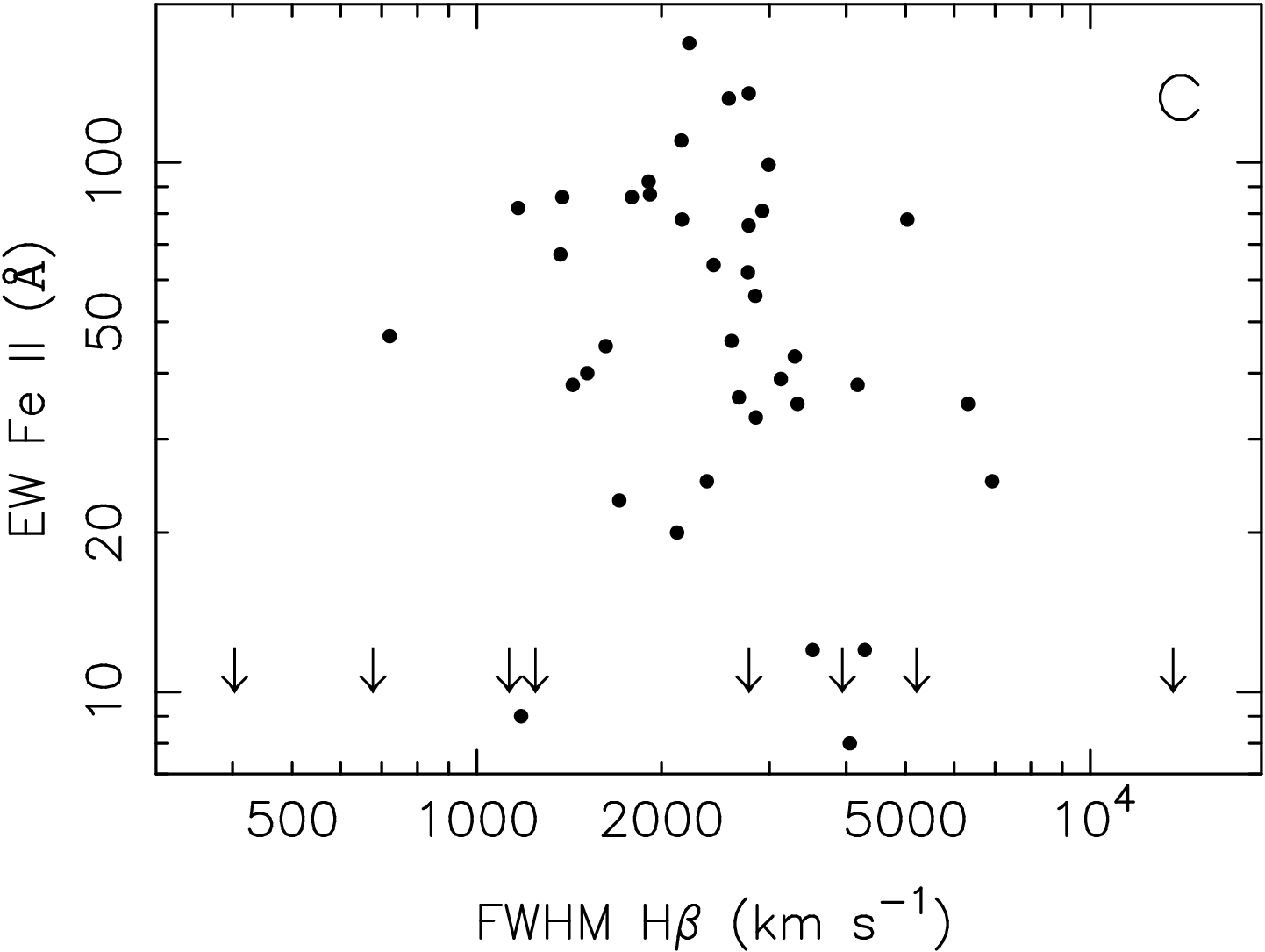}
}
\vspace{0.5 cm}
\hbox{
\includegraphics[width=4.8 cm, height=5.5 cm, angle=270]{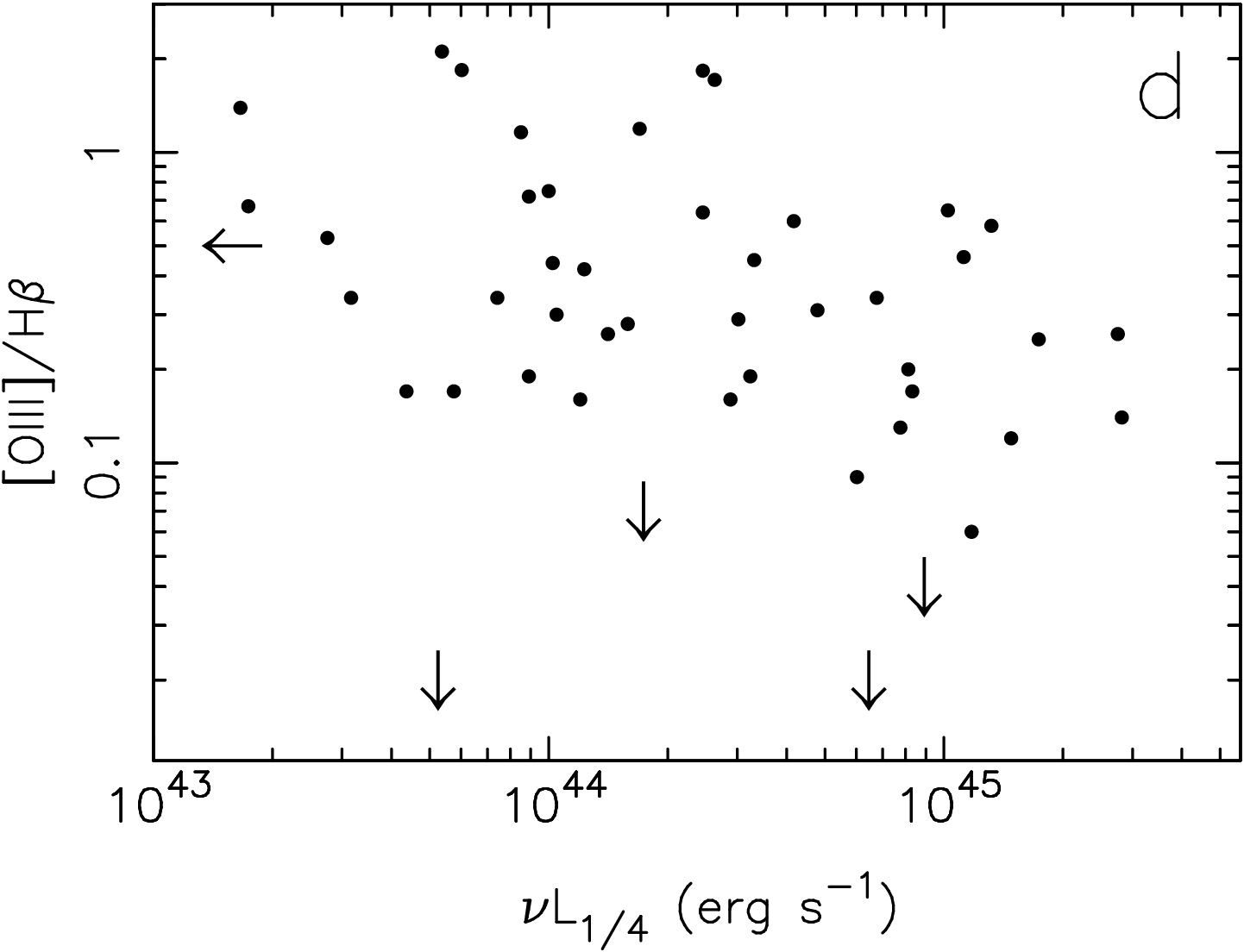}
\hspace{0.1 cm}
\includegraphics[width=4.8 cm, height=5.5 cm, angle=270]{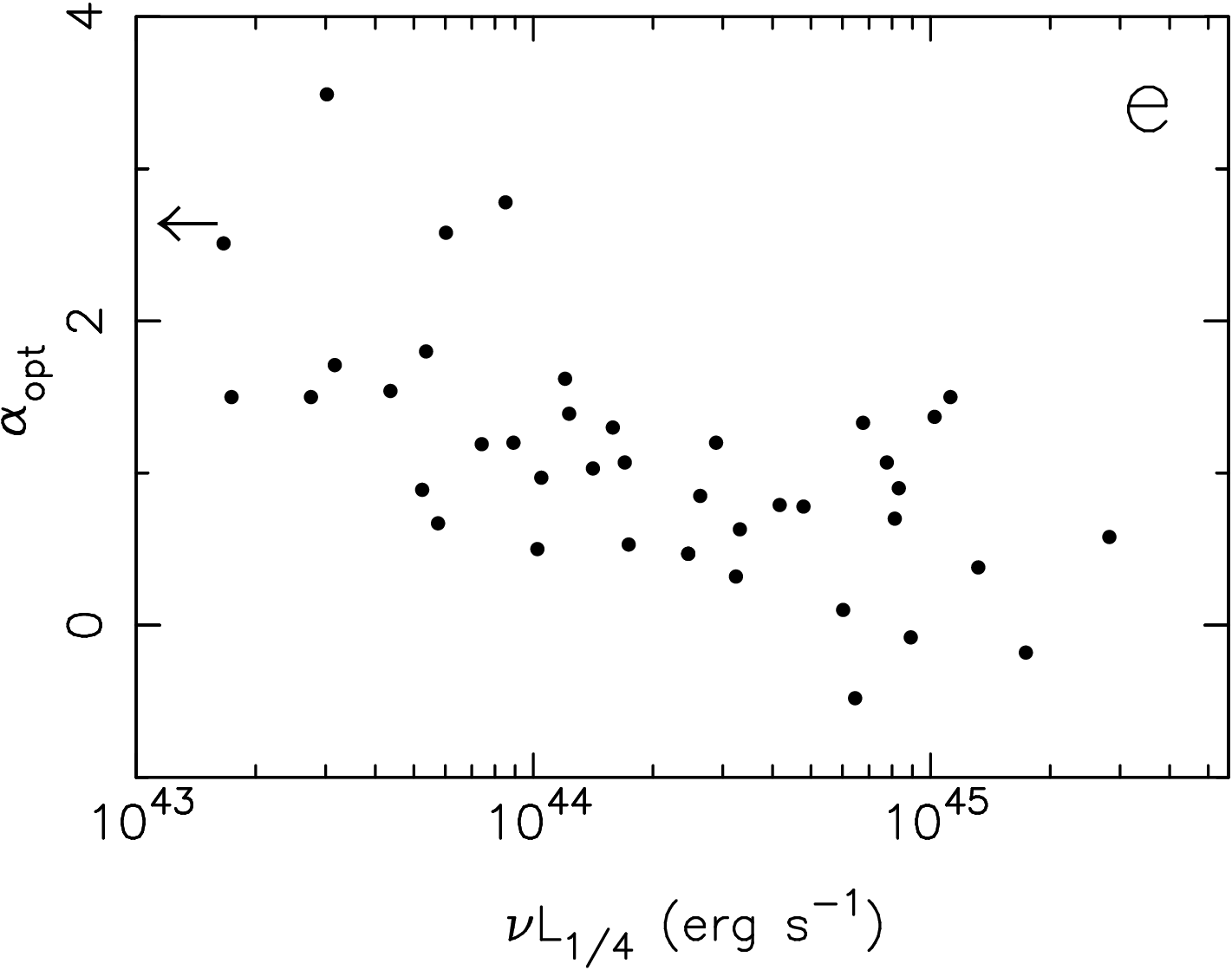}
\hspace{0.1 cm}
\includegraphics[width=4.8 cm, height=5.5 cm, angle=270]{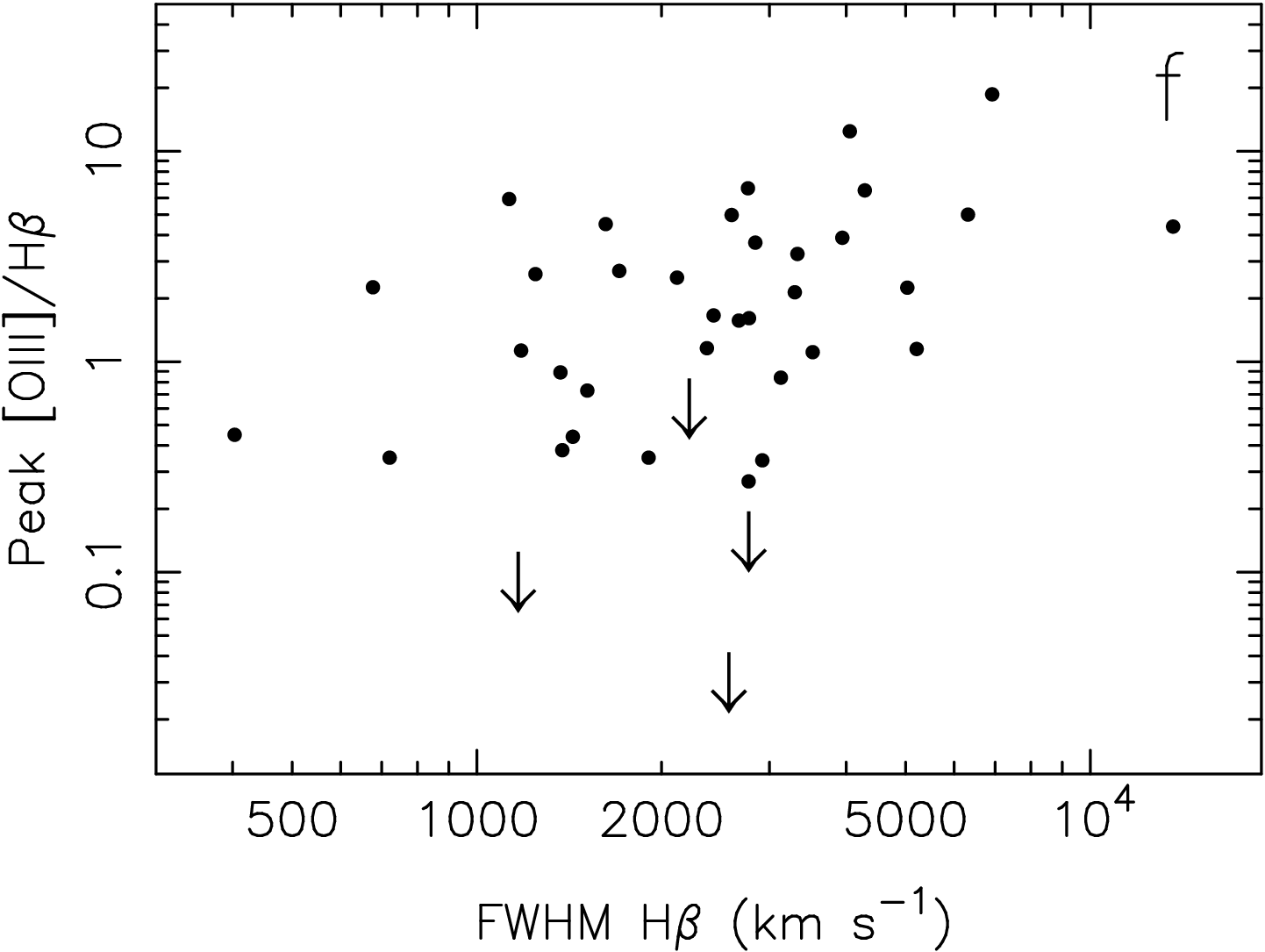}
}
\vspace{0.5 cm}
\hbox{
\includegraphics[width=4.8 cm, height=5.5 cm, angle=270]{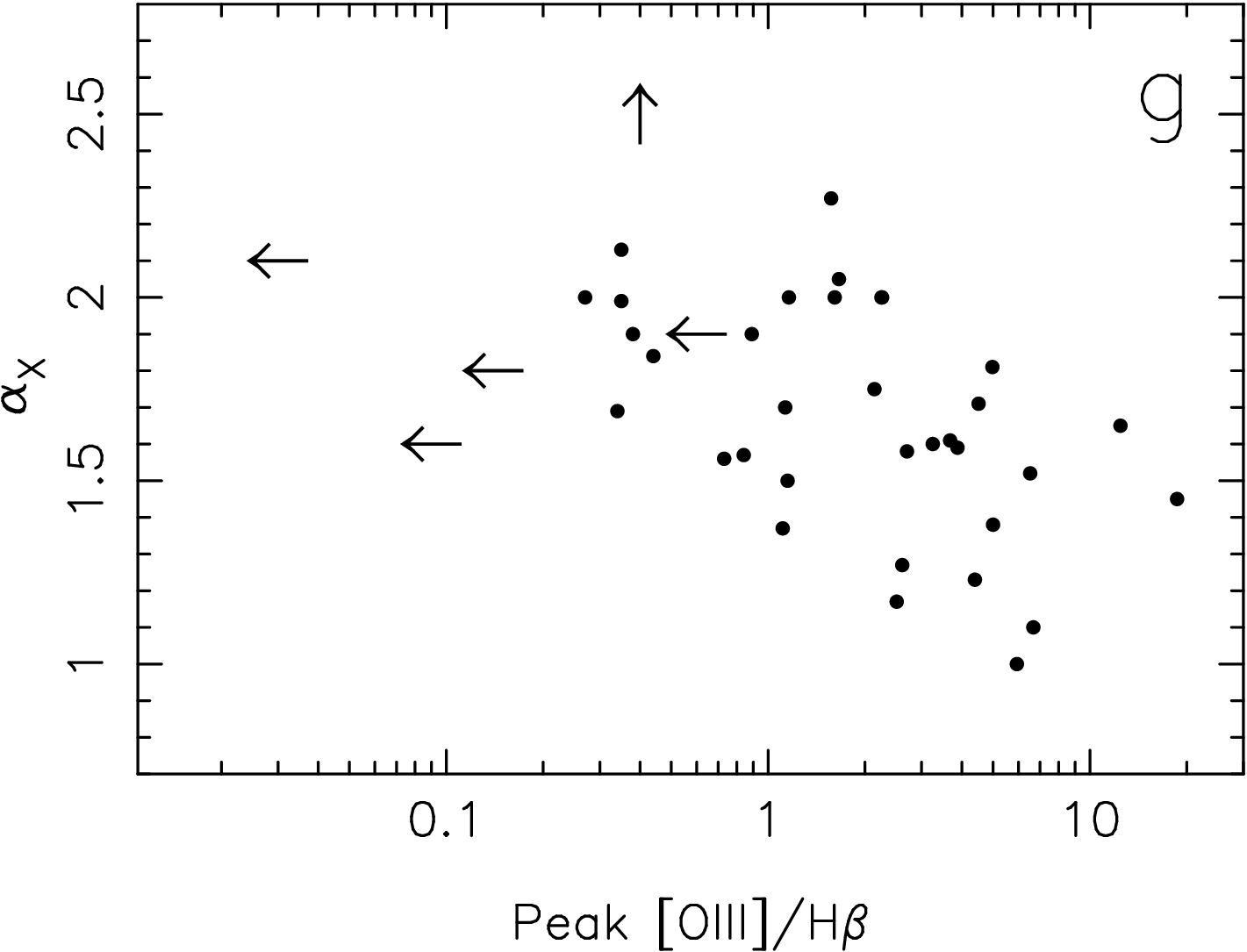}
\hspace{0.1 cm}
\includegraphics[width=4.8 cm, height=5.5 cm, angle=270]{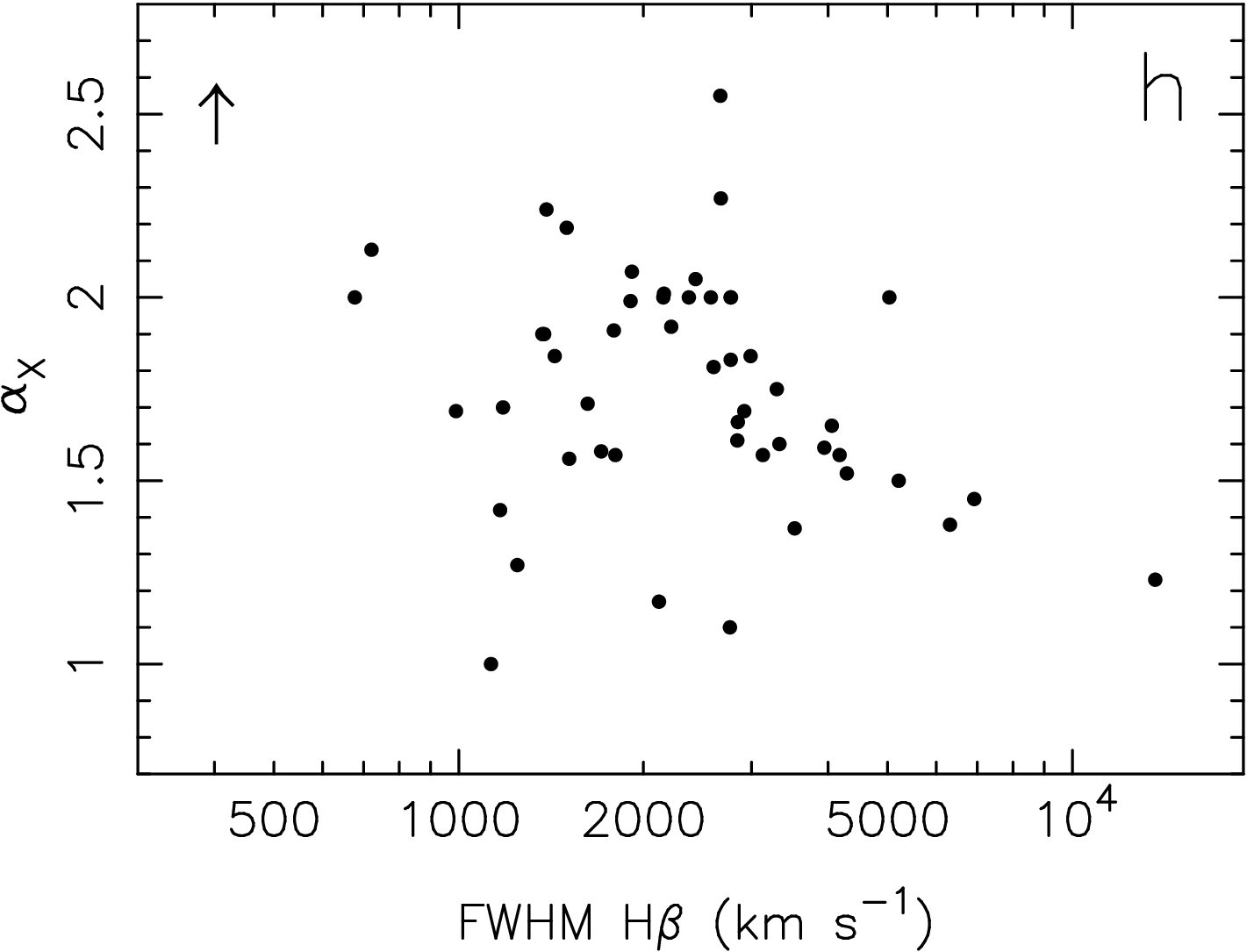}
\hspace{0.1 cm}
\includegraphics[width=4.8 cm, height=5.5 cm, angle=270]{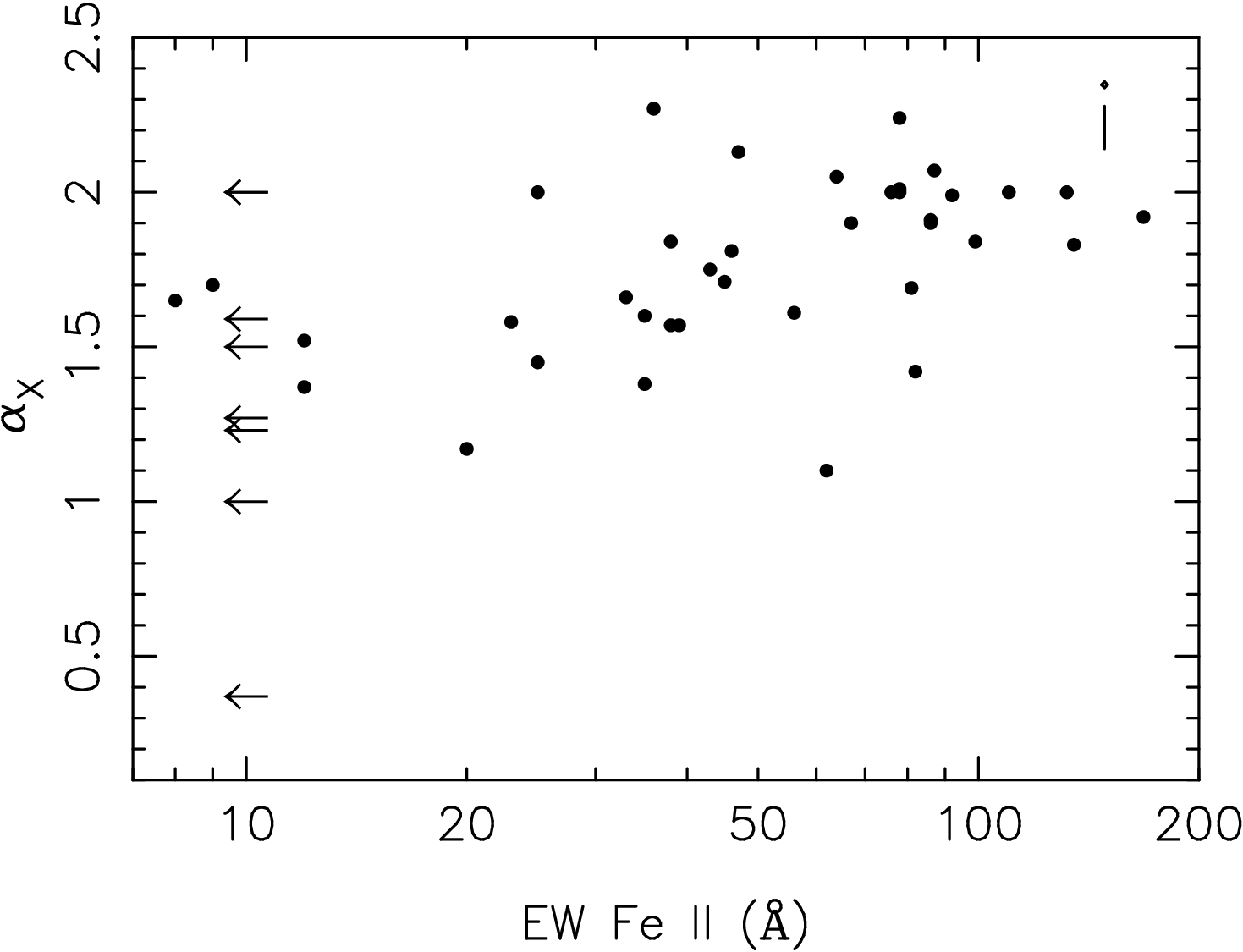}
}
\caption{
Correlation diagrams 
for various optical and X-ray parameters.  Arrows
indicate no detection in the case of \feii\ and 1$\sigma$ limits for
undetected \oiii. NGC 5905 has the steepest $\alpha_{X}$ and is
represented by an up arrow in panels g and h, it is also the lowest
luminosity object as is indicated by the left arrow in panels d and e.
}
\label{fig:correlation}
\end{figure*}

Correlation tests were first applied to pairs of parameters.  In each
trial two non-parametric correlation parameters were calculated,
namely the Spearman rank-order correlation coefficient $R_S$ and the
Kendall $\tau$ statistic (see \ncite{PRE92}); the results of the
Spearman rank-order tests are shown in Table~\ref{table:correlation}. Each entry
contains the $R_S$ (top) and the
number of objects (bottom)  included in the trial. Those with Spearman
rank probabilities $P_{S}<0.01$ are marked in {\bf bold}. At this
level of significance, $\ls 1$ spurious correlations might be
expected by chance from 55 trials.  Figures~\ref{fig:correlation} and
\ref{fig:outlyers} show correlation diagrams for various parameters.

In total, from 55 trials, 8 correlations were found with two-sided
probabilities $P_{S} < 0.01$. However, of these significant
correlations, two  merely represent trivial correlations between two
different measures of the same property (relative strength of \oiii\
to \hb, and \feii\ strength). All six non-trivial correlations are
shown in Figure~\ref{fig:correlation}.

Caution must be applied when interpreting the remaining correlations as
there is an inherent luminosity bias in the data. In the lowest
luminosity objects there is often significant contribution from the
host galaxy emission. The extreme cases of MCG+08-23-067, IC 3599, RX
J1258+23, Mrk 663 and NGC 5905 show Ca~\textsc{ii} H and K absorption
features from the host galaxy emission (see \ncite{VAU01}). As a
result these objects tend to show redder optical spectra (higher
$\alpha_{opt}$ values; see Figure~\ref{fig:correlation}e).  The other
potential difficulty with the current dataset is that the optical and
X-ray observation were separated by a time interval of roughly nine
years. As ultrasoft Seyferts are often highly variable this may introduce 
a large uncertainty in the derived values of $\alpha_{OX}$ for
individual sources, although the mean for the sample should
be robust. Also, the 
non-simultaneity of the X-ray and optical data may reduce the significance 
of some of the underlying X-ray/optical correlations.

The well-known anti-correlation between FWHM \hb\ and $\alpha_{X}$ is
only weak in the present data ($R_{S}=-0.31$; $P=0.03$). One reason
for the weakness of the correlation may be the non-simultaneity of the
data, as noted above. But it is also important to recognize that the
relation between \hb\ width and $\alpha_{X}$ is not necessarily
one-to-one, but probably results instead from a `zone of avoidance,'
i.e., virtually all ultrasoft Seyferts have relatively narrow \hb,
whereas  optically defined NLS1s can have a wide variety of X-ray
slopes.  This can be seen Figure~\ref{fig:correlation}h (see also
\ncite{BOL96}) where there is only one ultrasoft Seyfert (RX J1026+55)
with broad \hb. The absence of  ultrasoft Seyferts with broad \hb\ is
not the result of selection effects since the present sample is
complete in terms of soft X-ray flux and includes no obvious optical
selection bias.

It should also be noted that the correlation is stronger in the higher
luminosity sources, as was also the case in the \scite{GRU99} sample.
For the 25 objects with $\nu L_{1/4}< 10^{44.3}$~erg s$^{-1}$ the
correlation is weak ($R_{S}=-0.26; P_{S}=0.20$) whereas it for the 26
objects with $\nu L_{1/4} > 10^{44.3}$~erg s$^{-1}$ it is strong
($R_{S}=-0.58; P_{S}=0.002$).

There is no sigificant correlation between the width of \hb\ and the
strength of \feii. Such a correlation has been claimed by
\scite{WIL82}, \scite{ZHE90} and \scite{BOR92}, however, as discussed
in \scite{GAS85} and \scite{GAS00}, this is a result of how the
strength of \feii\ is measured. The quantity \feii/\hb\ has been seen
to correlate with FWHM \hb, but this is due to  EW \hb\ decreasing as
the line gets narrower. No correlation is seen if FWHM \hb\ is
compared directly with EW \feii. The lack of correlation between \hb\
width and \feii\ strength (when measured independently of \hb) has
been confirmed by \scite{GRU99}, \scite{VER01} (see their section
3.3.1) and the present work.

\subsubsection{Principal Component Analysis}

A Principal Component Analysis (PCA) was applied to the data in an attempt to 
separate independent sets of correlations. In essence, PCA defines a new 
coordinate system,
defined by a set of eigenvectors called the principal components,
which best describe the variance in the data. The first principal
component (PC1) explains the largest fraction of variance in the data,
the second (PC2) explains the largest fraction of the remaining
variance, and so on. The motivation behind PCA is to extract groups of
meaningful, independent correlations from a complex
dataset. \scite{FW99} and \scite{FRA92A} provide brief descriptions of
PCA as applied to quasar spectra.

Table~\ref{table:pca} reports the first three principal components of
an analysis of 8 parameters (as is standard in such analyses, the
ranked data were used to  reduce the effect of outlying objects and
allow for non-linear correlations). Only the 37 objects with
measurements  of all 8 parameters were used in the analysis  (this
necessarily excludes some of the most extreme objects with only upper
limits on \oiii\ emission).

The analysis was applied to various sets of input data, with the
number of input parameters varied from 7 to 12.  In each case the
first  two principal components explained greater than $50$\% of the total
variance. PC1 was most closely associated with $\alpha_{opt}$,
\oiii/\hb\ and $\nu L_{1/4}$ and least associated with FWHM \hb,
whereas PC2 appears dominated by the FWHM 
\hb--$\alpha_{X}$ relation.  PC3 appears to represent the variance in
$\alpha_{OX}$ and is the only remaining component to have an
eigenvalue above 1 (sometimes considered to be an indicator of the
significance of a component).  The ordering of PC2 and PC3 was
reversed in some of the tests, and they can be seen to contribute
almost equally to the total variance.  These results closely match
those of \scite{GRU99}. Indeed, when the PCA was repeated using
exactly the same input parameters as those used by \scite{GRU99} the
same PC1 was found. However, the non-simultaneity of the optical/X-ray
data, and the luminosity bias noted above, hamper the interpretation
of these principal components.

\begin{table}
  \centering
  \caption{Results of PCA. The relative significances of the
  first three principal components are listed, as well as their
  projections onto the original input parameters.}
  \begin{tabular}{@{}lrrr@{}}
                        &      PC1  &   PC2  &  PC3 \\
\hline
Eigenvalue              &     2.72  &  1.53  &  1.46  \\
Percentage variance     &    34.01  & 19.19  & 18.36  \\
Cumulative              &    34.01  & 53.21  & 71.56  \\
\hline
$\alpha_X$              &       0.52 &  0.59 &  -0.20   \\ 
$\nu L_{1/4}$           &       0.71 & -0.17 &   0.53   \\
FWHM \hb\               &       0.00 & -0.87 &  -0.18   \\
EW \hb\                 &       0.66 & -0.48 &  -0.18   \\
EW \feii\               &       0.62 &  0.24 &  -0.43   \\
\oiii/\hb\              &      -0.72 & -0.05 &   0.38   \\
$\alpha_{opt}$          &      -0.78 &  0.05 &  -0.25   \\
$\alpha_{OX}$           &       0.06 & -0.32 &  -0.83   \\
\hline
  \end{tabular}
\label{table:pca}
\end{table}

\subsection{Outlying Seyferts}
\label{sect:outlyers}

\begin{figure}
\centering
\includegraphics[width=6.2 cm, angle=270]{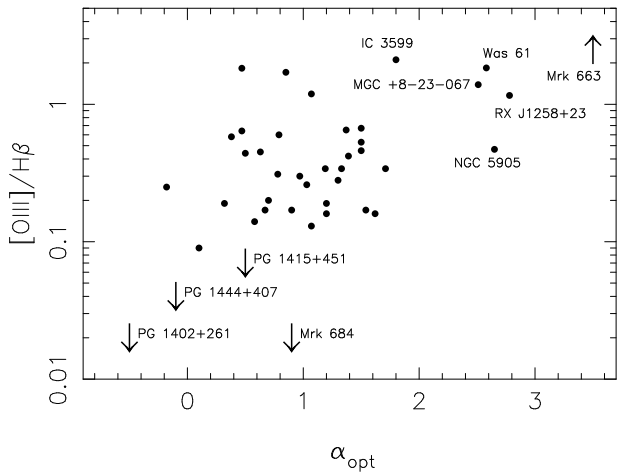}
\caption{Correlation diagram revealing objects which lie at the extremes
of the parameter space.} 
\label{fig:outlyers}
\end{figure}

Figure~\ref{fig:outlyers} shows a correlation diagram with individual
outlying objects marked. Four objects in particular stand out as
\feii-strong, \oiii-weak objects with blue optical spectra, these
objects are: PG 1402+261, PG 1415+451, Mrk 684 and PG 1444+407.  These
lie at the extreme negative end of the PC1 defined by BG92 (see BG92
and \ncite{KUR00}).  However, of these objects only Mrk 684 conforms
to the standard definition of an NLS1, with  FWHM \hb$<$2000~km
s$^{-1}$, although the other three have ultrasoft X-ray spectra.

At the other end of the parameter space there is a group of six
objects which stand out, these are: MCG+08-23-067, IC 3599,
Was 61, RX J1258+23, Mrk 663 and NGC 5905.  These all tend to possess
stronger forbidden than permitted line emission (\oiii/\hb$>$1)
and weak (and therefore redder, due to host galaxy emission) optical
continua when compared to soft X-rays ($\alpha_{OX}>1.5$).
However, with the exception of Mrk 663,
which shows no detectable \hb\ emission, none of these objects have
\oiii/\hb$>$3, a commonly used criterion for classifying Seyfert 2s.
Only two of these objects stand out from the $\alpha_{X}$
distribution; NGC 5905 had a remarkably steep soft X-ray spectrum
at the time of the \rosat\ survey, while Mrk 663 shows the hardest spectrum 
in the 1/4 keV sample. Mrk 663 is probably therefore the only candidate 
Seyfert 2 in the present sample besides the transient IC 3599.

Both IC 3599 and NGC 5905 have been identified as X-ray transient AGN
(e.g., \ncite{GRU95A}; \ncite{BRA95A}; \ncite{BAD96};
\ncite{KOM99B}). Both objects seem to have been detected by \rosat\
during some kind of nuclear outburst and have in recent years
`switched off,' showing a substantial drop in their X-ray and optical
continua as well as broad and high-excitation line emission.  IC 3599
was originally identified as a NLS1 (see \ncite{BRA95A}) but now shows
only very weak and narrow permitted lines, while NGC 5905 now
resembles an H~\textsc{ii} region more than a Seyfert galaxy. The
striking similarity of the optical spectrum of the other  weak-\hb\
objects to the two known transient AGN raises the interesting question
of whether these objects also exhibit X-ray transient behaviour
(as noted earlier the optical spectroscopy presented here was taken 9 years 
after the \rosat\ survey).  In this scenario the X-ray continuum has decreased
since the \rosat\ survey, along with the optical continuum and
permitted line emission\footnote{Mrk 663 was detected by the \rosat\
HRI during a  3.1~ksec pointed observation in 1995 at a count rate of
$\sim 0.2$~ct s$^{-1}$. This is comparable with the PSPC count rate
observed during the \rosat\ survey and suggests that Mrk 663 is a
persistent X-ray source.}.

\subsection{The Seyfert Galaxy Luminosity Function}
\label{sect:lum_function}

The information given in Tables~\ref{table:sample} and
\ref{table:opt_measurements} has been used to derive the  luminosity
function  of the 1/4~keV-selected Seyfert galaxy sample.  The
luminosity  function was calculated using the $1/V_{max}$ method of
\scite{SCH68}:

\[ \Phi(L) = \frac{1}{\Delta L} \sum_{i=1}^{N} \frac{1}{V_{max}}_{i},
\]

Here the luminosity $\nu L_{1/4}$ is taken to be the monochromatic
luminosity at 1/4~keV in the rest-frame of the source, in $\nu
L_{\nu}$ units (assuming 
spectral model comprising a $\alpha_{X}=2$ power-law modified by
Galactic absorption) .The $\nu L_{1/4}$ values are tabulated for
each source in  Table~\ref{table:opt_measurements}. For a survey at
1/4~keV, the $V_{max}$  calculation is greatly complicated by the
variation of the Galactic  foreground absorption across the sky.  A
similar problem was encountered by \scite{EDE99} when considering the
luminosity function of the AGN discovered in the \rosat\ WFC  survey
and here we employed the same approach in evaluating the effective
survey volume.

For each source, $V_{max}$ was estimated as:

\[ V_{max_{i}} = \int^{N_{Hlim}}_{0} \frac{1}{3}\Omega(N_{H}) (\frac{T(N_{H})}
{T(N_{H_{i}})})^{\frac{3}{2}} d_{i}^{3} 
(\frac{C_{i}}{C_{lim}})^{\frac{3}{2}}
 dN_{H}, \]

The function $\Omega(N_{H})$ represents the differential sky area as a
function of Galactic $N_{H}$. The
integral  $\int \Omega(N_{H}) dH_H$ represents the area of sky
actually surveyed whereas the integral  $\int \Omega(N_{H})
T(N_{H})^{\frac{3}{2}} dN_H$ gives the equivalent area
of  {\it unabsorbed} sky.  For the present survey the latter
integral is  $ 4.8  \times 10^{-2} $ steradians or 157 square  degrees
(whereas  integration  of  $\Omega(N_{H}$) between  the  same limits
gives $\sim$0.6 steradians). The integrations were carried out over the
northern hemisphere only, as the sample is restricted to $\delta \geq
0 \dg$.

\begin{figure}
\centering
\includegraphics[width=8.0 cm, angle=0]{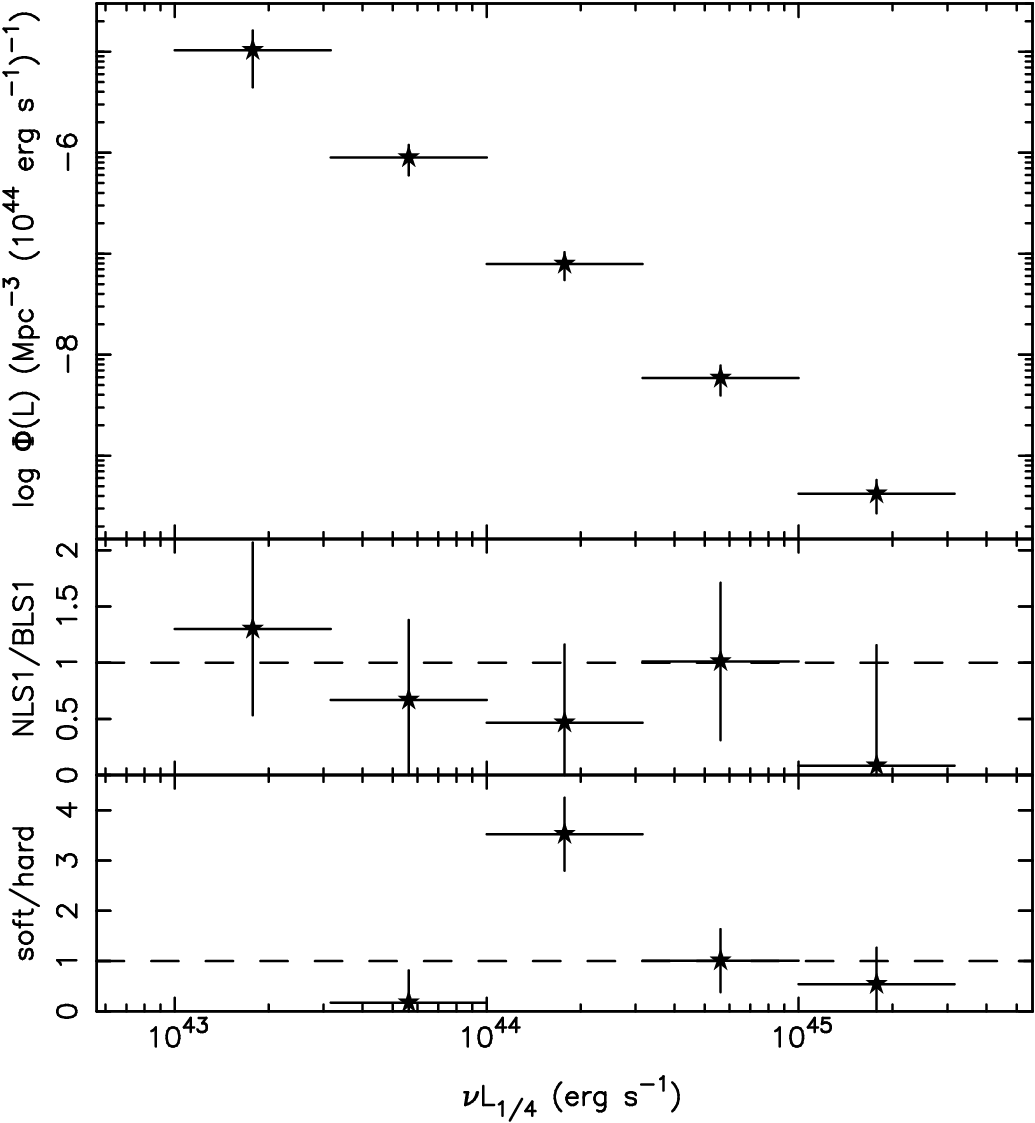}
\caption{
Luminosity function of the 1/4 keV-selected Seyfert galaxies. The top panel
shows the function for the complete sample.
The middle panel shows the ratio of luminosity functions when the
sample is divided into narrow- and broad-line Seyfert 1s.
The lower panel shows the ratio when the sample is divided on the
basis of X-ray spectral slope. The binning is every half-decade in luminosity.
}
\label{fig:lum_function}
\end{figure}

Figure~\ref{fig:lum_function} shows the resulting luminosity function binned 
into logarithmic luminosity intervals and and Table~\ref{table:lum_function}
lists the corresponding numerical information.  (Note this analysis
range excludes the sources NGC 4051 and NGC 5905 which have unusually low 
luminosities, $\nu L_{1/4} < 10^{42}$~erg s$^{-1}$.) 
The error bars were calculated using the prescription of
\scite{MAR85}.  The luminosity function we derive for full 
sample of 1/4 keV-selected Seyferts is entirely consistent with that 
previously calculated for the WFC-selected AGN (\ncite{EDE99}).

Next the sample was divided into subsamples and luminosity functions of
each subsample calculated. The ratio of the two luminosity functions
then gives the relative of space densities as a function of
luminosity. The middle panel of Figure~\ref{fig:lum_function} shows this
when the Seyferts were split into subsamples on
the basis of \hb\ width (i.e. NLS1 versus BLS1) or X-ray spectral 
slope (``ultrasoft'' versus ``normal'' spectrum objects) as discussed in
Section~\ref{sect:properties}.

\begin{table*}
\caption{Luminosity functions for the 1/4 keV sample. $\Phi(L)$ is in
units of Mpc$^{-3} (10^{44}$~erg s$^{-1})^{-1}$. \label{table:lum_function}}
\begin{tabular}{@{}lccccccc@{}}
\hline
$\log(\nu L_{1/4})$	&\multicolumn{3}{c}{All Seyferts}			&								&		\\
(erg s$^{-1}$)	&$\Phi(L)$			&\vvm		&	$N$	&				&		&		&	\\
(1)		&	(2)			&	(3)	&	(4)	&	(5)			&	(6)	&	(7)	&	(8)	\\
\hline																		
43.0-43.5	&$2.2\pm1.3\times10^{-6}$	&$0.58\pm0.13$	&	5	&				&		&		&	    	\\
43.5-44.0	&$6.1\pm2.0\times10^{-7}$	&$0.41\pm0.08$	&	12	&				&		&		&		\\
44.0-44.5	&$1.7\pm0.5\times10^{-7}$	&$0.53\pm0.08$	&	13	&				&		&		&		\\
44.5-45.0	&$4.0\pm1.5\times10^{-8}$	&$0.58\pm0.08$	&	13	&				&		&		&		\\
45.0-45.5	&$9.1\pm3.3\times10^{-9}$	&$0.58\pm0.10$	&	9	&				&		&		&		\\
\hline
                &\multicolumn{3}{c}{NLS1s}					&\multicolumn{3}{c}{BLS1s}		&	NLS1/	\\
              	&$\Phi(L)$			&	\vvm	&	$N$	&$\Phi(L)$			&\vvm		&	$N$	&BLS1	\\
\hline
43.0-43.5	&$6.4\pm5.1\times10^{-7}$	&$0.46\pm0.20$	&	2	&$4.9\pm3.6\times10^{-7}$	&$0.54\pm0.20$	&	2	&	$1.3\pm0.8$	\\
43.5-44.0	&$2.2\pm1.2\times10^{-7}$	&$0.19\pm0.11$	&	6	&$3.3\pm1.7\times10^{-7}$	&$0.57\pm0.13$	&	5	&	$0.7\pm0.7$	\\
44.0-44.5	&$5.4\pm3.2\times10^{-8}$	&$0.68\pm0.17$	&	3	&$1.2\pm0.4\times10^{-7}$	&$0.48\pm0.09$	&	10	&	$0.5\pm0.7$	\\
44.5-45.0	&$1.8\pm0.8\times10^{-8}$	&$0.52\pm0.12$	&	6	&$2.0\pm0.9\times10^{-8}$	&$0.59\pm0.12$	&	6	&	$1.0\pm0.7$	\\
45.0-45.5	&$7.1\pm7.1\times10^{-10}$	&$0.53\pm0.29$	&	1	&$8.4\pm3.2\times10^{-9}$	&$0.58\pm0.10$	&	8	&	$0.1\pm1.1$	\\
\hline		
                &\multicolumn{3}{c}{ultrasoft}					&\multicolumn{3}{c}{hard}		&	soft/	\\
              	&$\Phi(L)$			&	\vvm	&	$N$	&$\Phi(L)$			&\vvm		&	$N$	&hard	\\
\hline		
43.0-43.5	&	---		     	& ---         	&	0	&$2.2\pm1.3\times10^{-6}$	&$0.58\pm0.13$	&	5	&	---        	\\
43.5-44.0	&$9.0\pm4.7\times10^{-8}$	&$0.11\pm0.14$	&	4	&$5.2\pm2.0\times10^{-7}$	&$0.55\pm0.10$	&	8	&	$0.2\pm0.6$	\\
44.0-44.5	&$1.3\pm4.7\times10^{-7}$	&$0.57\pm0.10$	&	9	&$3.7\pm2.3\times10^{-8}$	&$0.54\pm0.17$	&	3 	&	$3.5\pm0.7$	\\
44.5-45.0	&$2.3\pm8.1\times10^{-8}$	&$0.52\pm0.09$	&	10	&$2.3\pm1.2\times10^{-8}$	&$0.61\pm0.14$	&	4	&	$1.0\pm0.6$	\\
45.0-45.5	&$3.2\pm1.8\times10^{-9 }$	&$0.56\pm0.14$	&	4	&$5.9\pm2.8\times10^{-9}$	&$0.59\pm0.13$	&	5	&	$0.5\pm0.7$	\\
\hline

\end{tabular}
\end{table*}

The ratio of the space densities of NLS1s and  BLS1s is formally consistent
with a constant over the range  of luminosities probed.   It is somewhat
more difficult to interpret the incidence of ultrasoft Seyferts. In the
present sample there are no such objects in the luminosity range
$\log(\nu L_{1/4}) = 43.0-43.5$ and the next highest luminosity bin
there is the suggestion of incompleteness (see 
Table~\ref{table:lum_function}).  If we focus on the 
$\log(\nu L_{1/4}) = 44.0-45.5$ range  the relative number of
ultrasoft objects appears to decrease with luminosity.
Interestingly a deficit of ultrasoft Seyferts at high luminosities
might be expected due to the effects of the `$K$-correction.'
Basically the steep spectrum objects will be harder to detect at higher 
redshifts due the decrease in flux {\it observer's frame} flux by a 
factor $(1+z)^{\alpha-1}$, where $\alpha$ is the spectral index; in effect
the bright soft X-ray emission is being redshifted out of the observing band.
Unfortunately the statistical limitations of the present sample
mitigate against a more detailed quantitative analysis of this point.

To date only two NLS1s  have been reported  at a redshift  $z>0.5$ namely the
quasar  E~1346+266 with $z=0.92$, which shows  optical and X-ray  spectra
characteristic  of NLS1s (\ncite{PUC94}), and RX~J2241-44 at $z=0.55$
(\ncite{GRU98}). Of the 69
spectroscopically  identified AGN in the deep   \rosat\   surveys   of
the   Lockman   Hole   (\ncite{HAS98}; \ncite{SCH98}) only one can be
postively identified with an NLS1 (at $z=0.462$; \ncite{HAS00}).

\section{discussion}
\label{sect:discussion}

This paper presents a new sample of soft X-ray bright Seyfert
galaxies. The sample is the first to be selected in the  
$\sim 1/4$~keV \rosat\ band and is statistically complete.

The analysis reported in Section~\ref{sect:correlation} reveals
correlations between the soft X-ray spectral slope $\alpha_{X}$ and
the optical \oiii\ and \feii\ line strengths, and the previously known
relation between 
$\alpha_{X}$ and FWHM \hb. These relationships appear more like zones
of avoidance than linear correlations (see also Section 4.2 of
\ncite{LAW97}). A steep X-ray slope (e.g., $\alpha_{X} \gs 1.7$) seems
to be a prerequisite for strong \feii\ (EW$>$100\AA) and is
almost invariably accompanied by strong, narrow \hb\ (FWHM \hb$\ls$3000~km
s$^{-1}$; Peak \oiii/\hb$\ls$0.5). Given the  non-simultaneity of
the optical/X-ray data and the significant soft X-ray variability 
exhibited by the source population, these relationships must be intrinsically
strong to remain apparent in the current analysis.

The link between $\alpha_{X}$ and the optical parameters
therefore seems to be driving the correlations responsible for the
first principal component of BG92 (which is dominated by the strength
of \oiii\ and \feii\ and to a lesser extent FWHM \hb). This is
supported by the strong correlation between BG92 PC1 and X-ray slope
found by \scite{BB98}. Thus it appears that the high energy continuum,
and the slope of the soft X-ray spectrum in particular, are driving many of the
relationships observed in optical correlation analyses.   This is
perfectly reasonable since the physical conditions of the line-emitting
plasma will be strongly effected by the incident high-energy continuum.  A
similar conclusion was reached by \scite{KUR00b}, who explained the
different ultraviolet emission properties of a small sample of NLS1s
as an effect of the different photoionising continuum in these objects.

The ultrasoft Seyferts, defined by their steep X-ray spectra, are also
characterised by extreme values of other observables. The high energy
continuum is more closely linked to the underlying accretion processes
than are the optical line ratios/widths. However, it is the
overlapping class of optically defined NLS1s that have received
particular attention, largely on the basis of exceptional X-ray
properties in many ultrasoft examples (e.g., \ncite{BOL93};
\ncite{POU95}). It is now clear that Seyferts selected on the basis of
\hb\ line width span a range of observed properties, many showing
spectra otherwise characteristic of `normal' Seyfert 1s; the present
analysis confirms that optically defined NLS1s do not show enhanced
\feii\ emission compared to BLS1s, as has been previously suggested.
The shape of the high energy continuum is therefore a much more direct
indicator of the primary driving processes in AGN than is the width of
\hb.  The present sample  then represents the ideal one in which to
search for extremes of behaviour.

Seyferts with weak permitted line emission (e.g., \hb\ and \feii)
sit at at one end of the observed correlations. These are generally
rare in soft X-ray surveys, yet the present sample contains a number
of such objects.  Two of these are known to be X-ray transient and it
is plausible that other \hb-weak Seyferts in the sample are transient
in nature.  This hypothesis clearly needs to be tested with repeat
X-ray and optical observations.  If a significant number of the 1/4
keV-selected sources are  confirmed as X-ray transients then this would 
suggest that, at least for a significant fraction of the population, the
ultrasoft Seyfert state may be a relatively short-lived one.

The alternative explanation, namely that these objects have not changed 
significantly since the \rosat\ survey, is equally interesting.   
\scite{PUC98A} and
\scite{GRU98A} discuss soft X-ray selected Seyferts which appear to
show optical reddening but no X-ray absorption. One possible solution
is that these objects all contain dusty, ionised gas along the
line-of-sight. The small dust grains are needed to redden the optical
continuum, while the surrounding gas is ionised and so produces no
soft X-ray absorption. However the lack of any correlations involving
the Balmer decrement (indicative of reddening by dust) suggests
this may not be the solution.  It is difficult to see how  Seyfert
galaxies that are bright in the 1/4 keV band, and so presumably
contain little or no absorption, yet have weak permitted lines, are
compatible with the standard Seyfert unification scheme
(e.g., \ncite{ANT93}).

\section{Conclusions}

This paper presents the first complete sample of AGN selected on the
basis of their $\sim 1/4$~keV flux.  The sample comprises 54 Seyfert
galaxies, 20 of which are identified as NLS1s on the basis that FWHM
\hb$\le$2000~km s$^{-1}$.

The well-known anti-correlation between the strengths of \feii\ and
\oiii\ emission is detected, along with correlations between the soft
X-ray slope and \oiii\ emission, \feii\ strength and \hb\ width. These
suggest that the so-called `primary eigenvector' found in other
samples is a direct result of these optical lines being correlated
with the soft X-ray spectral slope.  Particularly interesting are the
objects that lie at one end of the correlations, with relatively
strong \oiii\ emission and weak optical continua, as these are rare in
most other samples of soft X-ray selected AGN. Two of these objects
are known to be X-ray transients, future observations are needed to
confirm the nature of the other objects.  The luminosity function for
$\sim 1/4$~keV selected AGN is presented and these data are used to
examine the relative densities of Seyferts as a function of \hb\ line
width and X-ray slope.

\section*{ Acknowledgments }

The authors would like to thank Martin Ward and Paul O'Brien for
useful discussions throughout the course of this work, the referee,
Dirk Grupe, for a constructive referee's report, and the staff
of the ING and Lick Observatories.  This research made use of data
obtained  from the NASA/IPAC Extragalactic Database (NED), provided by
NASA/JPL under  contract with Caltech, from the Leicester Database and
Archive Service (LEDAS) at the Department  of Physics and Astronomy,
Leicester University, UK, and  from the Set of Identifications,
Measurements and Bibliography for  Astronomical Data (SIMBAD),
maintained by the Centre de Donnees astronomiques de Strasbourg.  SV
acknowledges support from PPARC.  The William Herschel Telescope is
operated on the island of La Palma by the Isaac Newton Group in the
Spanish Observatorio del Roque de los Muchachos of the Instituto de
Astrofisica de Canarias, and the Lick Observatory 120\arcs is operated
by the University of California.


\appendix

\section{Soft X-ray bright subsamples}
\label{append:sample}
Table~\ref{table:bl_sample} lists the subsamples of
BL Lacertae objects and Clusters and Table~\ref{table:star_sample} lists
the subsample of Galactic objects selected as described in 
Section~\ref{sect:sample}.
\begin{table*}
 \caption{The BL Lac and cluster samples. 
The columns provide  the following information: (2) The source
name;  (3)  the source type;  (4) the redshift; (5)  and  (6) the  right
ascension and declination as tabulated  in the RBSC; (7) the full-band
count rate;  (8) the hardness  ratio $HR1$; (9) the Galactic  column; (10)  
and (11) the derived C- and H-band count rates.}
 \begin{tabular}{@{}llrrrrrrrrr@{}}
\hline
 	&	 		&		&		&	R.A.			&	Dec.			&	$T$	 		&	 	 	&	$N_{H}$		&	$C$	&	$H$	\\
No	&	Name		&	Type	&	$z$	&	(J2000)			&	(J2000)			&	(ct/s)			&	$HR1$		&($10^{20}$ cm$^{-2}$)	&	(ct/s)	&	(ct/s)	\\
(1)	&	(2)		&	(3)	&	(4)	&	(5)			&	(6)			&	(7)			&	(8)		&	(9)		&	(10)	&	(11)	\\
\hline
1	&	1ES 0927+5	&	BL Lac	&	0.190	&	09	30	37	&	49	50	28	&	2.15$\pm$0.07	&	0.06$\pm$0.03	&	1.40		&	1.012	&	1.142	\\
2	&	RX J1008+47	&	BL Lac	&	0.343	&	10	08	11	&	47	05	26	&	1.10$\pm$0.05	&	-0.29$\pm$0.04	&	0.88		&	0.711	&	0.392	\\
3	&	B3 1009+427	&	BL Lac	&	0.364	&	10	12	44	&	42	29	58	&	0.66$\pm$0.04	&	0.02$\pm$0.05	&	1.09		&	0.325	&	0.339	\\
4	&	GB 1011+496	&	BL Lac	&	0.2	&	10	15	04	&	49	26	04	&	1.94$\pm$0.07	&	-0.38$\pm$0.03	&	0.79		&	1.340	&	0.602	\\
5	&	RX J1016+41	&	BL Lac	&	0.281	&	10	16	16	&	41	08	17	&	0.48$\pm$0.03	&	-0.08$\pm$0.06	&	1.14		&	0.259	&	0.221	\\
\hline
6	&	1ES 1028+511	&	BL Lac	&	0.361	&	10	31	18	&	50	53	40	&	4.46$\pm$0.09	&	-0.26$\pm$0.02	&	1.17		&	2.812	&	1.652	\\
7	&	87GB 10553	&	BL Lac	&	0.144	&	10	58	37	&	56	28	16	&	0.46$\pm$0.03	&	-0.48$\pm$0.05	&	0.67		&	0.340	&	0.119	\\
8	&	FIRST J110	&	BL Lac	&	--	&	11	00	21	&	40	19	33	&	0.61$\pm$0.04	&	-0.35$\pm$0.06	&	1.16		&	0.410	&	0.198	\\
9	&	Mrk 421		&	BL Lac	&	0.03	&	11	04	27	&	38	12	31	&	26.57$\pm$0.28	&	-0.21$\pm$0.01	&	0.71		&	16.075	&	10.495	\\
10	&	87GB 11051	&	BL Lac	&	--	&	11	07	48	&	15	02	17	&	0.45$\pm$0.04	&	-0.19$\pm$0.08	&	1.49		&	0.270	&	0.184	\\
\hline
11	&	87GB 11429	&	BL Lac	&	0.138	&	11	17	06	&	20	14	10	&	4.34$\pm$0.12	&	-0.01$\pm$0.02	&	1.36		&	2.191	&	2.147	\\
12	&	Mrk 180		&	BL Lac	&	0.046	&	11	36	26	&	70	09	32	&	4.53$\pm$0.08	&	-0.20$\pm$0.01	&	1.42		&	2.719	&	1.813	\\
13	&	87GB 11333	&	BL Lac	&	0.135	&	11	36	30	&	67	37	08	&	1.91$\pm$0.06	&	0.07$\pm$0.03	&	1.35		&	0.886	&	1.020	\\
14	&	Ton 116		&	BL Lac	&	--	&	12	43	12	&	36	27	42	&	1.30$\pm$0.05	&	-0.36$\pm$0.03	&	1.36		&	0.882	&	0.415	\\
15	&	PG 1246+586	&	BL Lac	&	--	&	12	48	18	&	58	20	31	&	0.52$\pm$0.04	&	-0.42$\pm$0.06	&	1.12		&	0.366	&	0.150	\\
\hline
16	&	1ES 1255+244	&	BL Lac	&	0.141	&	12	57	31	&	24	12	45	&	0.93$\pm$0.05	&	-0.10$\pm$0.05	&	1.26		&	0.513	&	0.420	\\
17	&	RX J1302+50	&	BL Lac	&	0.688	&	13	02	55	&	50	56	21	&	0.52$\pm$0.04	&	-0.05$\pm$0.06	&	1.18		&	0.274	&	0.248	\\
18	&	RX J1341+39	&	BL Lac	&	0.163	&	13	41	04	&	39	59	42	&	0.76$\pm$0.04	&	-0.10$\pm$0.04	&	0.80		&	0.416	&	0.340	\\
19	&	RX J1420+53	&	BL Lac	&	--	&	14	20	24	&	53	34	03	&	0.29$\pm$0.02	&	-0.92$\pm$0.03	&	1.18		&	0.275	&	0.011	\\
20	&	RX J1442+58	&	BL Lac	&	0.638	&	14	22	39	&	58	01	59	&	2.0$\pm$0.06		&	-0.09$\pm$0.02	&	1.32		&	1.090	&	0.910	\\
\hline
21	&	H 1426+428	&	BL Lac	&	0.129	&	14	28	32	&	42	40	28	&	4.20$\pm$0.09	&	-0.06$\pm$0.02	&	1.38		&	2.228	&	1.975	\\
22	&	PG 1437+398	&	BL Lac	&	--	&	14	39	17	&	39	32	48	&	1.43$\pm$0.05	&	-0.35$\pm$0.03	&	1.05		&	0.966	&	0.465	\\
23	&	[WB92] 144	&	BL Lac	&	--	&	14	48	01	&	36	08	33	&	0.60$\pm$0.04	&	-0.54$\pm$0.05	&	1.05		&	0.460	&	0.137	\\
24	&	1ES 1533+535	&	BL Lac	&	0.89	&	15	35	01	&	53	20	42	&	1.43$\pm$0.04	&	0.00$\pm$0.02	&	1.32		&	0.716	&	0.716	\\
25	&	RX J1631+42	&	BL Lac	&	0.468	&	16	31	24	&	42	16	56	&	0.53$\pm$0.03	&	-0.07$\pm$0.04	&	1.04		&	0.282	&	0.245	\\
\hline
1	&	ABELL 1656	&	Cluster	&	0.023	&	12	59	47	&	27	56	35	&	11.70$\pm$0.13	&	0.33$\pm$0.05	&	0.92		&	3.920	&	7.781	\\
2	&	ABELL 1795	&	Cluster	&	0.062	&	13	48	52	&	26	35	40	&	3.77$\pm$0.09	&	0.28$\pm$0.02	&	1.19		&	1.358	&	2.415	\\
3	&	ABELL 1914	&	Cluster	&	0.171	&	14	26	01	&	37	49	35	&	0.91$\pm$0.04	&	0.25$\pm$0.04	&	0.95		&	0.342	&	0.571	\\
4	&	ABELL 2129	&	Cluster	&	0.030	&	16	28	37	&	39	32	48	&	4.5$\pm$0.08		&	0.30$\pm$0.01	&	0.86		&	1.575	&	2.925	\\
\hline		
\end{tabular}
\label{table:bl_sample}
\end{table*}

\begin{table*}
 \caption{The Galactic sample.
The columns  provide the following  information: (2) the source name;
(3) and (4) the right ascension and declination as tabulated in the
RBSC; (5) the full-band count rate; (6) hardness ratio; (7) Galactic
column; (8) and (9) derived C- and H-band count rates; (10 and (11)
the source type and optical magnitude (V-band where available). 
}
 \begin{tabular}{@{}llrrrrrrrrr@{}}
\hline
 	& 	 		& R.A.			&	Dec			&	$T$		&			&	$N_{H}$	&	$C$	&	$H$		&		&	\\
No.	& 	Name 		& (J2000)		&	(J2000)			&	(ct/s)		&	$HR1$		&(10$^{20}$ cm$^{-2}$)	&(ct/s)	&	(ct/s)		&	Type	&	Mag	\\
(1) 	& 	(2) 		& (3)			&	(4)			&	(5)		&	(6)		&	(7)	&	(8)	&	(9)		&	(10)	&	(11)\\
\hline
1	&	HD76943 B	& 09	00	38	&	41	47	02	&	0.68$\pm$0.05	&	-0.46$\pm$0.05	&	1.29	&	0.493	&	0.183	&	F5V		&	4.0\\
2	&	HR 3922		& 09	57	12	&	57	25	12	&	0.62$\pm$0.03	&	0.09$\pm$0.05	&	1.08	&	0.280	&	0.336	&	G5III		&	6.0\\
3	&	G 196-3		& 10	04	21	&	50	23	17	&	0.69$\pm$0.04	&	-0.26$\pm$0.05	&	0.82	&	0.434	&	0.255	&	M3Ve		&	13.3\\
4	&	RE J1032+53	& 10	32	10	&	53	29	40	&	5.07$\pm$0.10	&	-1.00$\pm$0.00	&	1.16	&	5.069	&	0.000	&	WD/DA		&	14.5\\
5	&	RE J1043+49	& 10	43	11	&	49	02	27	&	0.93$\pm$0.05	&	-1.00$\pm$0.00	&	1.32	&	0.930	&	0.000	&	WD		&	16.1\\
\hline
6	&	FH Uma		& 10	47	10	&	63	35	22	&	0.26$\pm$0.02	&	-0.97$\pm$0.01	&	1.06	&	0.253	&	0.004	&	CV/AM Her	&	19.4\\
7	&	EK Uma		& 10	51	35	&	54	04	37	&	1.11$\pm$0.05	&	-0.99$\pm$0.00	&	0.98	&	1.103	&	0.006	&	CV/AM Her	&	18.0\\
8	&	DM Uma		& 10	55	43	&	60	28	10	&	0.90$\pm$0.04	&	-0.19$\pm$0.04	&	0.68	&	0.534	&	0.363	&	K0/RSCVN	&	9.3\\
9	&	LB 1919		& 10	59	16	&	51	24	52	&	4.18$\pm$0.09	&	-0.99$\pm$0.00	&	1.06	&	4.159	&	0.021	&	WD/DA		&	16.8\\
10	&	HD 95559	& 11	02	02	&	22	35	46	&	1.01$\pm$0.07	&	0.01$\pm$0.07	&	1.41	&	0.498	&	0.509	&	G5III		&	8.9\\
\hline
11	&	AN Uma		& 11	04	25	&	45	03	19	&	1.77$\pm$0.07	&	-0.94$\pm$0.01	&	1.11	&	1.718	&	0.053	&	CV/AM Her	&	15.5\\
12	&	Ton 61		& 11	12	38	&	24	09	09	&	0.61$\pm$0.05	&	-0.96$\pm$0.03	&	1.23	&	0.596	&	0.012	&	WD/DA		&	15.1\\
13	&	PG 1234+48	& 12	36	45	&	47	55	30	&	1.45$\pm$0.06	&	-1.00$\pm$0.00	&	1.19	&	1.451	&	0.000	&	WD/sd:B		&	14.4\\
14	&	31 Com		& 12	51	42	&	27	32	21	&	1.11$\pm$0.09	&	0.31$\pm$0.07	&	0.90	&	0.382	&	0.724	&	G0IIIp		&	4.9\\
15	&	Gl 490		& 12	57	40	&	35	13	34	&	0.84$\pm$0.04	&	-0.22$\pm$0.04	&	1.23	&	0.511	&	0.327	&	M0.5e		&	10.3\\
\hline
16	&	RS CVn		& 13	10	36	&	35	56	04	&	0.89$\pm$0.04	&	0.04$\pm$0.04	&	1.03	&	0.425	&	0.461	&	F4v+RSCVn	&	8.2\\
17	&	HZ 43		& 13	16	21	&	29	05	55	&	72.84$\pm$0.32&	-0.99$\pm$0.00	&	1.08	&	72.470	&	0.364	&	WD/DA		&	12.7\\
18	&	HD 116204	& 13	21	32	&	38	52	49	&	1.32$\pm$0.05	&	0.22$\pm$0.03	&	0.97	&	0.513	&	0.803	&	G8III/RSCVn	&	7.3\\
19	&	GJ 3789		& 13	31	46	&	29	16	31	&	0.41$\pm$0.03	&	-0.32$\pm$0.07	&	1.16	&	0.271	&	0.139	&	M4		&	12.0\\
20	&	BD +23 258	& 13	32	41	&	22	30	07	&	0.84$\pm$0.05	&	-0.13$\pm$0.05	&	1.44	&	0.477	&	0.367	&	KV		&	9.6\\
\hline
21	&	HR 5110		& 13	34	47	&	37	10	59	&	2.77$\pm$0.08	&	-0.07$\pm$0.02	&	0.92	&	1.481	&	1.287	&	F2IV/RSCVn	&	4.9\\
22	&	RX J1342+28	& 13	42	10	&	28	22	50	&	0.58$\pm$0.04	&	-0.95$\pm$0.02	&	1.14	&	0.561	&	0.014	&	SSS/Glob Cl	&	--\\
23	&	HD 123351	& 14	06	26	&	30	50	51	&	0.81$\pm$0.05	&	0.00$\pm$0.05	&	1.25	&	0.405	&	0.405	&	K0 		&	7.6\\
24	&	HR 5404		& 14	25	11	&	51	51	09	&	2.00$\pm$0.06	&	0.02$\pm$0.02	&	1.29	&	0.978	&	1.018	&	F7V		&	4.1\\
25      &       RX J1605+54     & 16    05      18      &       54      21      01      &       0.50$\pm$0.02 &       -0.11$\pm$0.04  &       1.23    &       0.278   &       0.223   &       M      &                19.1\\ 
\hline
26	&	HD 146696	& 16	15	43	&	44	33	10	&	0.58$\pm$0.04	&	0.07$\pm$0.06	&	1.15	&	0.271	&	0.312	&	G0 		&	8.9\\
27	&	GJ 9557A	& 16	19	55	&	39	42	23	&	0.64$\pm$0.03	&	-0.16$\pm$0.04	&	0.93	&	0.371	&	0.269	&	F0V		&	5.5\\
\hline
\end{tabular}
\label{table:star_sample}
\end{table*}

\section{New optical observations of the Seyfert sample}
\label{append:optical_data}

The new optical spectra   were   obtained   with   the  4.2m   William
Herschel Telescope    (WHT)    at    the Observatario Roque de  los
Muchachos on the Island of  La Palma, on the nights of  1999 March 21
and 22, and  with the 3.0m Shane  Telescope at Lick Observatory,
Mt. Hamilton, California from 1999 May 21--23.

The  WHT  observations used  the  ISIS spectrograph\footnote{See  {\tt
http://www.ing.iac.es/$\sim$bgarcia/isis\_new/isis\_home.html}} with
the R300B  grating and a  EEV 42  CCD camera on  the blue arm and the
R158R grating with a TEK 2  CCD camera on the red arm. The combination
of blue and red spectra provided coverage from 3600\AA~to 9000\AA.
The Shane data were obtained  with the Kast spectrograph\footnote{See
{\tt http://www.ucolick.org/$\sim$mountain/mthamilton/techdocs/
instruments/kast/kast\_index.html}} using grism \#2 and grating \#6 in
the blue and red arms respectively, both with Reticon 1200x400  CCD
cameras.  This set-up provided similar wavelength coverage
(3500\AA--9000\AA) to the WHT data.

The  data were  extracted following standard  procedures using
IRAF\footnote{IRAF is the  Image Reduction and Analysis Facility and
is written and  supported by  the IRAF  programming group  at the
National   Optical   Astronomy   Observatories   (NOAO)   in   Tucson,
Arizona.   See    {\tt   http://iraf.noao.edu/iraf/web/}}.
Individual frames were bias subtracted and flat-field
corrected. Spectra were traced on the CCD using a low-order polynomial
and then optimally extracted (\ncite{HOR86}) using a variable
extraction slit-width (typically 4\arcs). Background regions for sky
subtraction  were located where possible either side of the target
spectrum. Sky-line and cosmic ray removal occurs during the extraction
procedure. Extracted spectra were wavelength calibrated, corrected for
atmospheric extinction and  flux calibrated by comparison with a
photometric standard.

Multiple exposures of  the same  object were combined  where possible
prior to extraction  in order to  increase the  signal-to-noise ratio
and help remove cosmic ray contamination.  However, this was not
possible in cases where only one exposure was taken or when target
source image moved across the CCD chip between exposures, in which
case  the data were extracted separately from each frame and then
combined.

The  absolute fluxes of  the separate  blue and  red spectra  for each
object generally match to within 10 per cent. One source, RX J1619+40, 
was observed  at both  observatories to allow a
test of the  flux calibration.  The two sets of spectra match to
within 20 per cent over the useful wavelength range.  Given that  the
source  may have varied  between the  two observations (separated  by
2  months) it  seems  reasonable to  conclude that  the absolute  flux
calibration of  these data  are good  to within 20 per cent, a value
which is  typical of optical spectroscopy obtained under
non-photometric conditions.  The spectral  resolution of the data were
estimated from  fits to the  intrinsically narrow arc lamp  lines. For
the WHT  data the  blue spectra have  a FWHM $\sim 3.5$~\AA\ and  the
red have FWHM $\sim 5$~\AA, the Shane data  have FWHM of 4~\AA\ and
8~\AA\ in the blue and red respectively. These correspond to a
velocity width $\sim 200$~km/s at \oiii\ $\lambda 5007$.

Before any measurements were taken  from the optical data, the red and
blue  spectra for  each object  were combined  into one  spectrum.
The data were convolved with a Gaussian (of width smaller than the
spectral resolution) in order to smooth out any remaining bad pixels
(but  without degrading   the   spectral  resolution).  The blue data
were then  scaled to match the  flux of the red data in the region of
overlap, which was typically 350~\AA\ wide for the Shane data but only
$\ls  100$~\AA\ for  the WHT  data.  (The red data were chosen as the 
flux norm simply because the signal-to-noise is higher in the red arm 
than it is in the blue.) The data were then  combined and averaged  in the
overlap region to produce a single, continuous blue-red spectrum.

Due  to the  small overlap  between blue  and red  spectra in  the WHT
observations, the  normalisation between blue  and red is  rather less
accurate  than  in the  case  of the  Shane  data,  and forcing
continuity between blue and red  may introduce a systematic error in
the shape of the spectrum in the overlapping region. For most sources
this has no  effect on the derived  spectral properties, but if,  as
in the case of RX J1054+48, the \hb\ line falls in the overlap region
then the detailed  profile of  the  line  will be  distorted
somewhat. This  is unavoidable as the  line falls at the far end  of
each spectrum, where the calibration  is worst,  and the small
overlap means  that neither spectrum contains the complete line
profile.

\subsection{Optical measurements}
\label{section:opt_measurements}

The optical spectra were used to measure the basic optical properties
of each object.  Redshifts were obtained from the identification papers
or derived from fitting a Gaussian to the upper half of the observed \oiii\
$\lambda$5007 line in each spectrum. (The centroid of a Gaussian fit
and the line peak were generally consistent within the limits set by 
the spectral resolution of the data.)

\subsubsection{ {\rm Fe~\textsc{ii}} subtraction}

In many of these spectra there  is a clear contribution from blends of
\feii\  line  emission   on  both  the  blue  and   red  sides
of  the \hb-\oiii\ complex.  An automatic fitting routine  developed by
\scite{GOA00} (based on the method of BG92) was used to
estimate the strength of these  lines, and remove any \feii
~contamination from the \hb\ region.

The method  involves comparing a template optical \feii\ line spectrum
with the  observed spectra. The template used  in the present work was
the  same as that of BG92, namely  the \feii\ lines of the
bright NLS1  I Zw 1, which shows very  strong and narrow permitted
\feii\ emission (e.g., \ncite{PHI78}; \ncite{OKE79}).  The  template was
redshifted to match  the source, smoothed (by convolving with a
Gaussian), and scaled to fit  the data either side of \hb.  (The
convolution  was carried  out over  a wavelength  range broad enough
that edge  effects are not significant). The  four free parameters in
the  fit were: the strength  of the  \feii\ emission,  the width  of
the convolution kernel (broadening parameter),  and two parameters
describing  a  1st  order polynomial used  to  model  the continuum
local to  \hb. (Where possible the region around He~\textsc{ii}
$\lambda 4686$ was ignored in the fit, although in some cases the
He~\textsc{ii} emission may be broad and this will effect the fit.)
The best-fit optical \feii\ spectrum was then subtracted from the data.
The \feii\ flux was measured between $\lambda$4434 and $\lambda$4684 as
in BG92. (Note that when \feii\ measurements were taken from
the literature, they were scaled by a factor, derived from the I Zw 1
template, to take into account the different in wavelength ranges used
to define \feii\ flux.)

\begin{figure*}
\centering
\hbox{
\includegraphics[width=4.1 cm, angle=270]{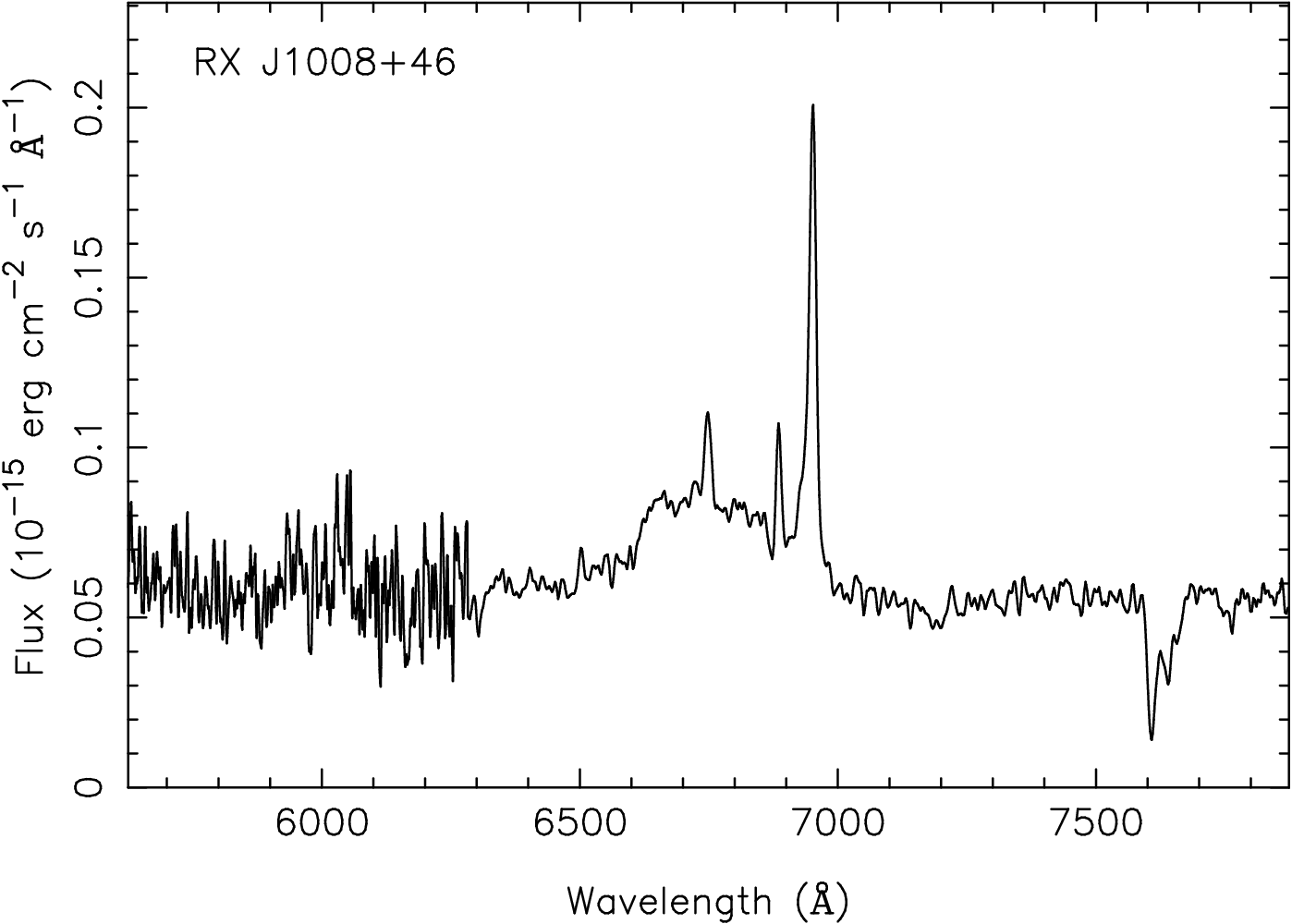} 
\includegraphics[width=4.1 cm, angle=270]{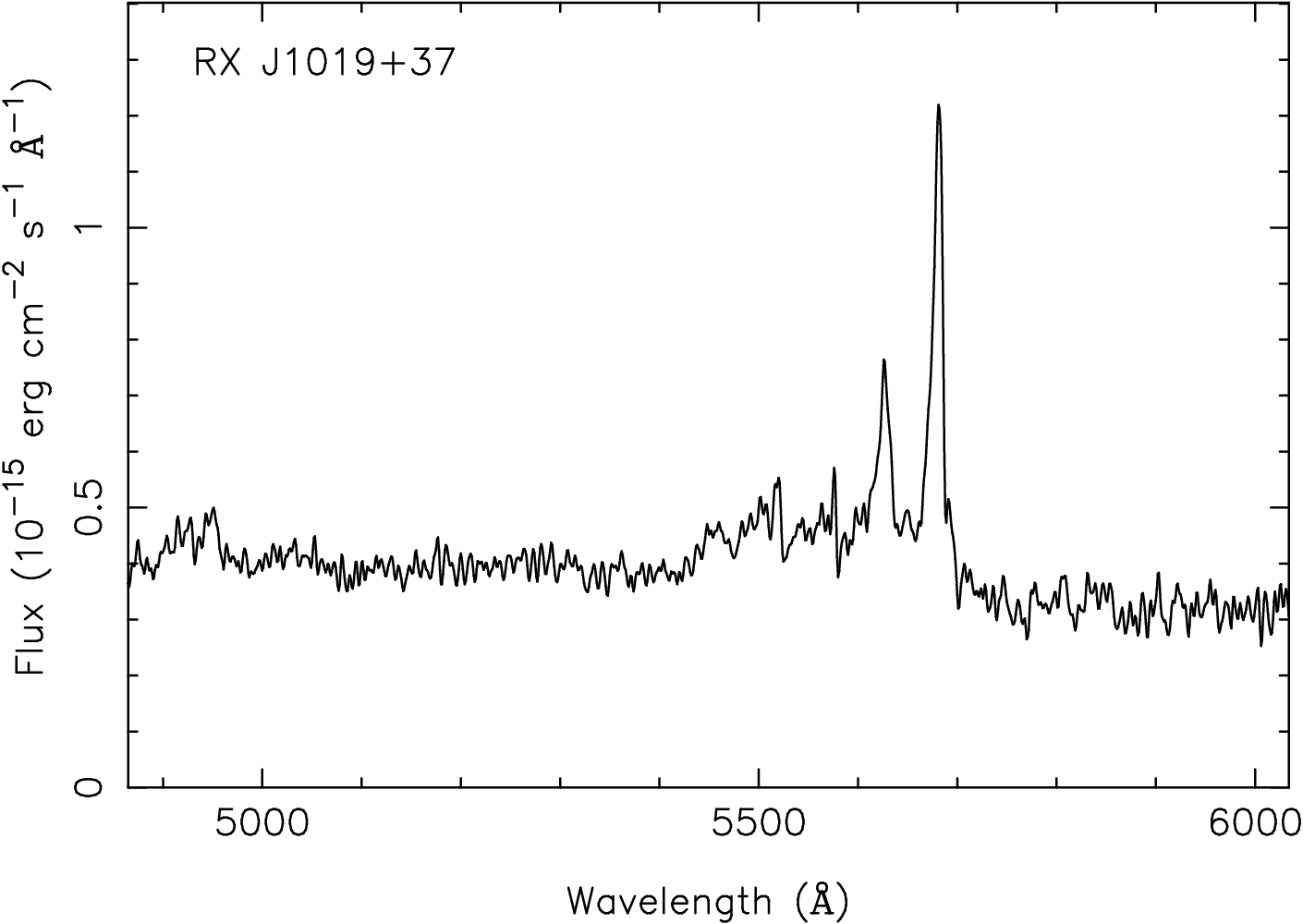}
\includegraphics[width=4.1 cm, angle=270]{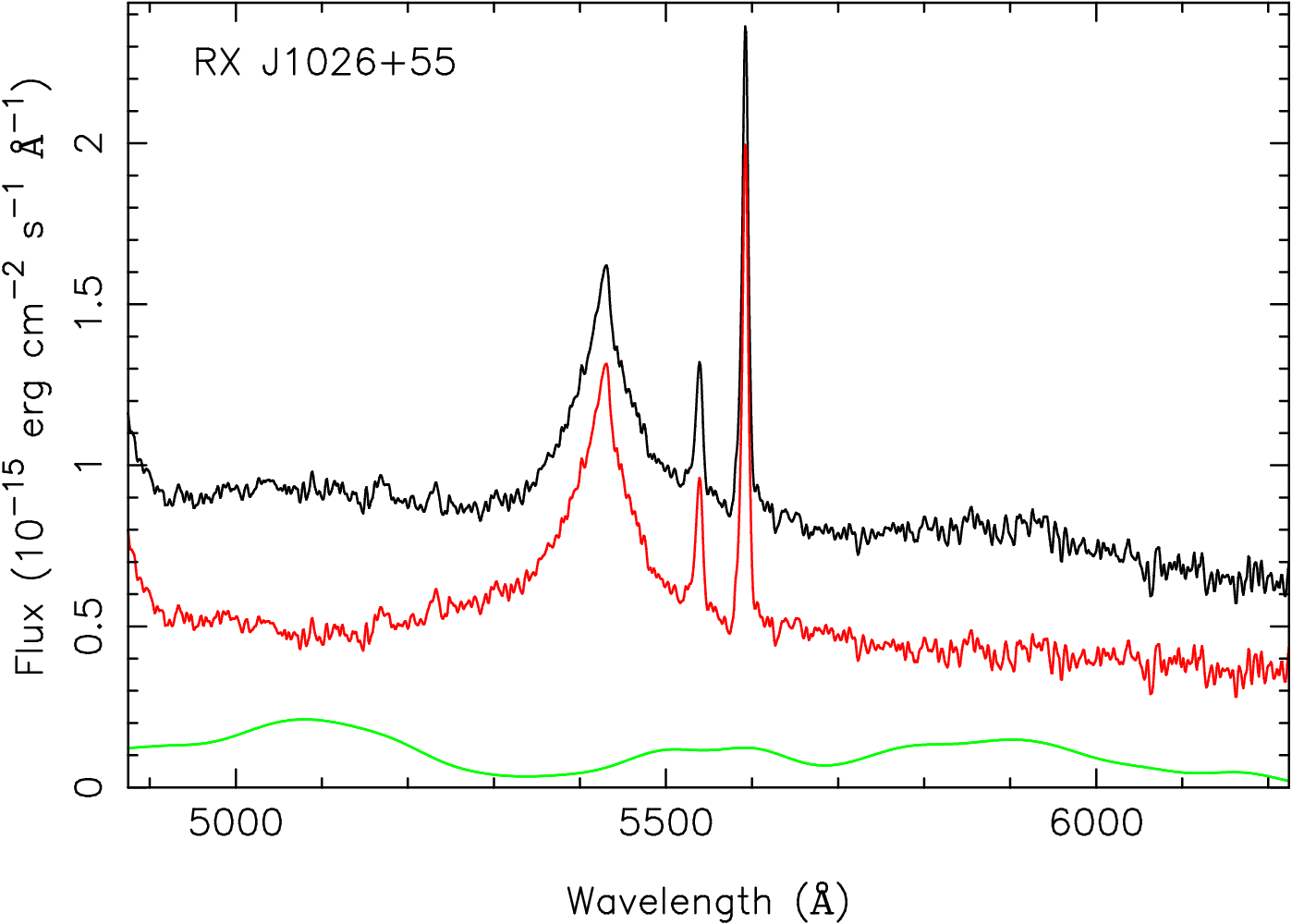}
}
\hbox{
\includegraphics[width=4.1 cm, angle=270]{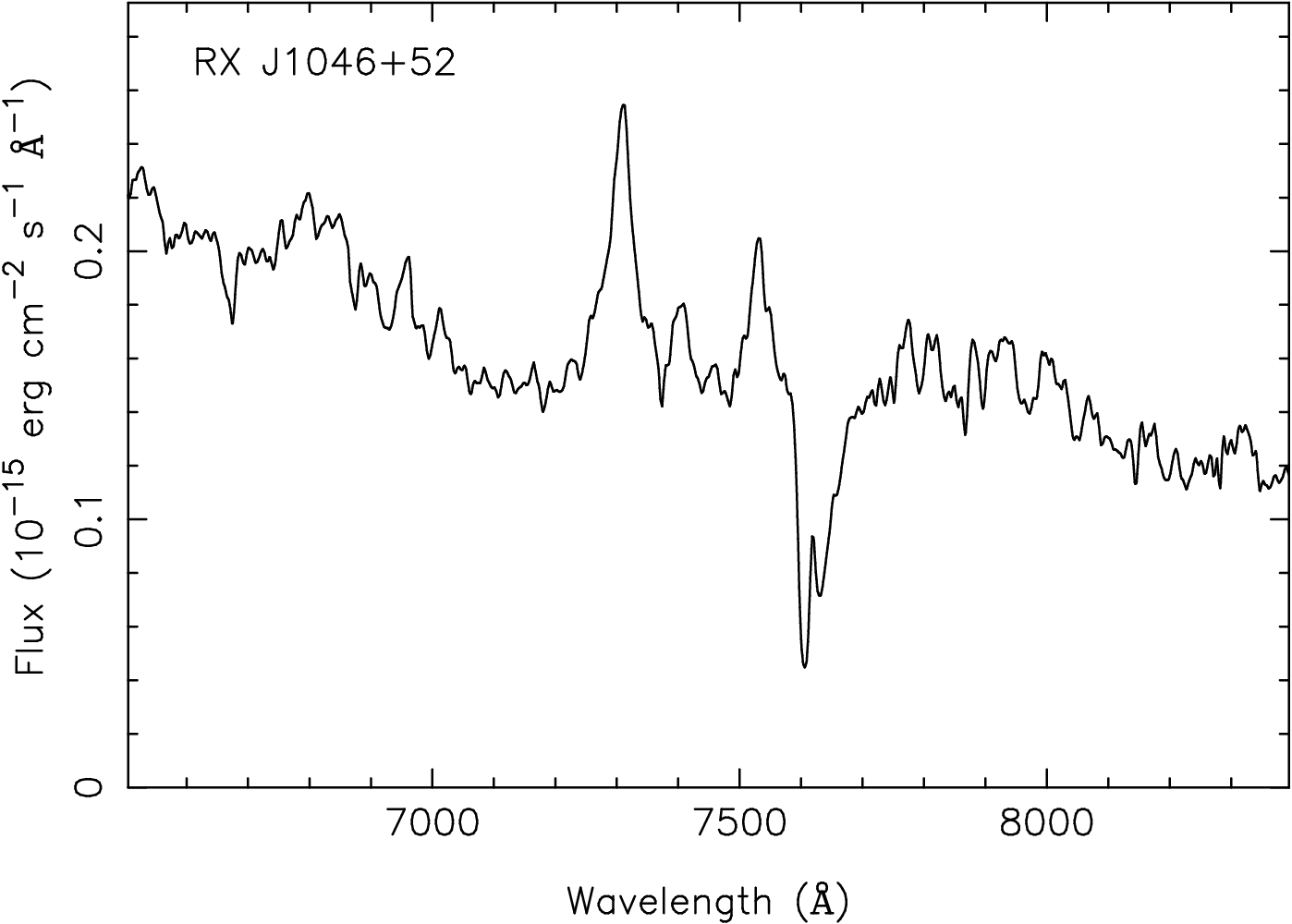}
\includegraphics[width=4.1 cm, angle=270]{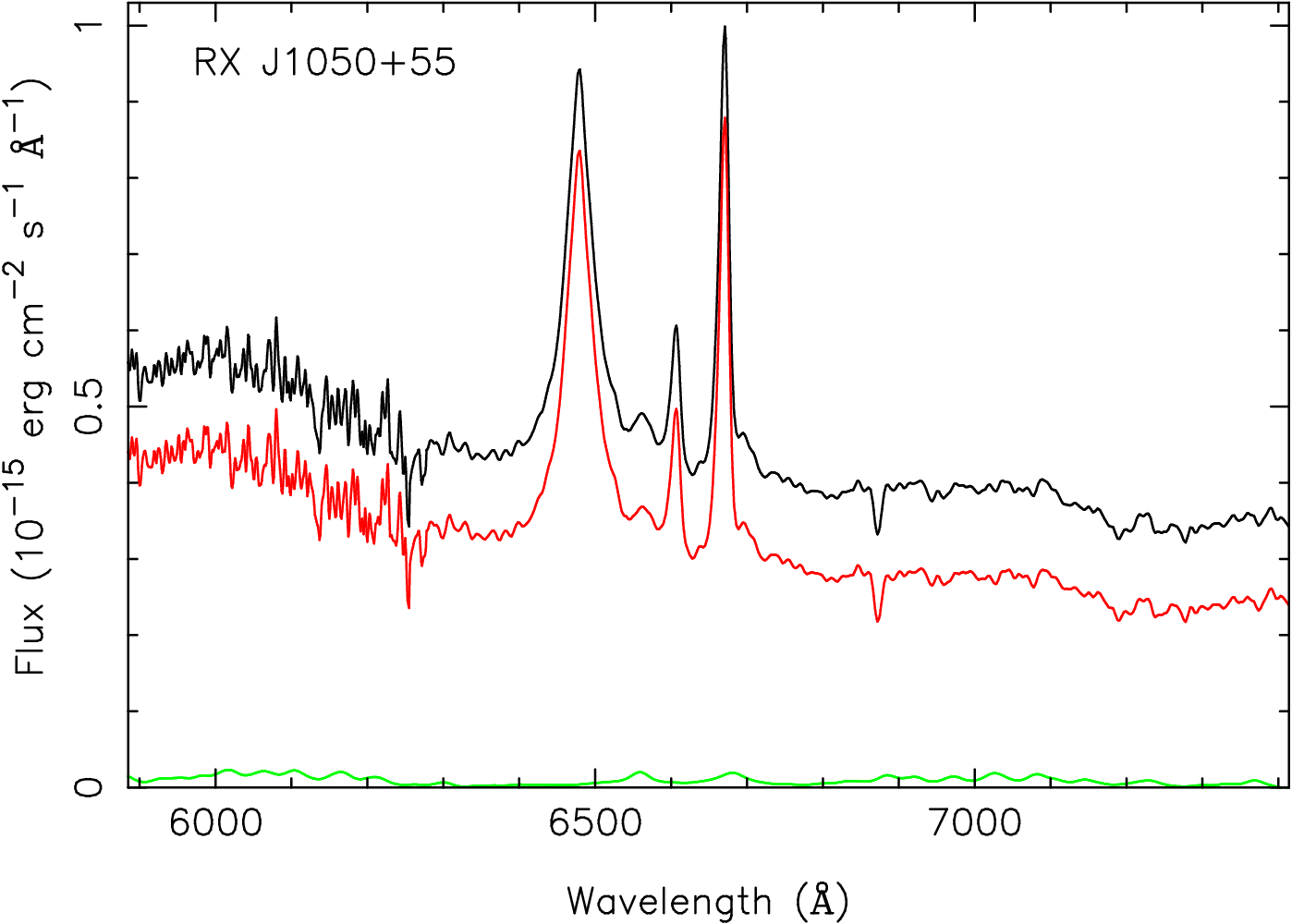}
\includegraphics[width=4.1 cm, angle=270]{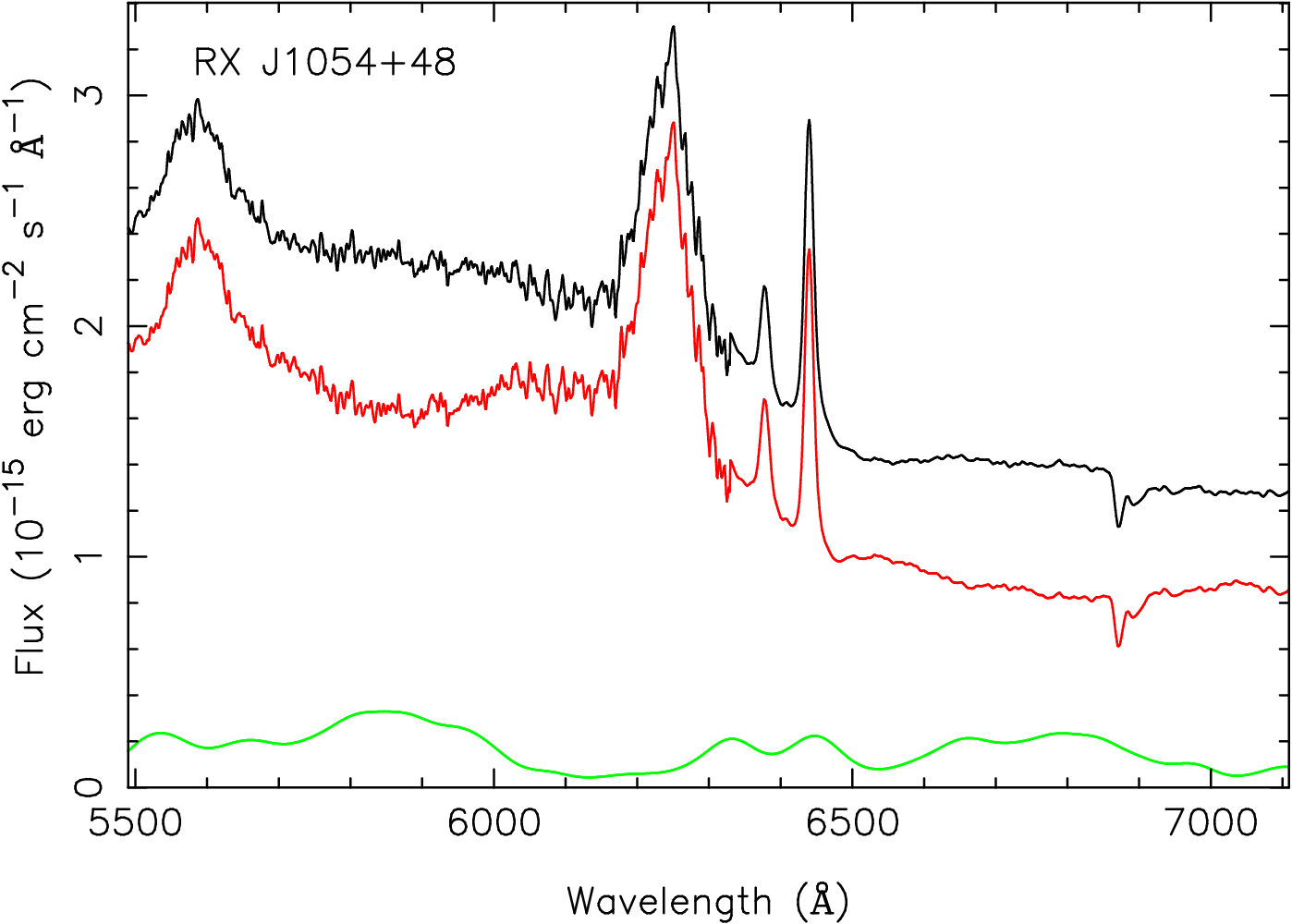}
}
\hbox{
\includegraphics[width=4.1 cm, angle=270]{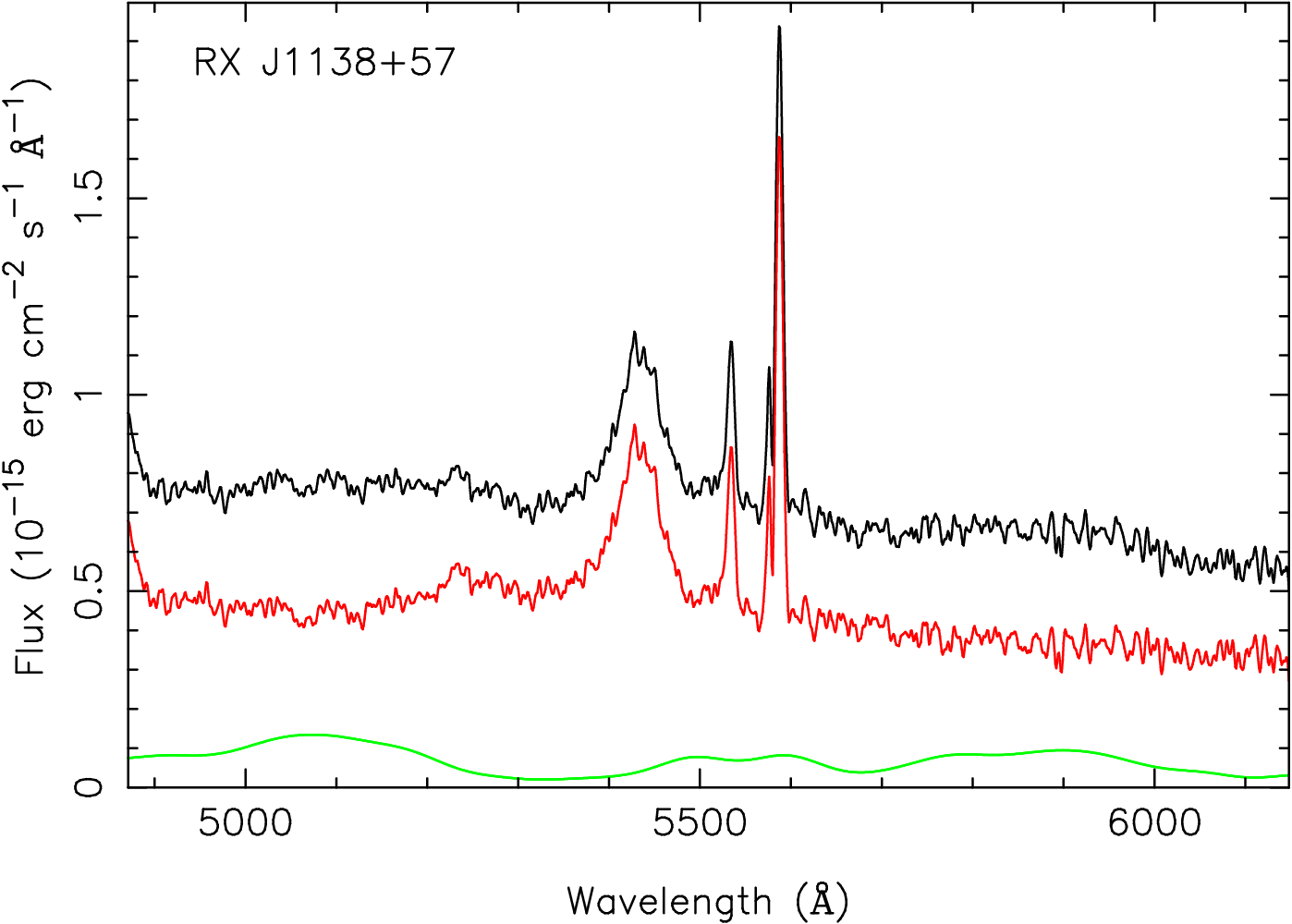}
\includegraphics[width=4.1 cm, angle=270]{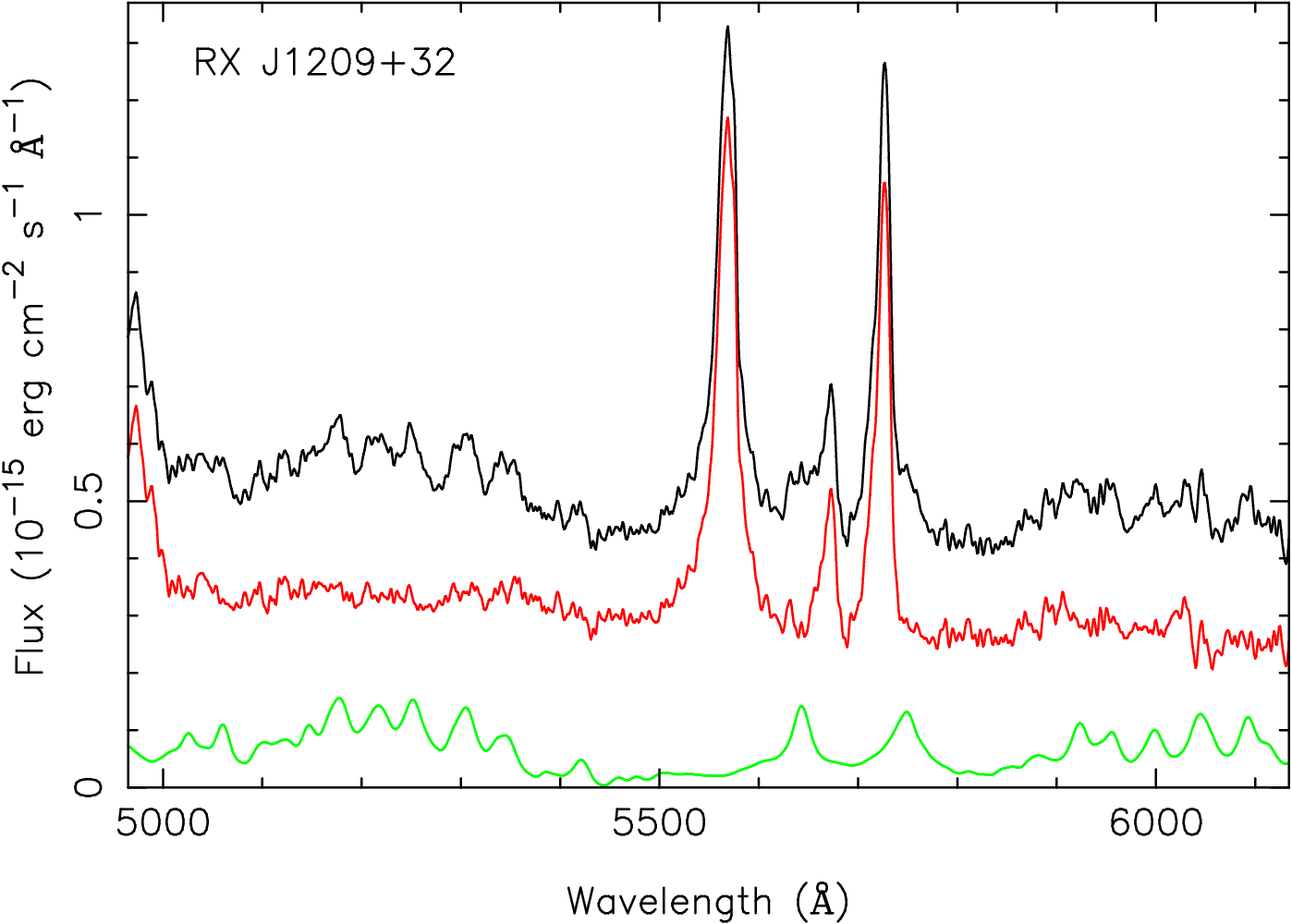}
\includegraphics[width=4.1 cm, angle=270]{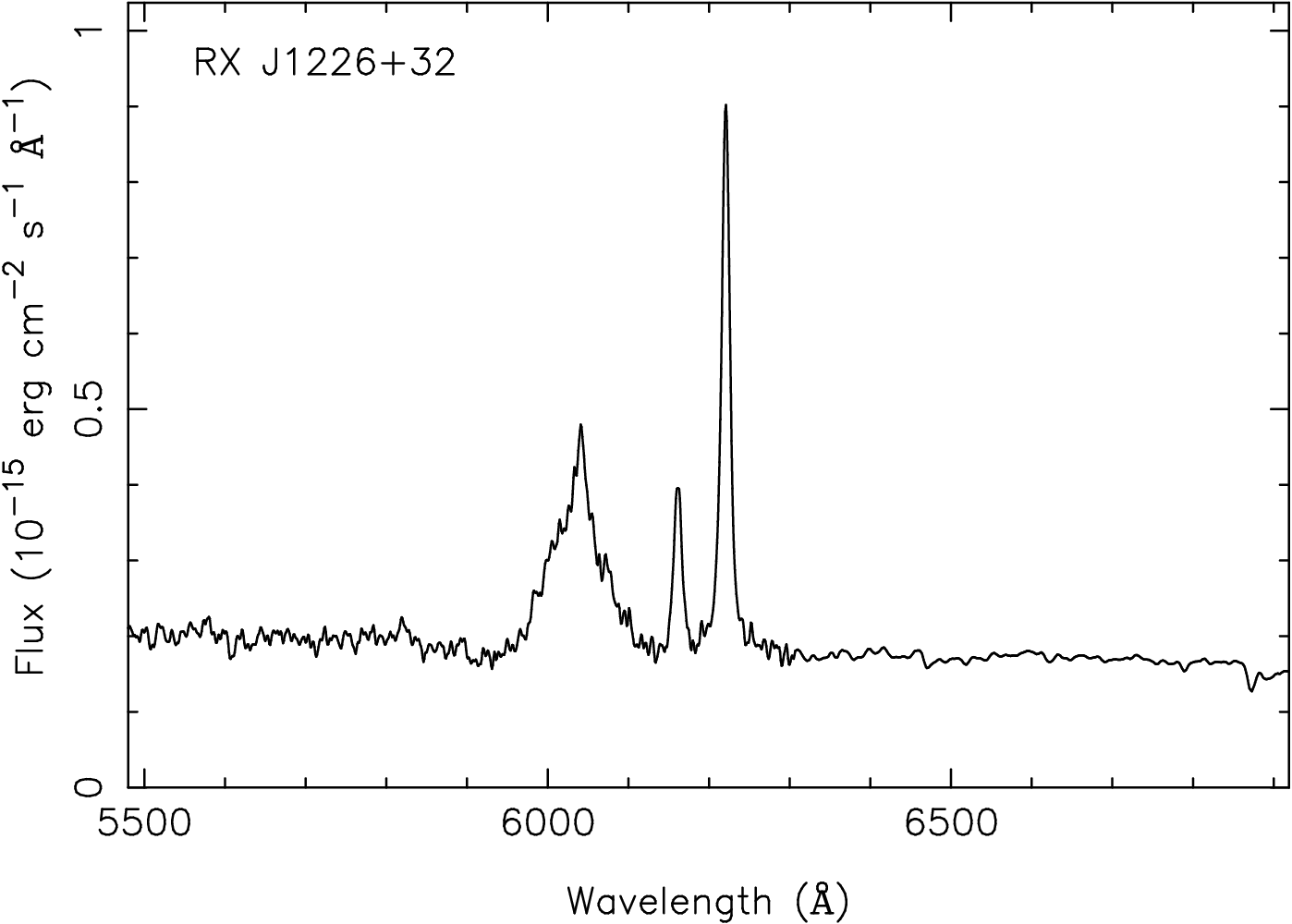}
}
\hbox{
\includegraphics[width=4.1 cm, angle=270]{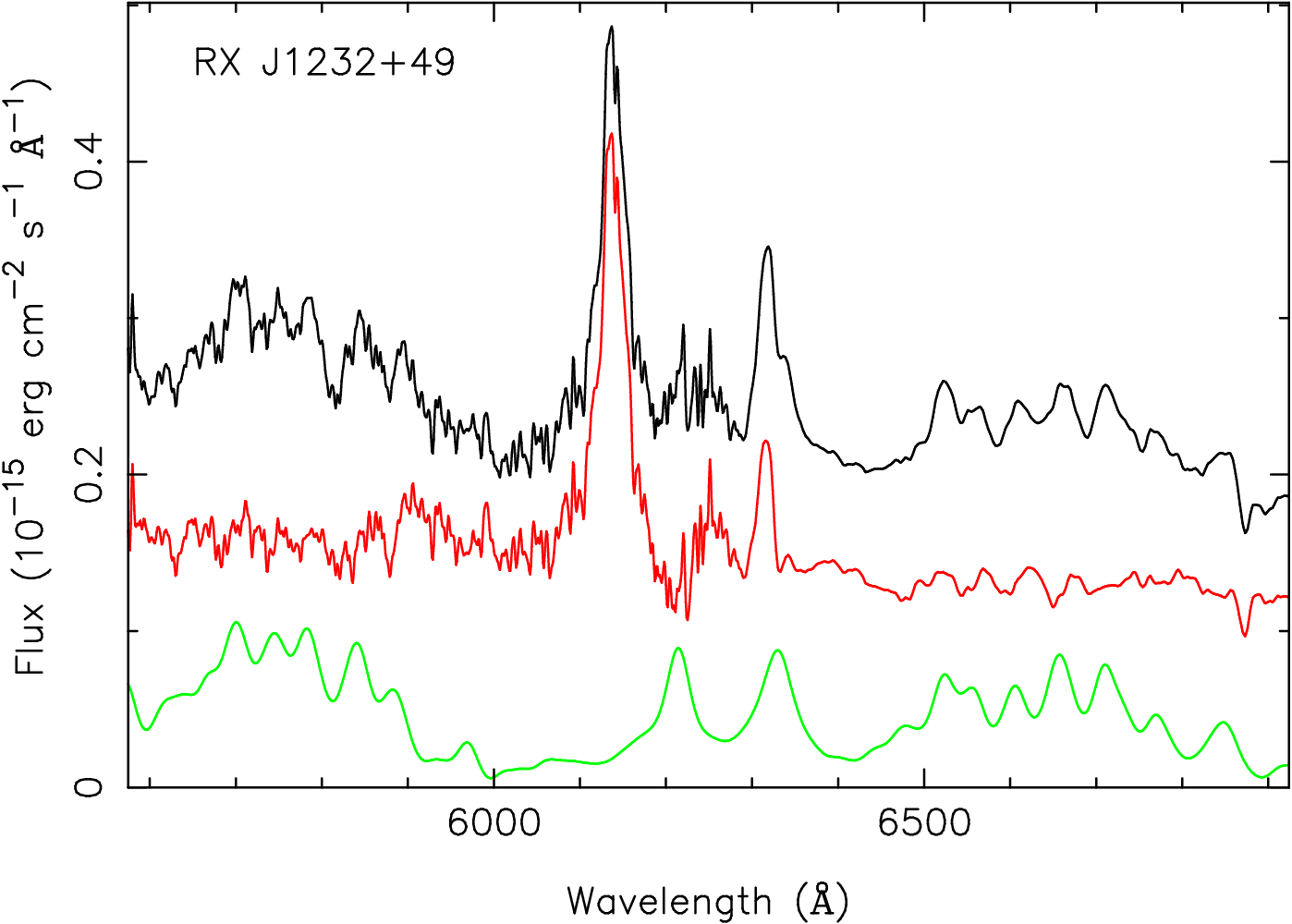}
\includegraphics[width=4.1 cm, angle=270]{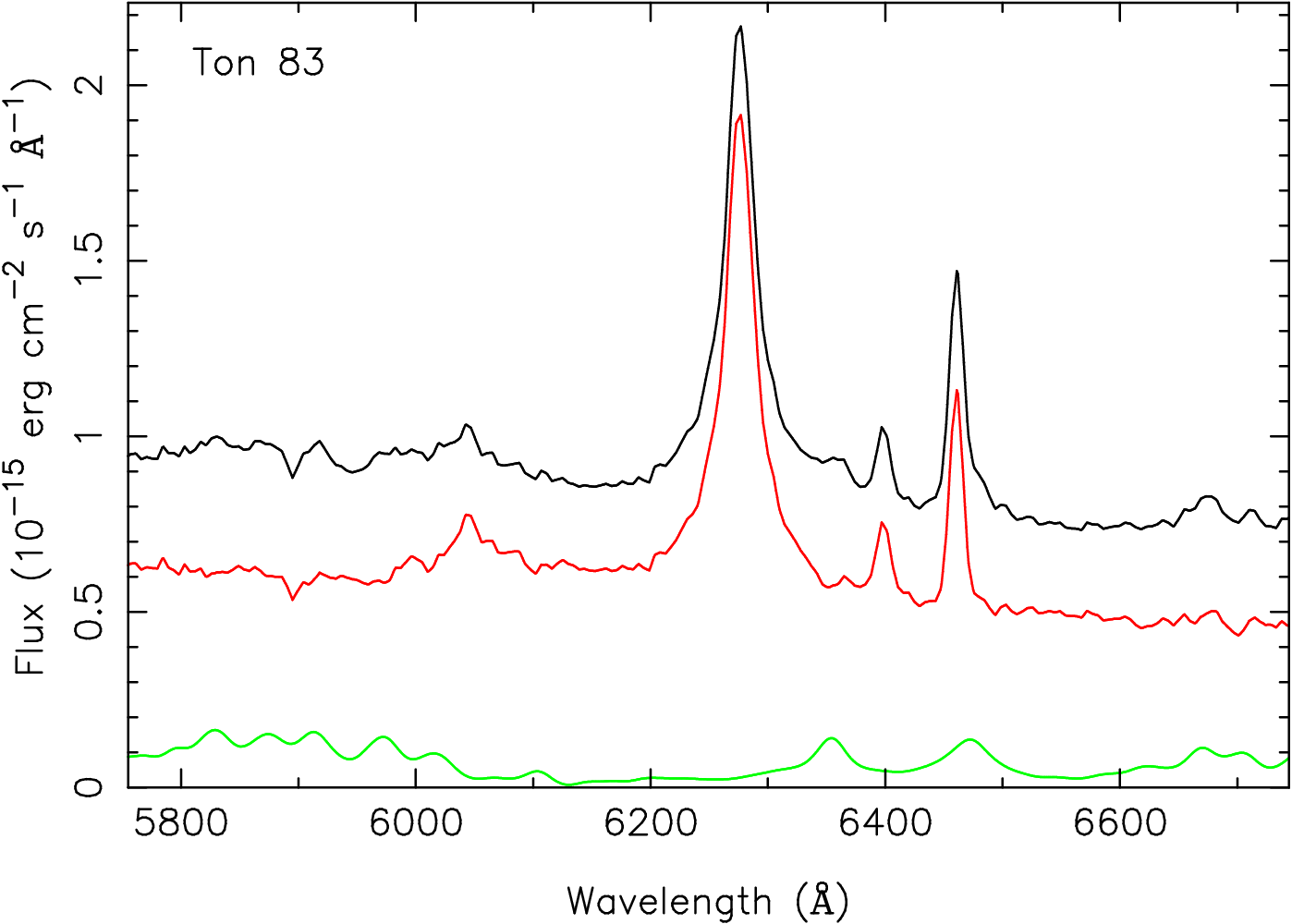}
\includegraphics[width=4.1 cm, angle=270]{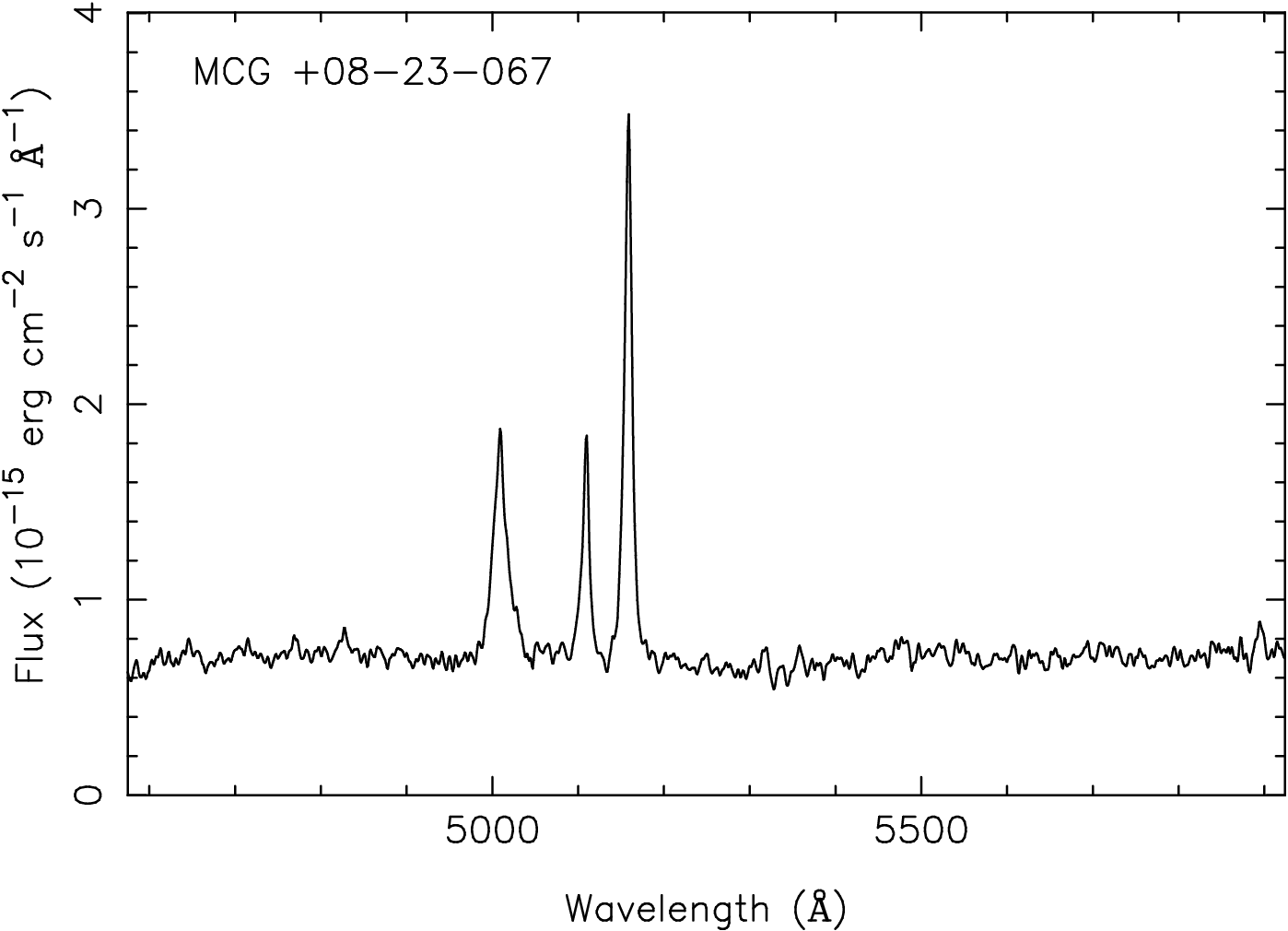} 
}
\hbox{
\includegraphics[width=4.1 cm, angle=270]{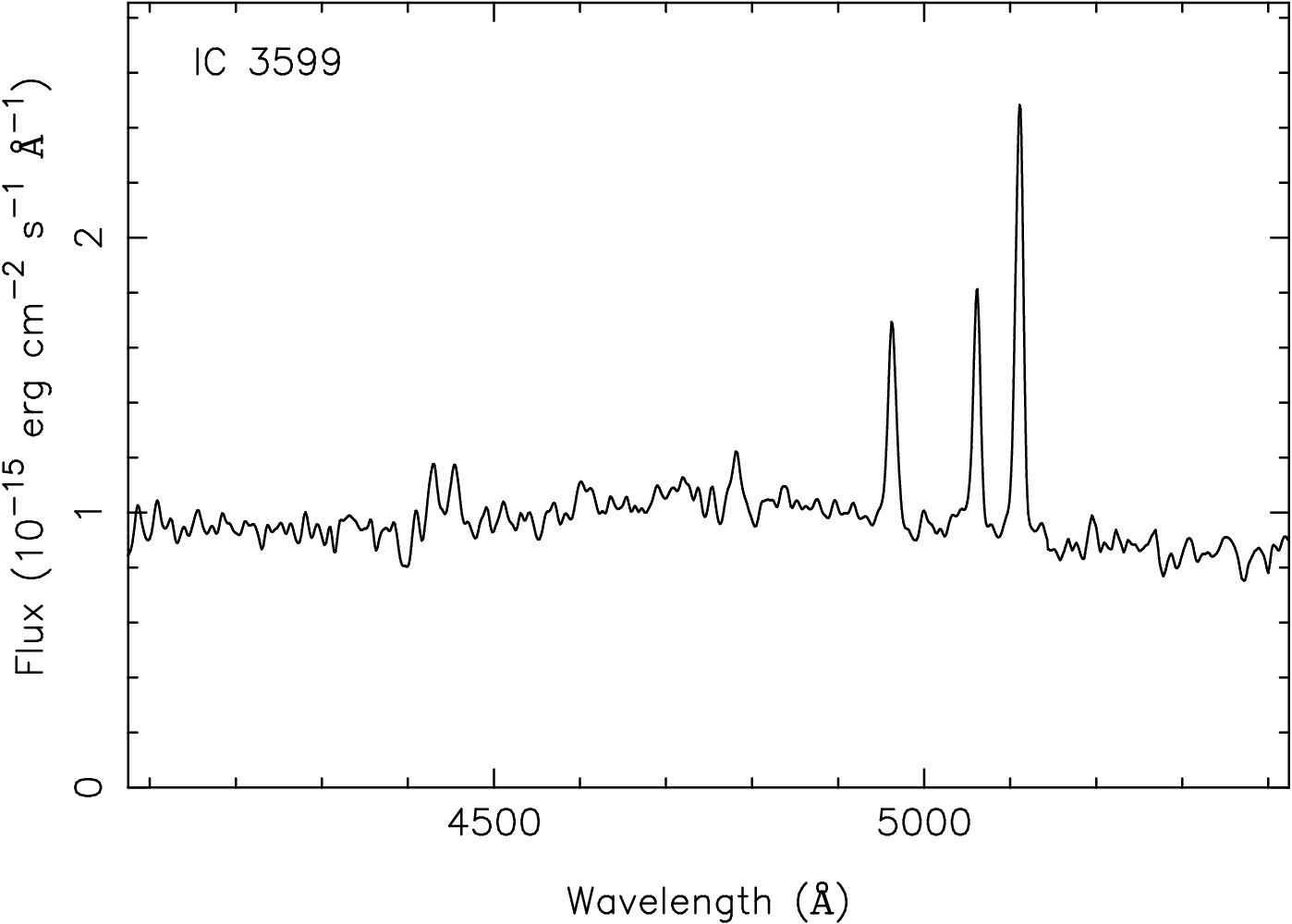}
\includegraphics[width=4.1 cm, angle=270]{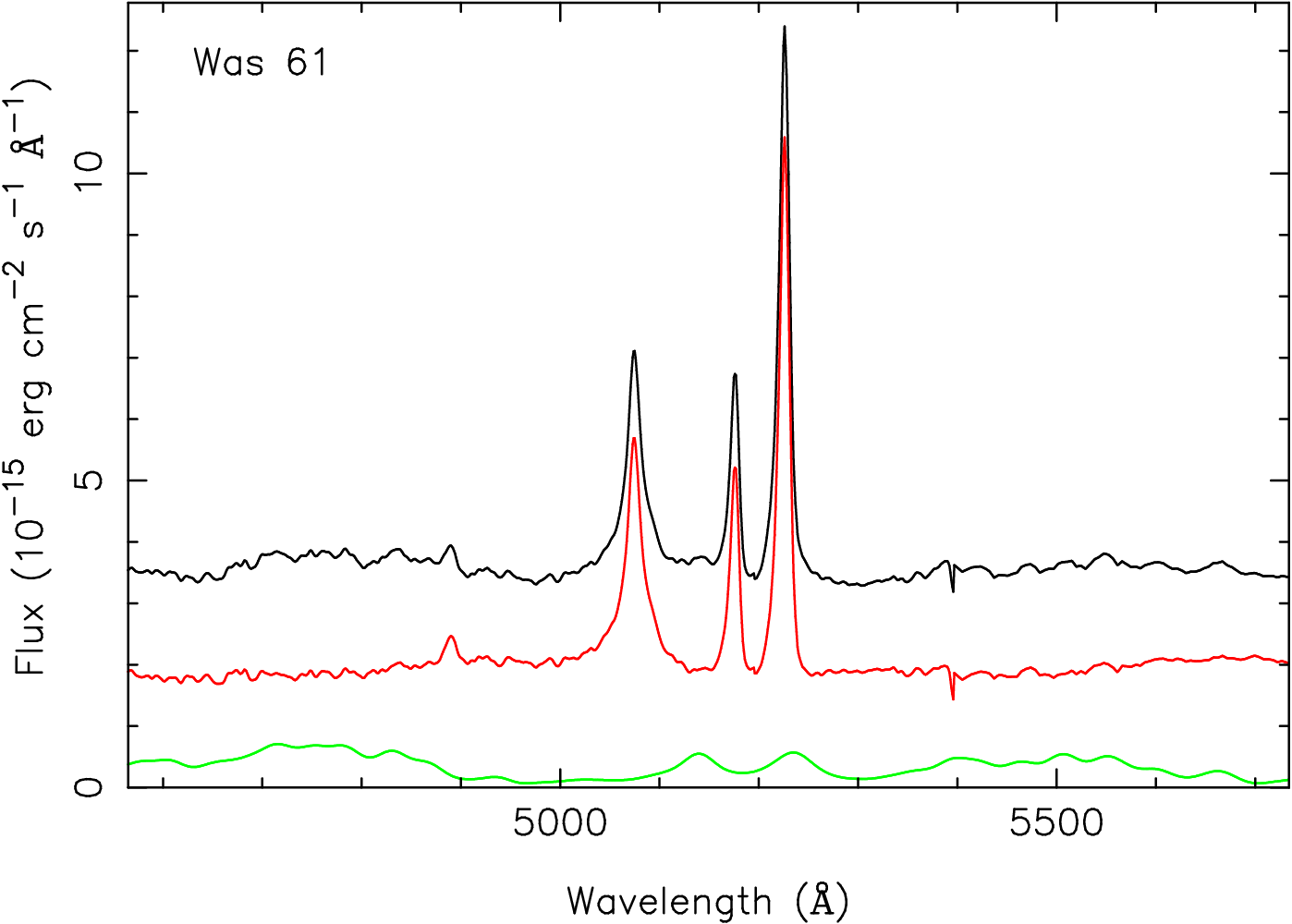}
\includegraphics[width=4.1 cm, angle=270]{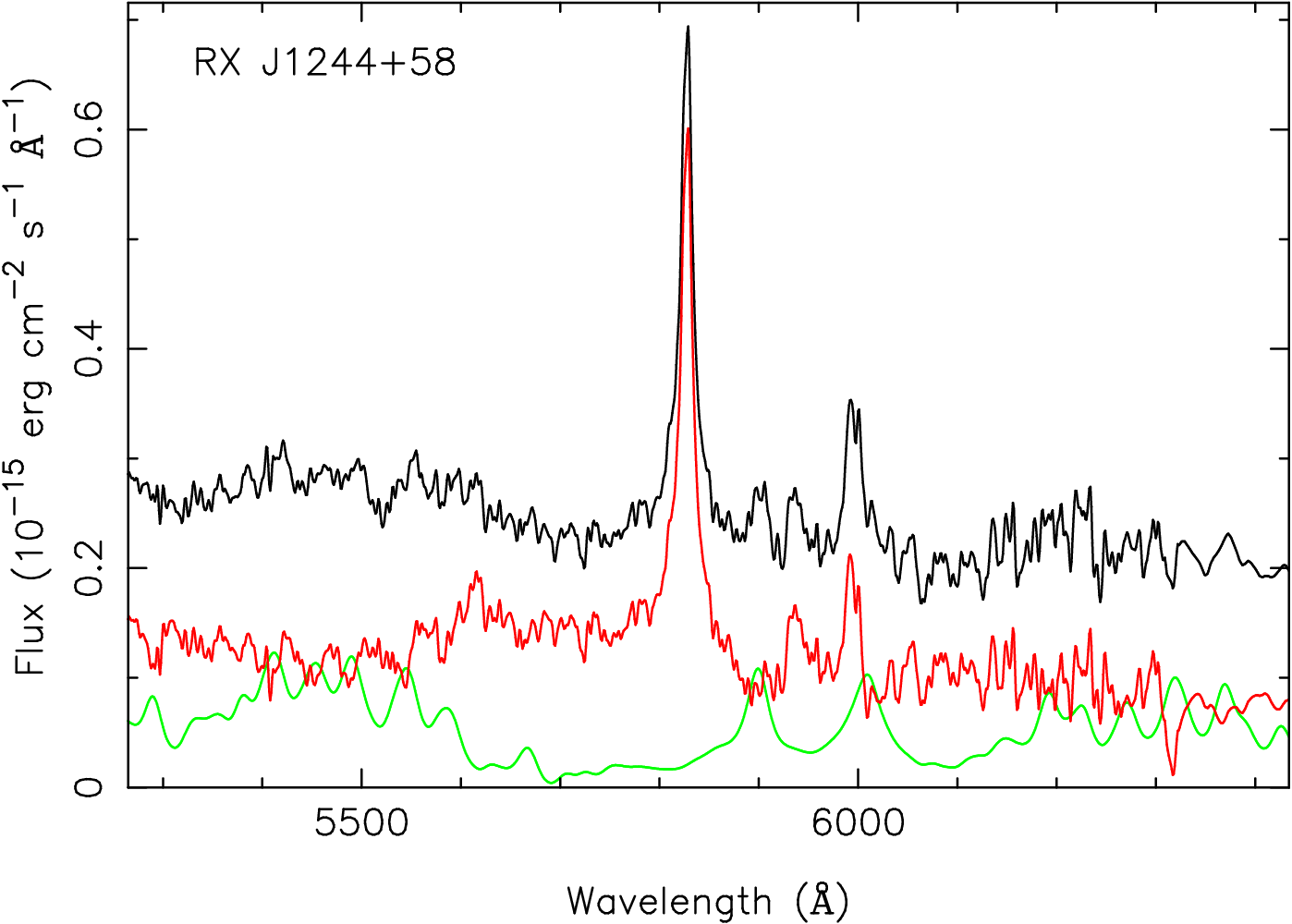}
}
\caption{Close-up of the \hb--\oiii\ region showing the fitted \feii\
emission. In each panel the topmost curve shows the original spectrum,
the middle curve shows the spectrum after \feii\ subtraction and the
lower curve shows the blurred \feii\ template. Objects with no
measurable \feii\ emission are shown with only one curve. The topmost
curve is shifted upwards by an arbitrary amount for clarity.\label{fig:optical_plots}}
\end{figure*}

The \hb--\oiii\ regions of all the observed objects, before and after
\feii\ subtraction, are shown in Figure~\ref{fig:optical_plots}.  The
procedure described above generally did a reasonable job of estimating
the strength of the \feii\ emission and decontaminating the optical
spectrum. There are three obvious exceptions however. In both RX
J1050+55 and RX J1054+48 the part of the spectrum containing the \hb
~line  is redshifted into the region between the blue and red WHT
spectra, and as the overlap between the two spectra is not well
determined it is difficult to accurately measure the surrounding \feii
~emission.  In the highest redshift member of the sample, RX J1046+52,
the \hb--\oiii\ region is affected by strong A-band telluric
absorption.

As mentioned above, the width of the \feii\ lines in the fit was left
as a free parameters. I Zw 1 has FWHM \hb$\sim 900$~km s$^{-1}$,
and has \feii\ lines of comparable width. This means that it was not
possible to accurately model optical \feii\ emission narrower than $900$~km
s$^{-1}$. As there are very few Seyfert 1s with FWHM \hb$\ls$ 900~km
s$^{-1}$ this posed no serious problem.  (The results of this fitting
agree with the claim of BG92 that the \hb\ and optical \feii\ lines
have similar widths.)  The other limitation of this method is that it
assumes the ratios of the optical \feii\ lines within and between
blends are the same as those in I Zw 1. This appears to be a reasonable
approximation for most objects, but a few (e.g., PG 1415+451) may show
slightly different ratios.

\subsubsection{Other line measurements}

The other optical properties were measured from the dereddened, \feii
~subtracted spectra using the {\tt DIPSO} software package. 
To measure line fluxes
local straight-line continua were fitted and subtracted from underneath
the lines and the remaining line flux was integrated. In
many cases there is a clear contribution to \hb\ from a separate narrow
component. In order to measure the properties of only the broad \hb
~line, a narrow \hb\ component was constructed, with a width determined from
fitting the narrow \oiii\ lines, and included in a multiple Gaussian
fit to the \hb\ line. In most spectra the narrow component could be
isolated and removed before the properties of the broad \hb\ line were measured.
The H$\alpha$ line contains contributions from [N~\textsc{ii}]
$\lambda$6548, 6584 which in general could not be isolated in these data. Here, as in
\scite{GRU99}, the  [N~\textsc{ii}] flux is subtracted by assuming
that these lines contribute  35\% of the flux in \oiii\ $\lambda$5007,
following \scite{FER86}.  

\subsubsection{Continuum measurements}

The  continuum   level  at  the   positions  of  \hb\   $\lambda  4861$
and H$\alpha$~$\lambda$6563 were measured from the \feii\ subtracted
spectra, and further  continuum fluxes were measured at 4000, 5500 and
7000~\AA (in the rest frame of the source). The  equivalent widths of
the \hb, \oiii\  and \feii\ lines were calculated  with respect to the
continuum  underneath \hb\ to allow direct  comparison   with  the
measurements   of  BG92   and \scite{GRU99}. Two spectral
indices ($S_{\nu} \propto \nu^{-\alpha}$) were constructed from these
flux measurements, namely:

\[ \alpha_{opt} = 4.11 ~ \log (\frac{f_{7000\rm{\AA}}}{f_{4000\rm{\AA}}}) \]

\[ \alpha_{OX} = 0.489 ~ \log (\frac{f_{5500\rm{\AA}}}{f_{0.25 \rm{keV}}}) \]

These indices are almost identical to those used in \scite{GRU98}.
In seven cases where the $\lambda$5500 flux is not available it was
derived in an approximate way from $m_{V}$ (column 13 of
Table~\ref{table:sample}). 
Assuming that the uncertainty in flux ratios is $\sim 10$\% leads to an
uncertainty of $\sim 0.27$ in $\alpha_{opt}$ and $\sim 0.03$ in $\alpha_{OX}$ and 

Table~\ref{table:opt_measurements} lists the derived parameters. The columns list
the following information: (3) FWHM of broad \hb; (4) FWHM of \oiii\
$\lambda$5007; (5), (6) and (7) list the local equivalent widths of
the \hb, \feii\ and \oiii\ $\lambda$5007 lines; (8) Balmer decrement,
i.e., the ratio of fluxes in H$\alpha$/\hb; (9) ratio of the peak
fluxes of \oiii\ $\lambda$5007 and broad \hb; (10) and (11) ratio of
fluxes of \oiii\ and \feii\ to broad \hb; (12) and
(13)  spectral indices defined above; (14) the monochromatic 0.25~keV
luminosity in $\nu L_{\nu}$ units (in the rest frame of the source).

\begin{table*}
\caption{Properties of the sources which comprise the 1/4 keV-selected
Seyfert galaxy sample. \label{table:opt_measurements}}
 \begin{tabular}{@{}llrrrrrrrrrrrrr@{}}
\hline
 	&	 	&	\hb	&	\oiii 	& 	\hb\ 	&	\feii\ 	&	\oiii  	& 	 	&peak		&		&			&		&		&	\\
 	&	 	&	FWHM	&	FWHM	& 	EW	&	EW	&	EW 	& H$\alpha$/ 	&	\oiii/	&  \oiii/	&	\feii/		&		&  		&$\log(\nu L_{1/4})$	\\
No	&	Name	&	(km/s)	&(km/s)		&(\AA)		&(\AA)		&(\AA)		&	\hb	& 	\hb	&	\hb	&	\hb		&$\alpha_{opt}$	&$\alpha_{OX}$	&erg/s	\\
(1)	&	(2)	&	(3)	&	(4)	&	(5)	&	(6)	&	(7)	&	(8)	&	(9)	&	(10)	&	(11)		&	(12)	&(13)		&	(14)\\
\hline																					
1  &1E 0919+515		&	1390	&	 -	&	38	&	78	&	 -	&	 -	&	 -	&	 -	&	2.05		&	 -	&	0.9	&	44.5	\\
2  &Mrk 110		&	2120	&	 -	&	145	&	20	&	3	&	 -	&	2.5	&	0.7	&	0.14		&	 -	&	0.9	&	43.9	\\
3  &US 0656		&	 -	&	 -	&	 -	&	 -	&	 -	&	 -	&	 -	&	 -	&	-		&	 -	&	1.0	&	44.8	\\
4  &PG 0953+414		&	3130	&	 -	&	156	&	39	&	19	&	 -	&	0.8	&	0.1	&	0.25		&	 -	&	1.4	&	45.2	\\
5  &IRAS 10026+4347	&	2990	&	825	&	55	&	99	&	5	&	 -	&	 -	&	0.1	&	1.81		&	0.1	&	1.1	&	44.8	\\
\hline																					
6  &RX J1008+46		&	13645	&	765	&	120	&	0	&	31	&	 -	&	4.4	&	0.3	&	0.00		&	 -	&	0.8	&	45.4	\\
7  &Ton 1187		&	2850	&	370	&	76	&	33	&	22	&	3.5	&	 -	&	0.3	&	0.43		&	 -	&	1.2	&	44.5	\\
8  &RX J1019+37		&	1130	&	675	&	75	&	0	&	26	&	2.8	&	5.9	&	0.3	&	0.00		&	1.3	&	0.8	&	44.8	\\
9  &Mrk 141		&	4175	&	395	&	30	&	38	&	16	&	3.8	&	 -	&	0.5	&	1.25		&	1.5	&	1.3	&	43.4	\\
10  &Mrk 142 		&	1790	&	280	&	52	&	86	&	10	&	3.3	&	 -	& 	0.2	&	1.65		&	1.2	&	0.9	&	44.0	\\
\hline																					
11  &RX J1026+55	&	5035	&	530	&	183	&	78	&	29	&	1.7	&	2.2	&	0.2	&	0.43		&	1.6	&	1.1	&	44.1	\\
12  &RE J1034+396	&	1500	&	900	&	60	&	 -	&	45	&	3.0	&	 -	&	0.8	&	-		&	 -	&	1.0	&	44.0	\\
13  &RX J1046+52	&	2670	&	 -	&	29	&	-	&	 -	&	 -	&	 -	&	 -	&	-		&	 -	&	1.1	&	45.4	\\
14  &RX J1050+55	&	2780	&	760	&	51	&	0	&	23	&	5.0	&	1.6	&	0.5	&	0.00		&	1.5	&	1.2	&	45.1	\\
15  &RX J1054+48	&	5210	&	1115	&	71	&	0	&	18	&	3.6	&	1.2	&	0.3	&	0.00		&	-0.2	&	1.3	&	45.2	\\
\hline																					
16  &EXO 1055+60	&	2155	&	540	&	90	&	110	&	25	&	 -	&	 -	&	0.3	&	1.27		&	1.3	&	1.2	&	44.2	\\
17  &RX J1117+65	&	2160	&	880	&	65	&	78	&	11	&	 -	&	 -	&	0.2	&	1.20		&	1.2	&	1.1	&	44.5	\\
18  &PG 1116+21		&	2920	&	 -	&	175	&	81	&	11	&	 -	&	0.3	&	0.1	&	0.46		&	 -	&	1.2	&	45.1	\\
19  &EXO 1128+691	&	1800	&	 -	&	10	&	 -	&	 -	&	 -	&	 -	&	 -	&	-		&	 -	&	0.8	&	44.0	\\
20  &RX J1138+57	&	2845	&	485	&	39	&	56	&	25	&	3.3	&	3.7	&	0.6	&	1.43		&	0.5	&	0.9	&	44.4	\\
\hline																					
21  & NGC 4051		&	990	&	 -	&	 -	&	 -	&	 -	&	 -	&	 -	&	 -	&	-		&	 -	&	1.8	&	41.8	\\
22  &RX J1209+32	&	1370	&	860	&	63	&	67	&	28	&	3.5	&	0.9	&	0.4	&	1.06		&	0.6	&	0.9	&	44.5	\\
23  &RX J1226+32	&	3940	&	640	&	66	&	0	&	43	&	5.0	&	3.9	&	0.6	&	0.00		&	1.4	&	0.9	&	45.0	\\
24  &RX J1232+49	&	1905	&	895	&	55	&	92	&	9	&	4.5	&	0.3	&	0.2	&	1.67		&	0.9	&	0.9	&	44.9	\\
25  &Ton 83		&	1435	&	670	&	92	&	38	&	13	&	3.5	&	0.4	&	0.1	&	0.41		&	0.6	&	1.0 	&	45.4	\\
\hline																					
26  &MCG +8-23-067	&	1245	&	645	&	30	&	0	&	41	&	3.6	&	2.6	&	1.4	&	0.00		&	2.5	&	1.0	&	43.2	\\
27  &IC 3599 		&	675	&	615	&	9	&	0	&	18	&	3.7	&	2.3	&	2.1	&	0.00		&	1.8	&	0.7	&	43.7	\\
28  &Was 61		&	2765	&	740	&	30	&	62	&	55	&	6.8	&	6.7	&	1.8	&	2.09		&	2.6	&	1.1	&	43.8	\\
29  &RX J1244+58	&	720	&	785	&	56	&	47	&	11	&	3.3	&	0.3	&	0.2	&	0.84		&	0.3	&	0.9	&	44.5	\\
30  &RX J1258+23	&	6920	&	420	&	29	&	25	&	34	&	4.0	&	18.6	&	1.2	&	0.86		&	2.8	&	1.0	&	43.9	\\
\hline																					
31  & RX J1312+26	&	3530	&	535	&	20	&	12	&	3	&	2.7	&	1.1	&	0.2	&	0.58		&	1.5	&	1.1	&	43.6	\\
32  & Ton 1571		&	1330	&	650	&	59	&	86	&	10	&	2.7	&	0.4	&	0.2	&	1.47		&	0.7	&	1.2	&	43.8	\\
33  & RX J1319+52	&	1620	&	415	&	29	&	45	&	35	&	4.0	&	4.5	&	1.2	&	1.53		&	1.1	&	0.9	&	44.2	\\
34  & RX J1328+24	&	2370	&	610	&	65	&	25	&	20	&	4.4	&	1.2	&	0.3	&	0.39		&	0.8	&	1.1	&	44.7	\\
35  & IRAS 13349+243	&	2775	&	1340	&	54	&	76	&	7	&	4.8	&	0.3	&	0.1	&	1.41		&	1.1	&	1.3	&	44.9	\\
\hline																					
36  & RX J1339+40	&	1180	&	425	&	21	&	9	&	9	&	3.5	&	1.1	&	0.4	&	0.43		&	1.4	&	0.9	&	44.1	\\
37  & RX J1342+38	&	4290	&	385	&	75	&	12	&	45	&	3.7	&	6.5	&	0.6	&	0.16		&	0.8	&	0.8	&	44.6	\\
38  & PG 1341+25	&	3300	&	440	&	60	&	43	&	18	&	3.9	&	2.1	&	0.3	&	0.72		&	1.0	&	1.2	&	44.0	\\
39  & Mrk 663		&	 -	&	770	&	0	&	0	&	11	&	 -	&	 -	&	 -	&	-		&	3.5	&	1.1	&	43.4	\\
40  & RX J1355+56	&	2600	&	935	&	44	&	46	&	46	&	4.8	&	5.0	&	1.8	&	1.05		&	0.5	&	1.1	&	44.4	\\
\hline																					
41  & PG 1402+261	&	2220	&	 -	&	74	&	168	&	$<$1.8	&	 -	&	$<$0.6	&	$<$0.02	&	2.25		&	-0.5	&	1.1	&	44.8	\\
42  & PG 1415+451	&	2575	&	 -	&	69	&	132	&	$<$5.6	&	2.6	&	$<$0.03	&	$<$0.07	&	1.91		&	0.5	&	1.3	&	44.2	\\
43  & RX J1426+39	&	 -	&	 -	&	 -	&	 -	&	 -	&	 -	&	 -	&	 -	&	-		&	 -	&	1.3	&	43.8	\\
44  & Mrk 684		&	1170	&	 -	&	45	&	82	&	$<$1.1	&	2.0	&	$<$0.09	&	$<$0.02	&	1.81		&	0.9	&	1.3	&	43.7	\\
45  & Mrk 478		&	1915	&	610	&	65	&	87	&	12	&	3.7	&	 -	&	0.2	&	1.34		&	0.7	&	0.9	&	44.9	\\
\hline																					
46  & PG 1444+407	&	2775	&	 -	&	141	&	135	&	$<$7.3	&	2.5	&	$<$0.14	&	$<$0.04	&	0.96		&	-0.1	&	1.2	&	45.0	\\
47  & RX J1448+35	&	2430	&	625	&	54	&	64	&	24	&	3.1	&	1.7	&	0.4	&	1.18		&	0.5	&	1.3	&	44.0	\\
48  & NGC 5905 		&	405	&	415	&	3	&	0	&	2	&	10.9	&	0.4	&	0.5	&	0.00		&	2.7	&	1.7	&	41.7	\\
49  & RX J1529+56	&	4055	&	540	&	100	&	8	&	170	&	3.8	&	12.5	&	1.7	&	0.08		&	0.9	&	1.1	&	44.4	\\
50  & MCG +06-36-003	&	6320	&	420	&	100	&	35	&	34	&	3.3	&	5.0	&	0.3	&	0.35		&	1.2	&	1.2	&	43.9	\\
\hline																					
51  & RX J1618+36	&	1705	&	390	&	20	&	23	&	13	&	2.9	&	2.7	&	0.6	&	1.14		&	1.7	&	1.0	&	43.5	\\
52  & RX J1619+40	&	3330	&	650	&	44	&	35	&	29	&	2.9	&	3.2	&	0.7	&	0.81		&	1.5	&	1.1	&	43.2	\\
53  & RX J1629+40	&	2675	&	750	&	55	&	36	&	32	&	3.0	&	1.6	&	0.6	&	0.64		&	0.4	&	0.9	&	45.1	\\
54 & RX J1646+39	&	1515	&	800	&	84	&	40	&	22	&	2.0	&	0.7	&	0.3	&	0.47		&	1.0	&	1.1	&	44.2	\\
\hline		
\end{tabular}
\end{table*}

\begin{figure*}
\centering
\hbox{
\includegraphics[width=4.1 cm, angle=270]{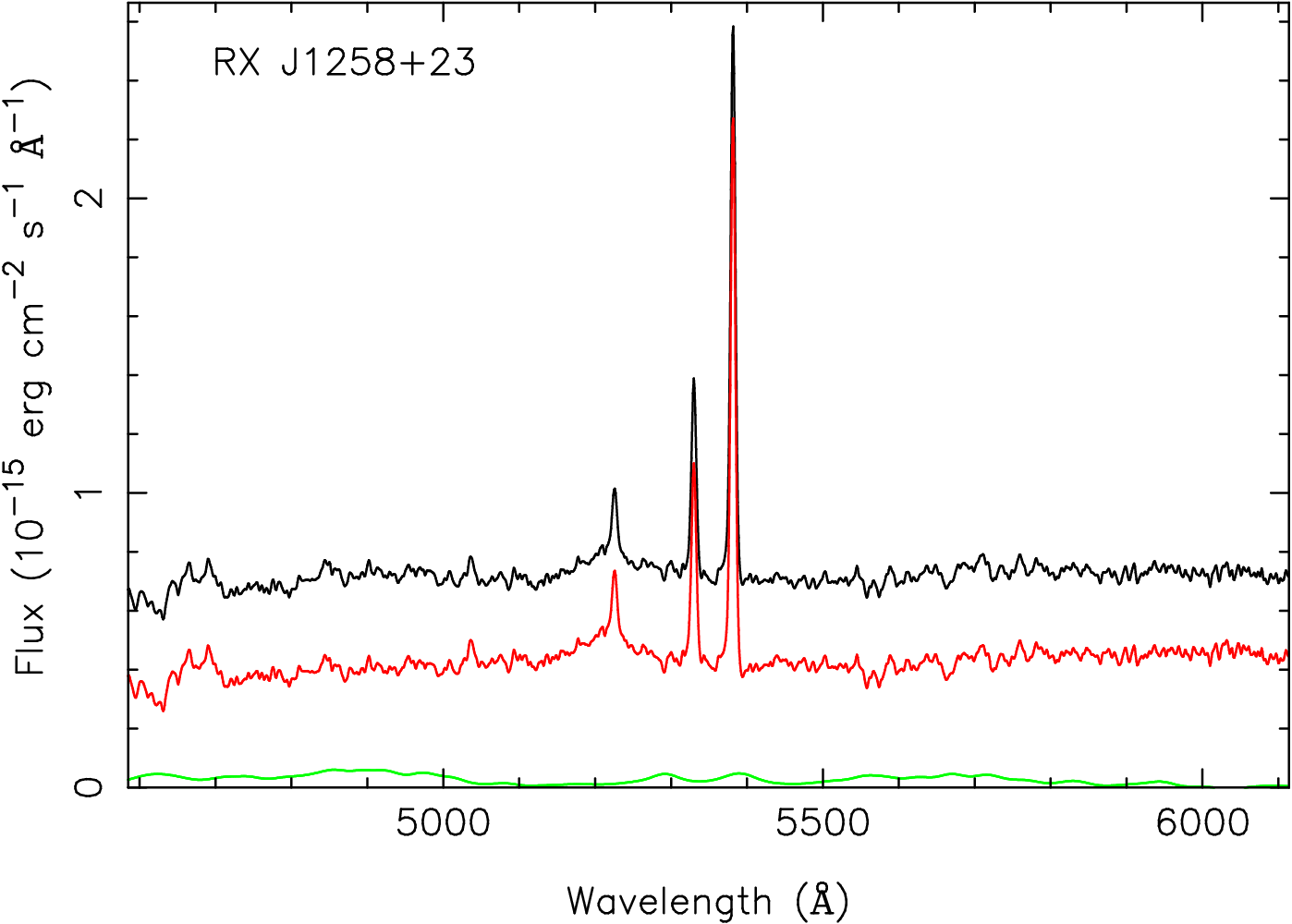}
\includegraphics[width=4.1 cm, angle=270]{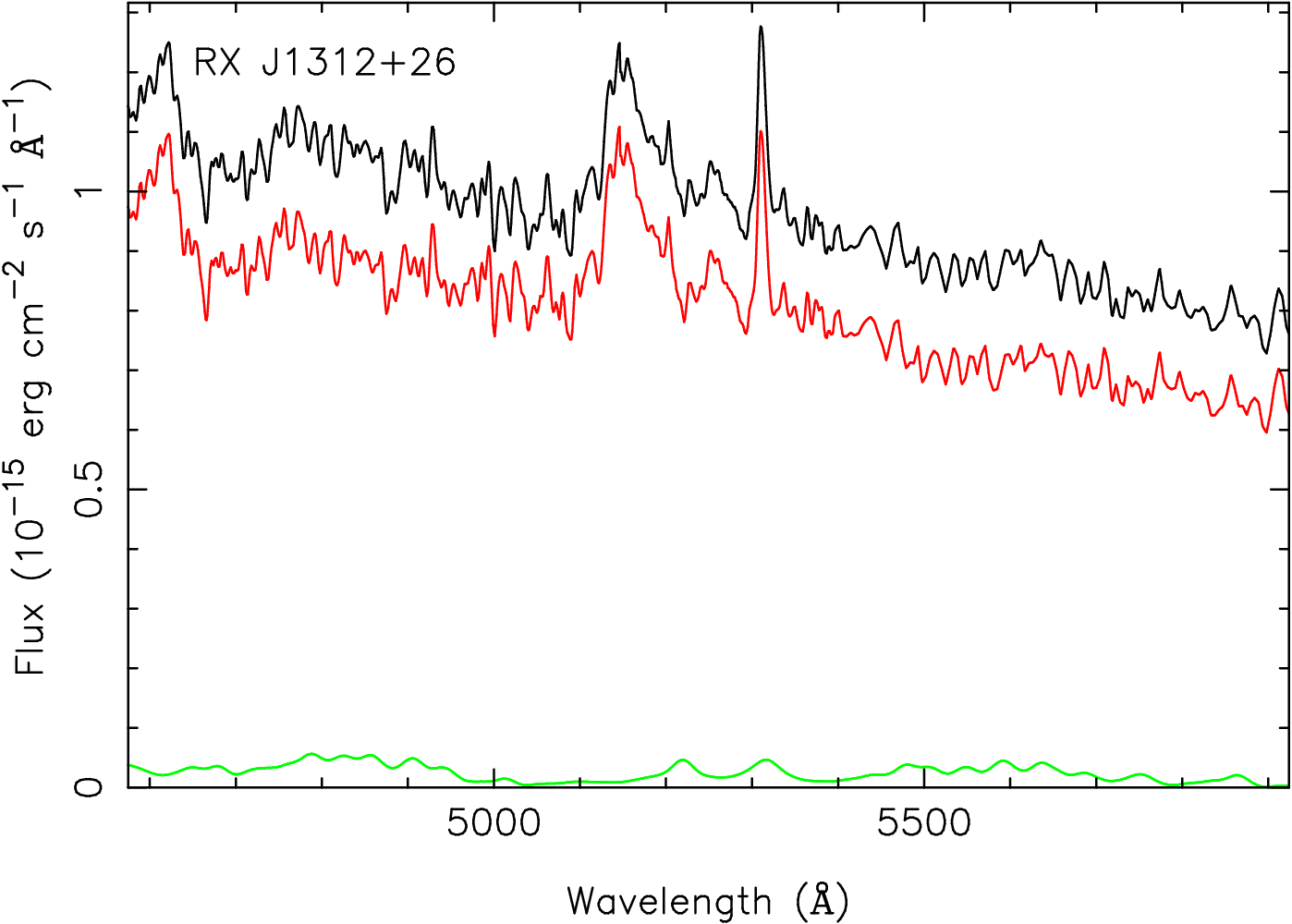}
\includegraphics[width=4.1 cm, angle=270]{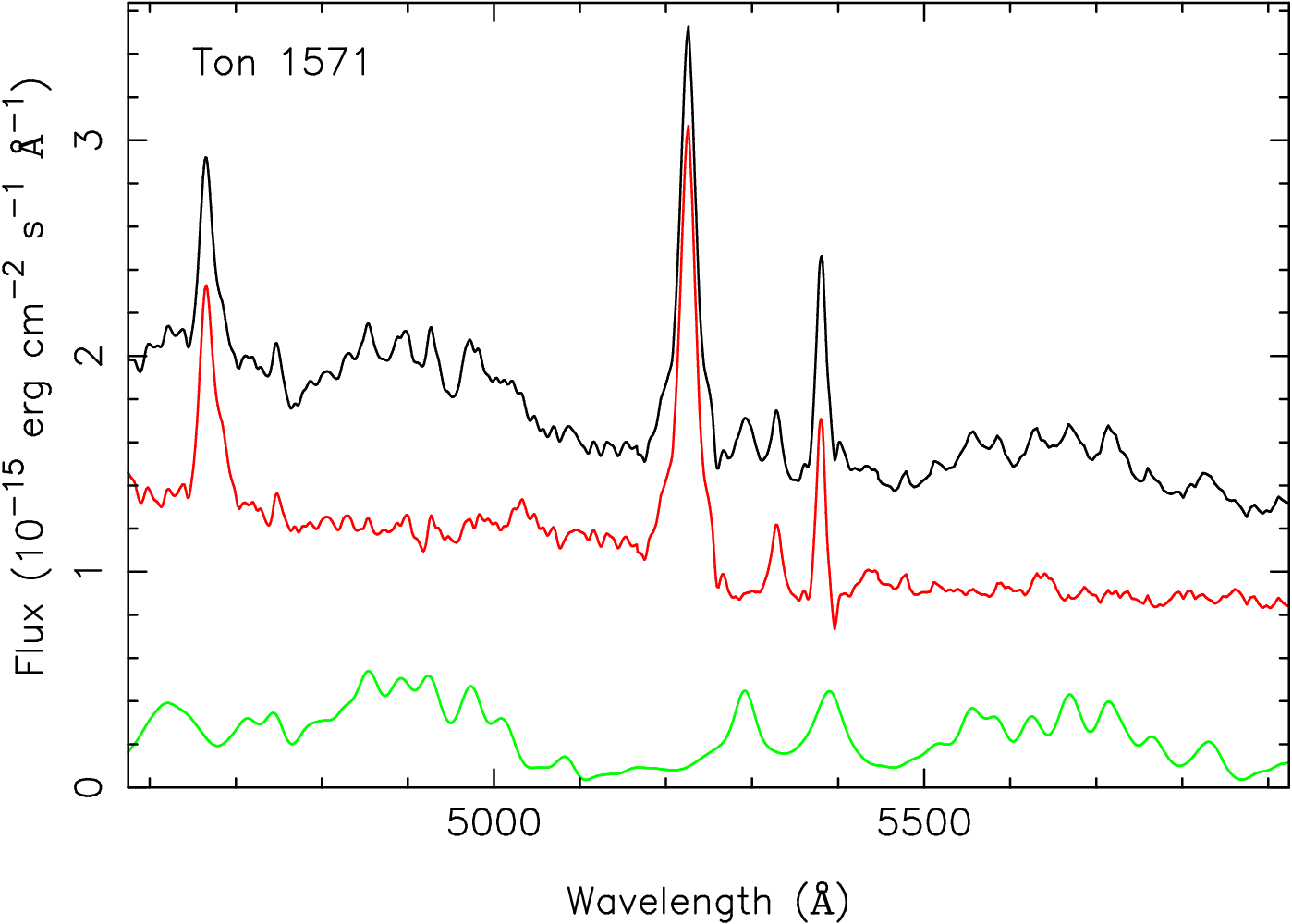}
}
\hbox{
\includegraphics[width=4.1 cm, angle=270]{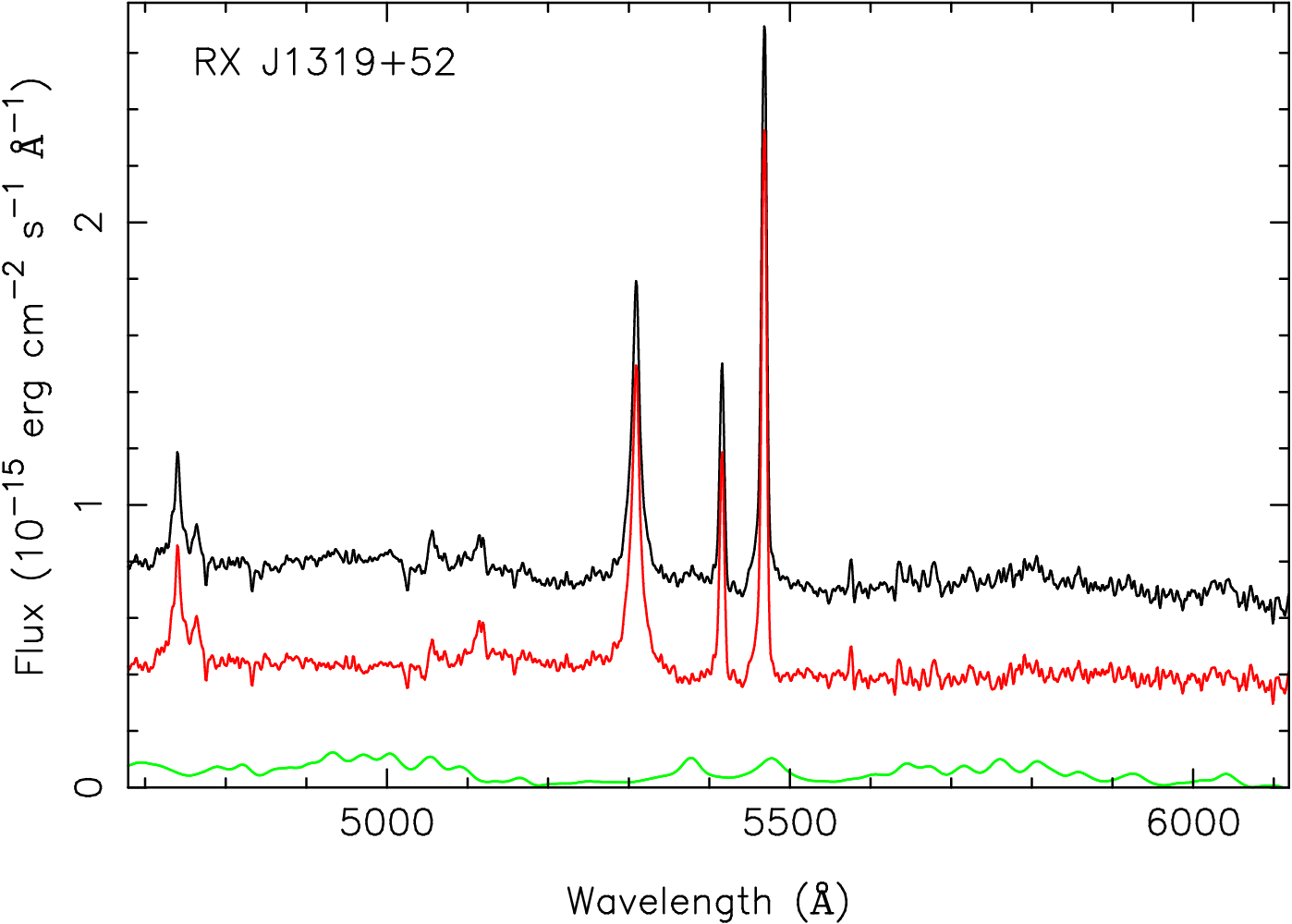} 
\includegraphics[width=4.1 cm, angle=270]{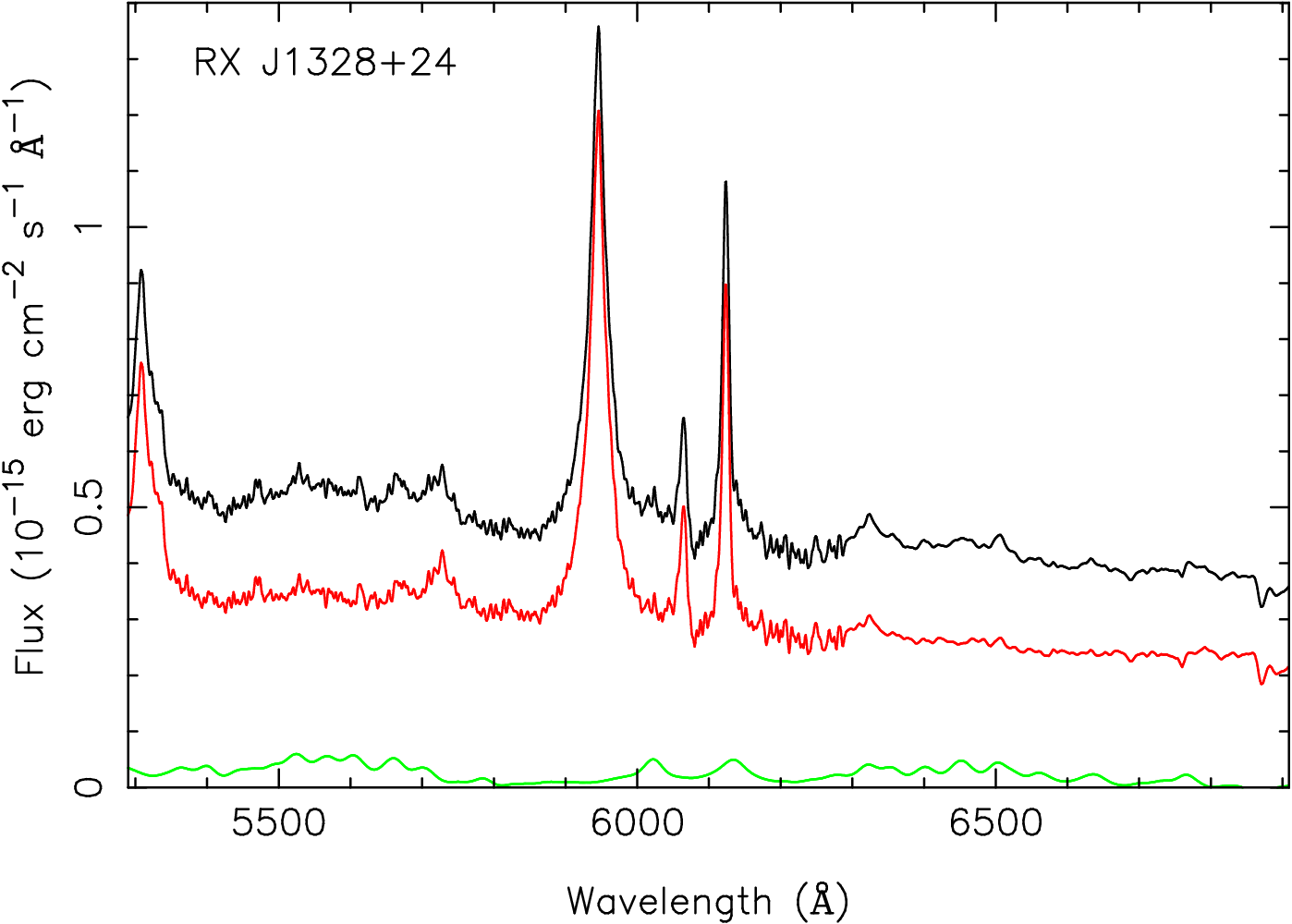}
\includegraphics[width=4.1 cm, angle=270]{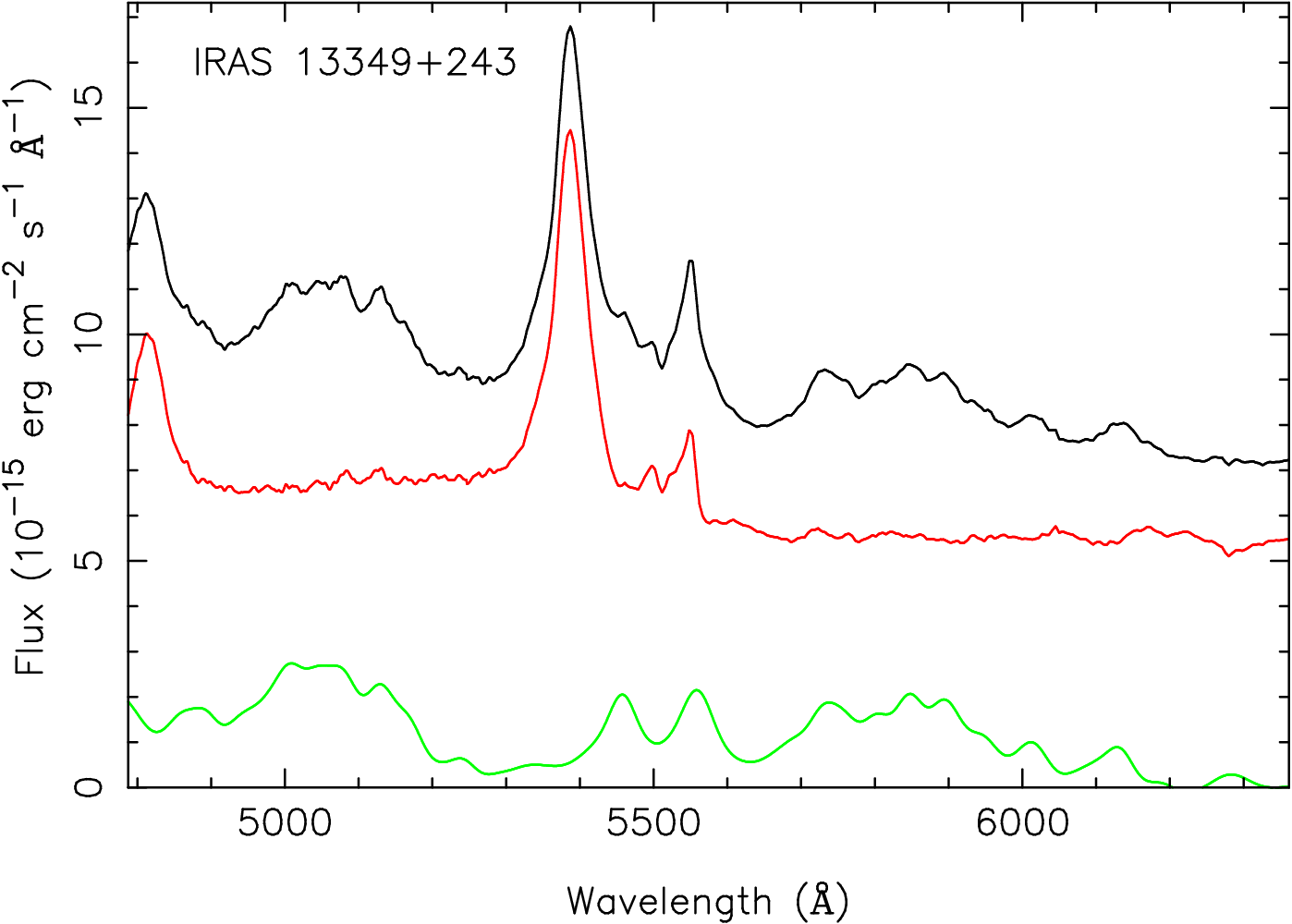}
}
\hbox{
\includegraphics[width=4.1 cm, angle=270]{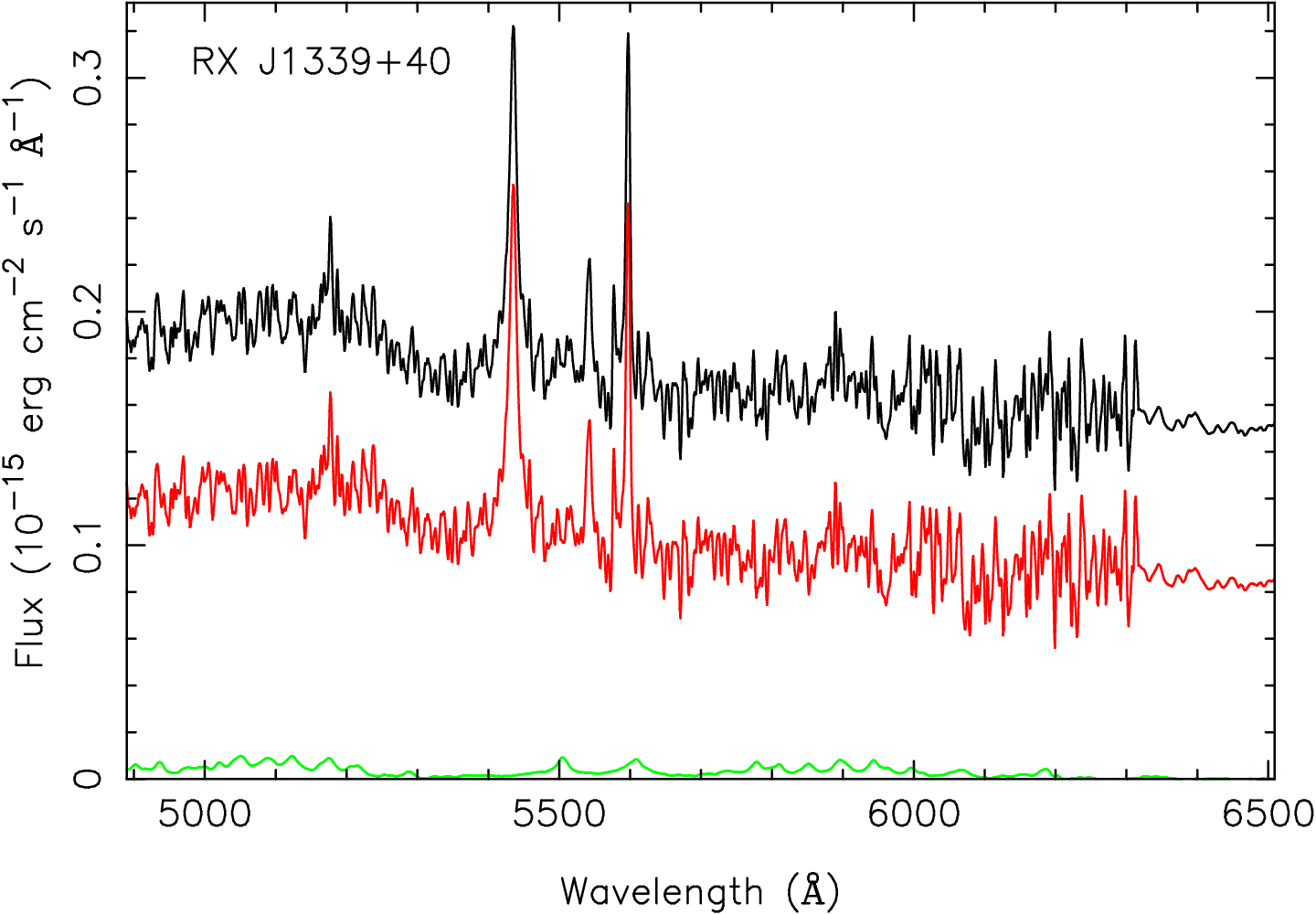}
\includegraphics[width=4.1 cm, angle=270]{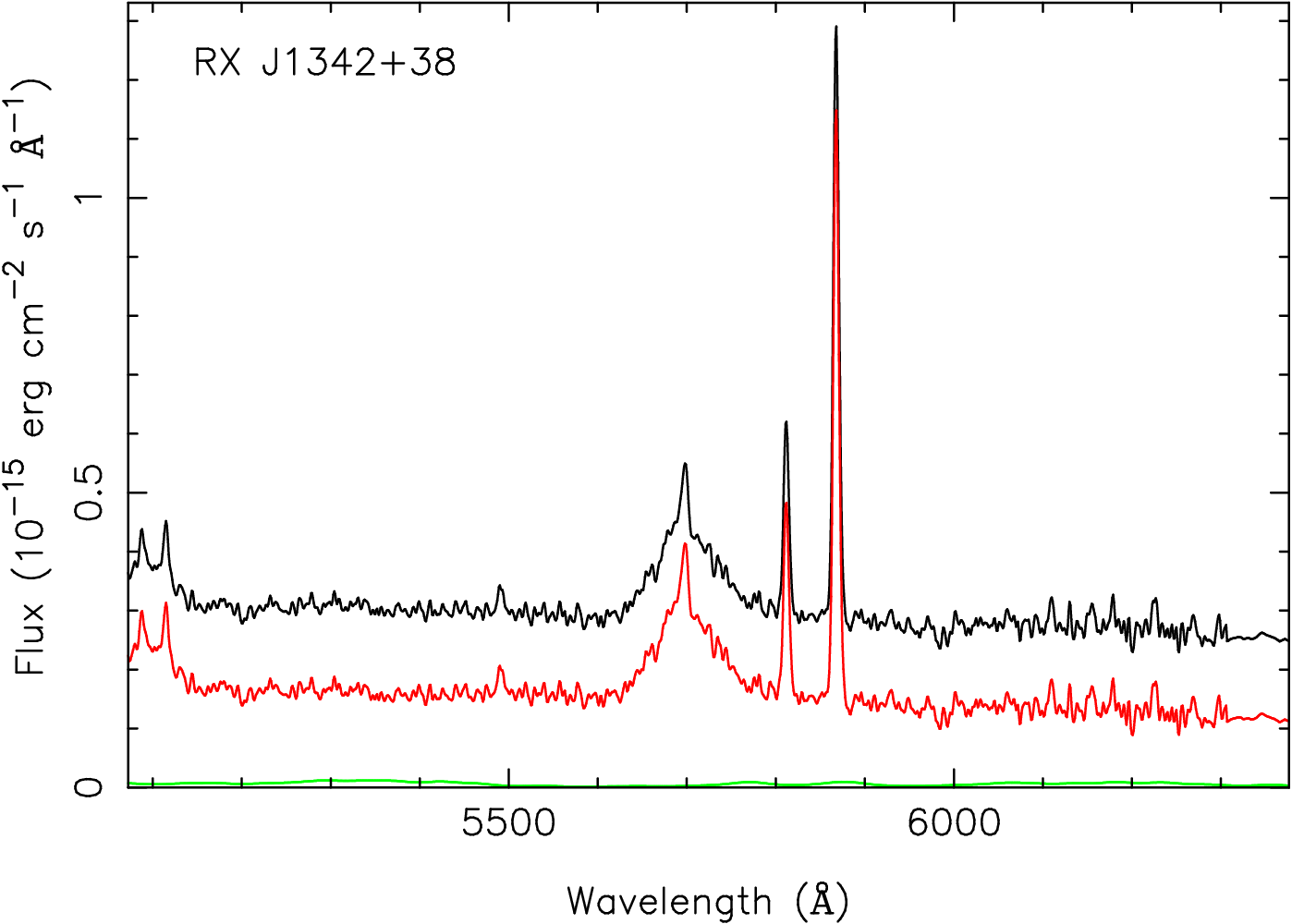}
\includegraphics[width=4.1 cm, angle=270]{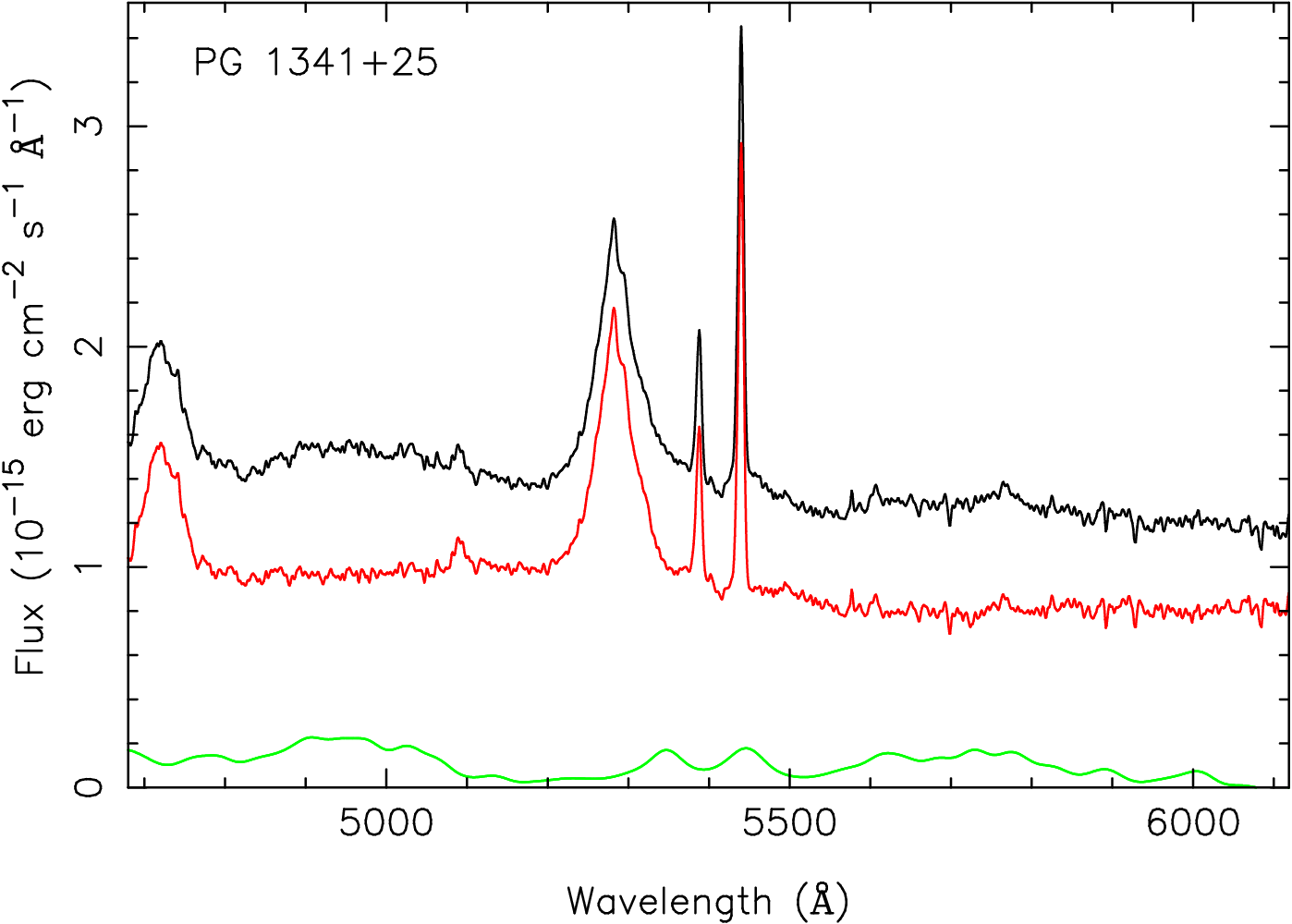}
}
\hbox{
\includegraphics[width=4.1 cm, angle=270]{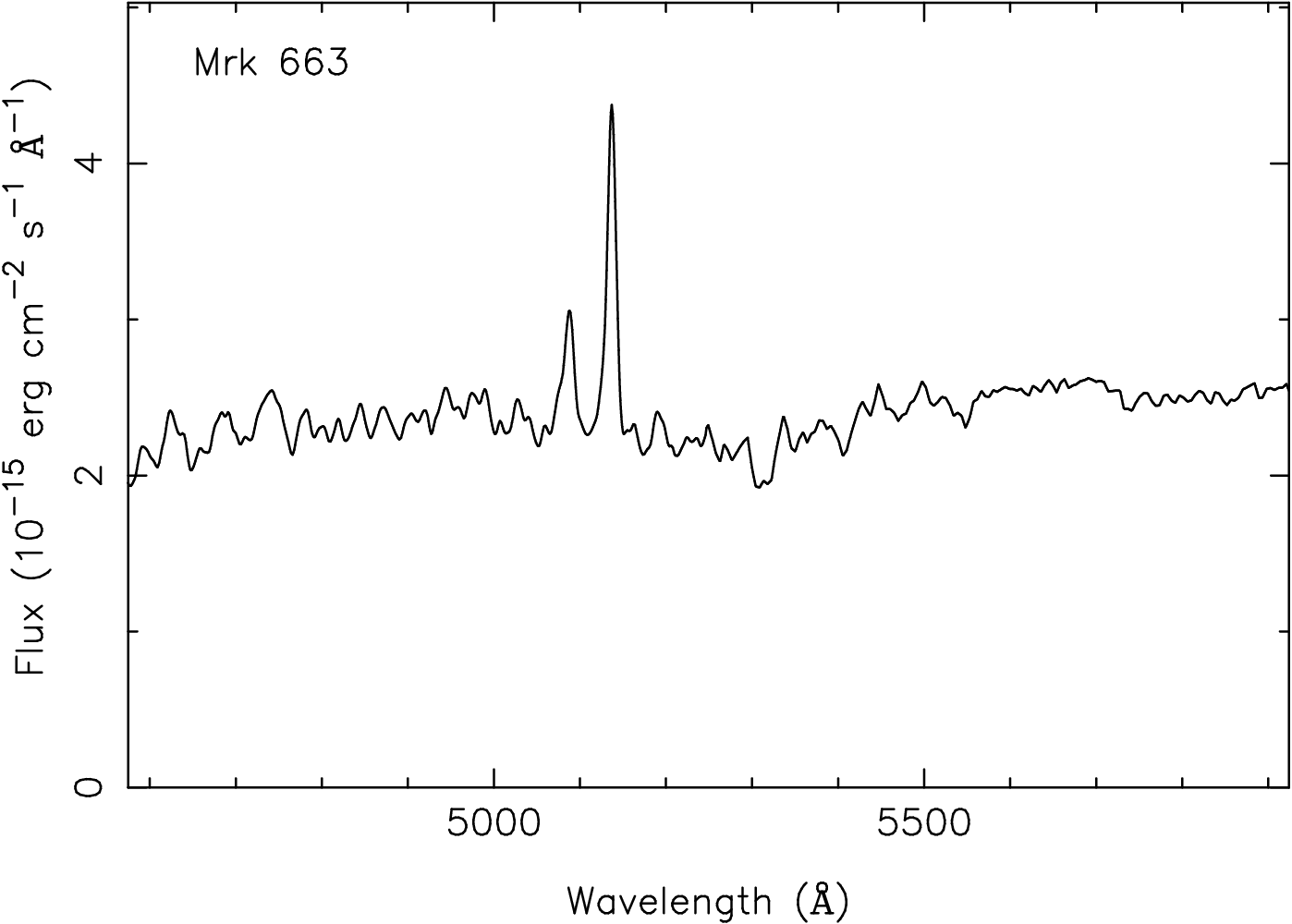}
\includegraphics[width=4.1 cm, angle=270]{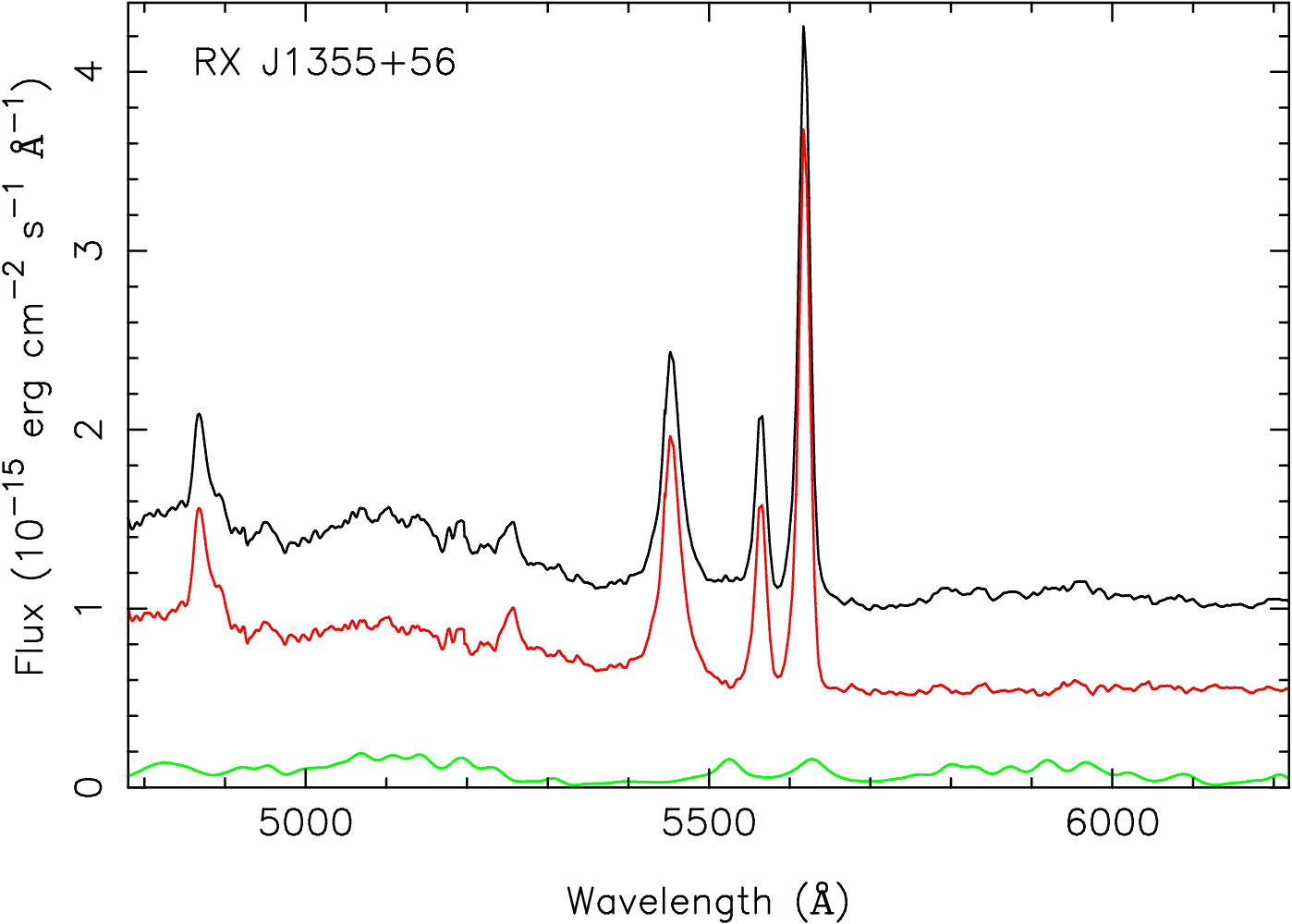}
\includegraphics[width=4.1 cm, angle=270]{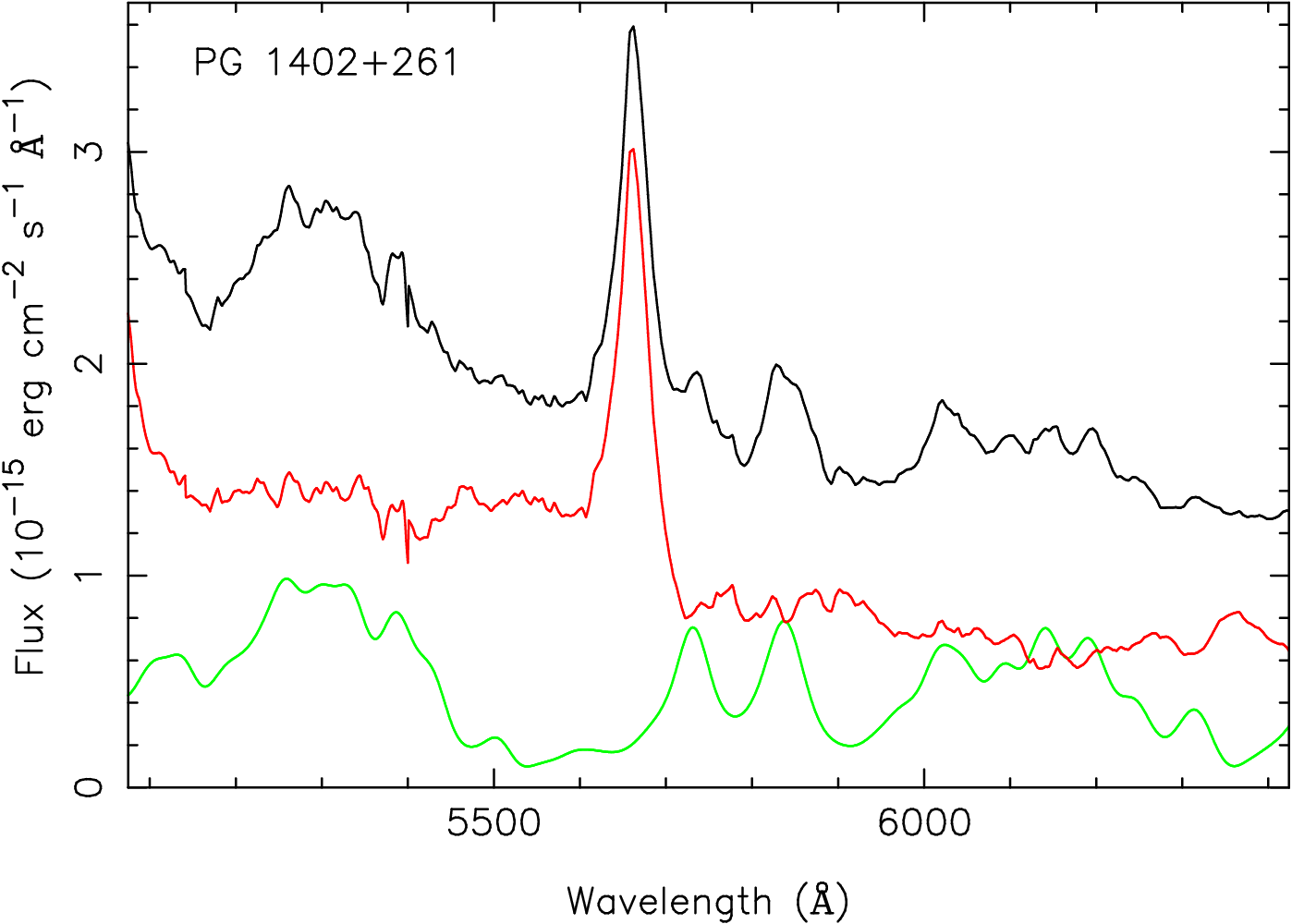}
}
\hbox{
\includegraphics[width=4.1 cm, angle=270]{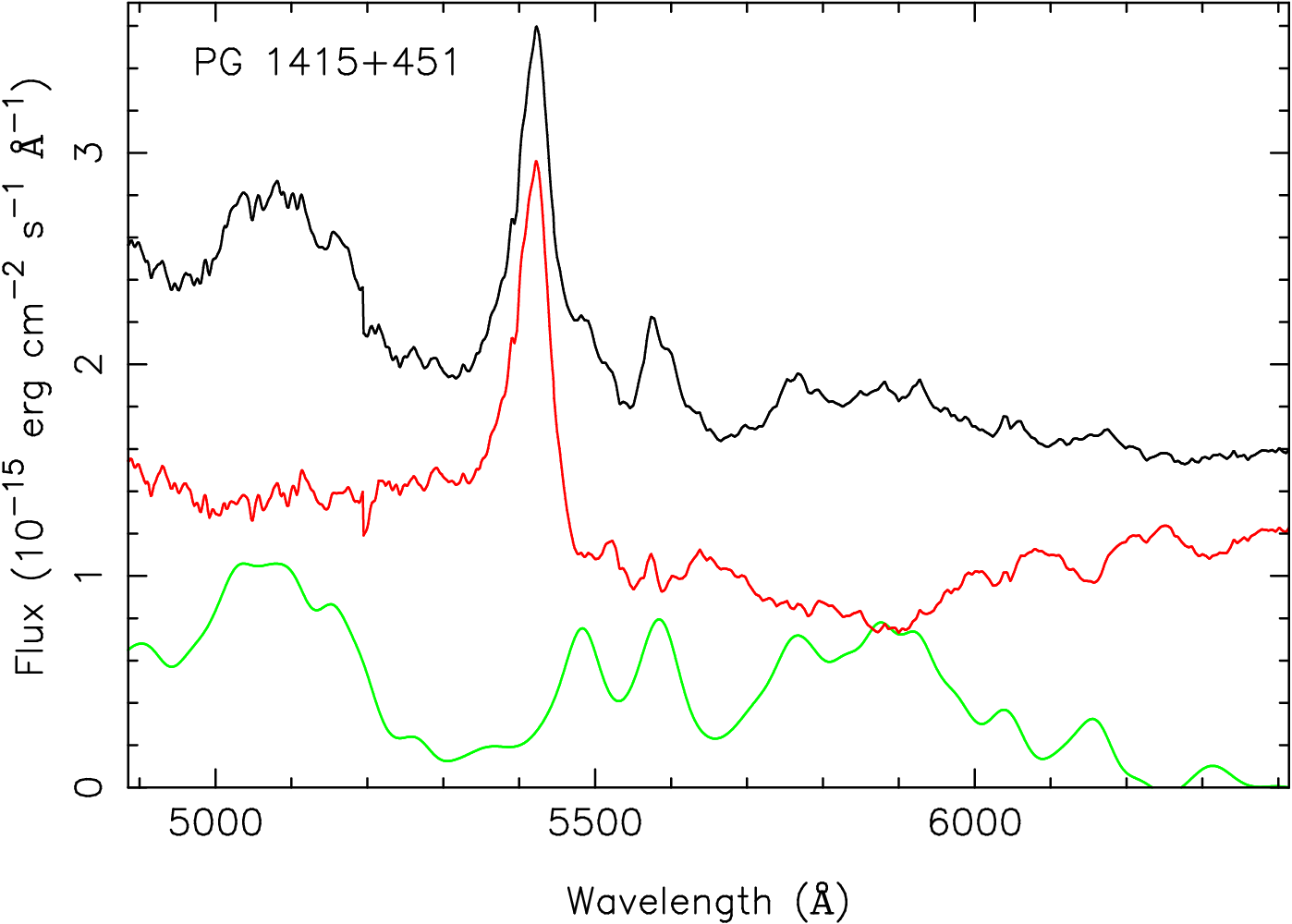}
\includegraphics[width=4.1 cm, angle=270]{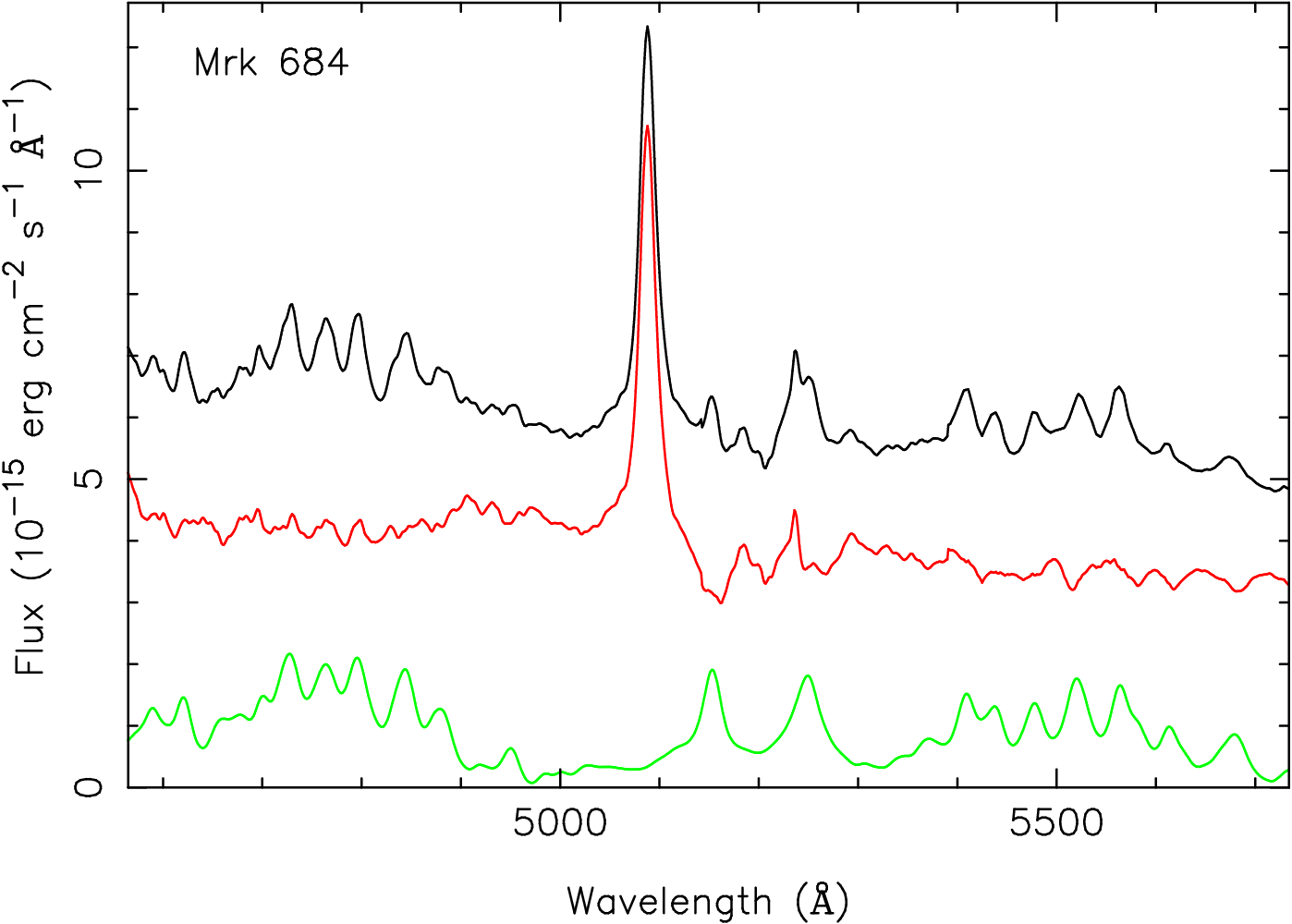}
\includegraphics[width=4.1 cm, angle=270]{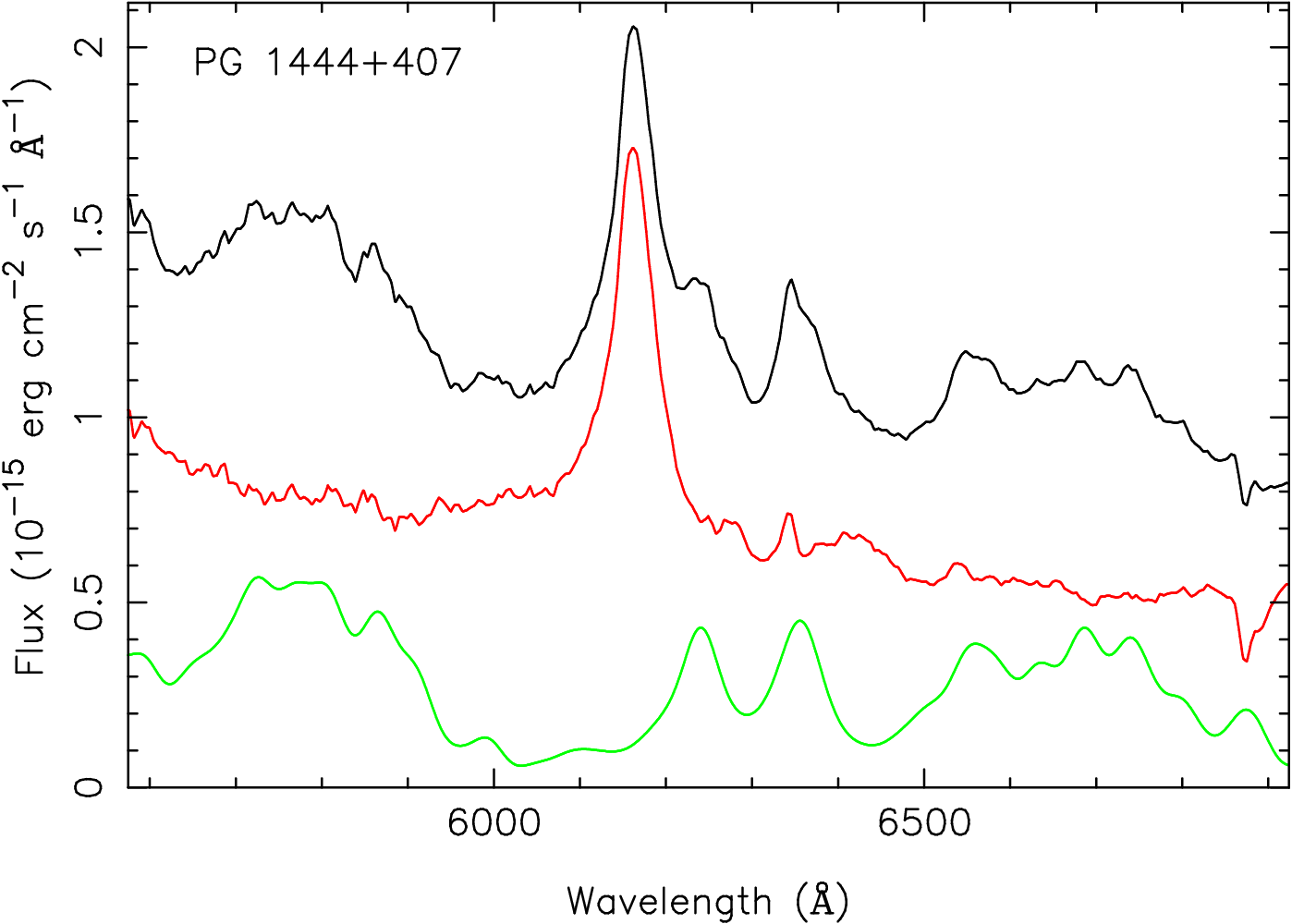}
}
\end{figure*}

\begin{figure*}
\centering
\hbox{
\includegraphics[width=4.1 cm, angle=270]{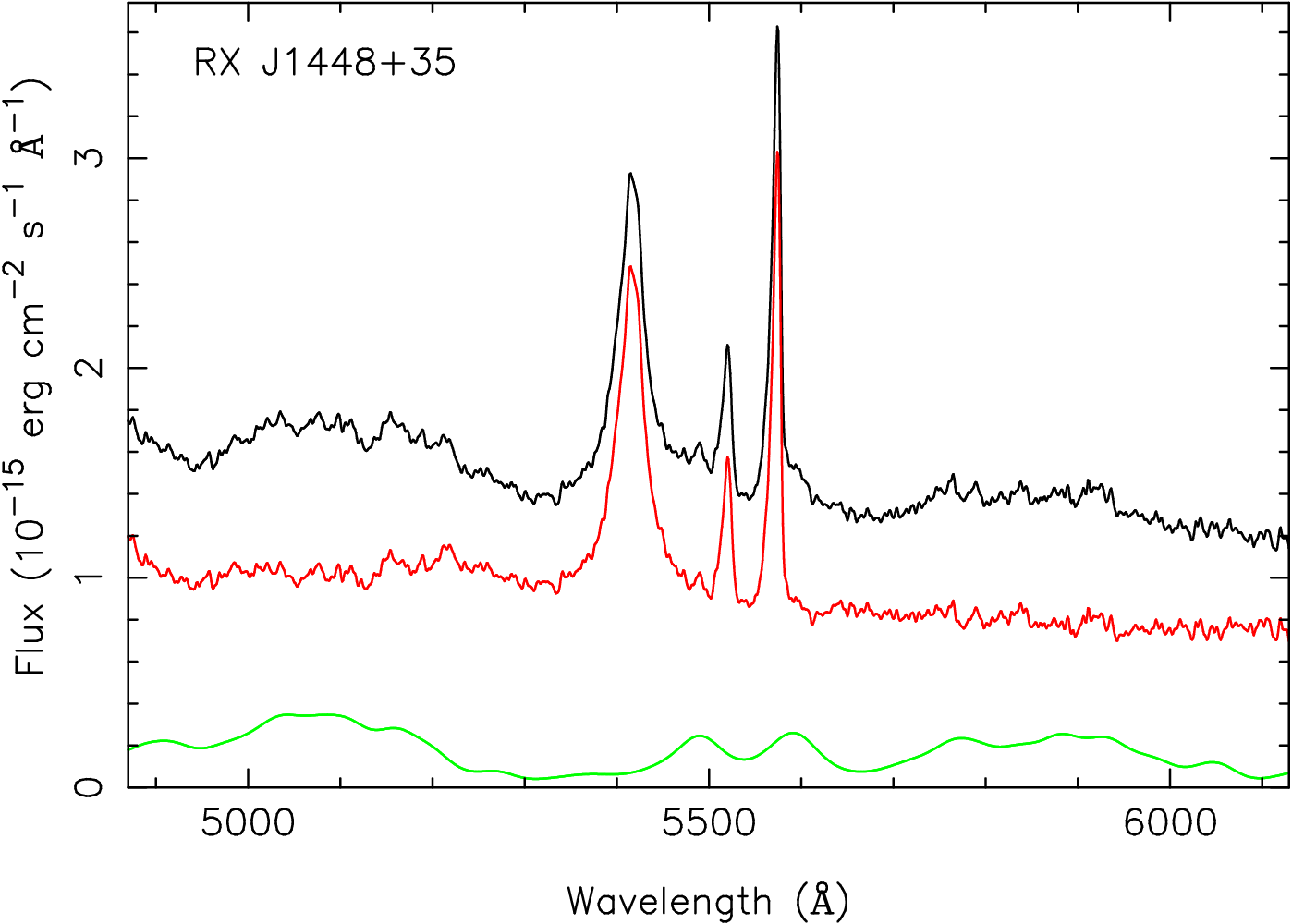}
\includegraphics[width=4.1 cm, angle=270]{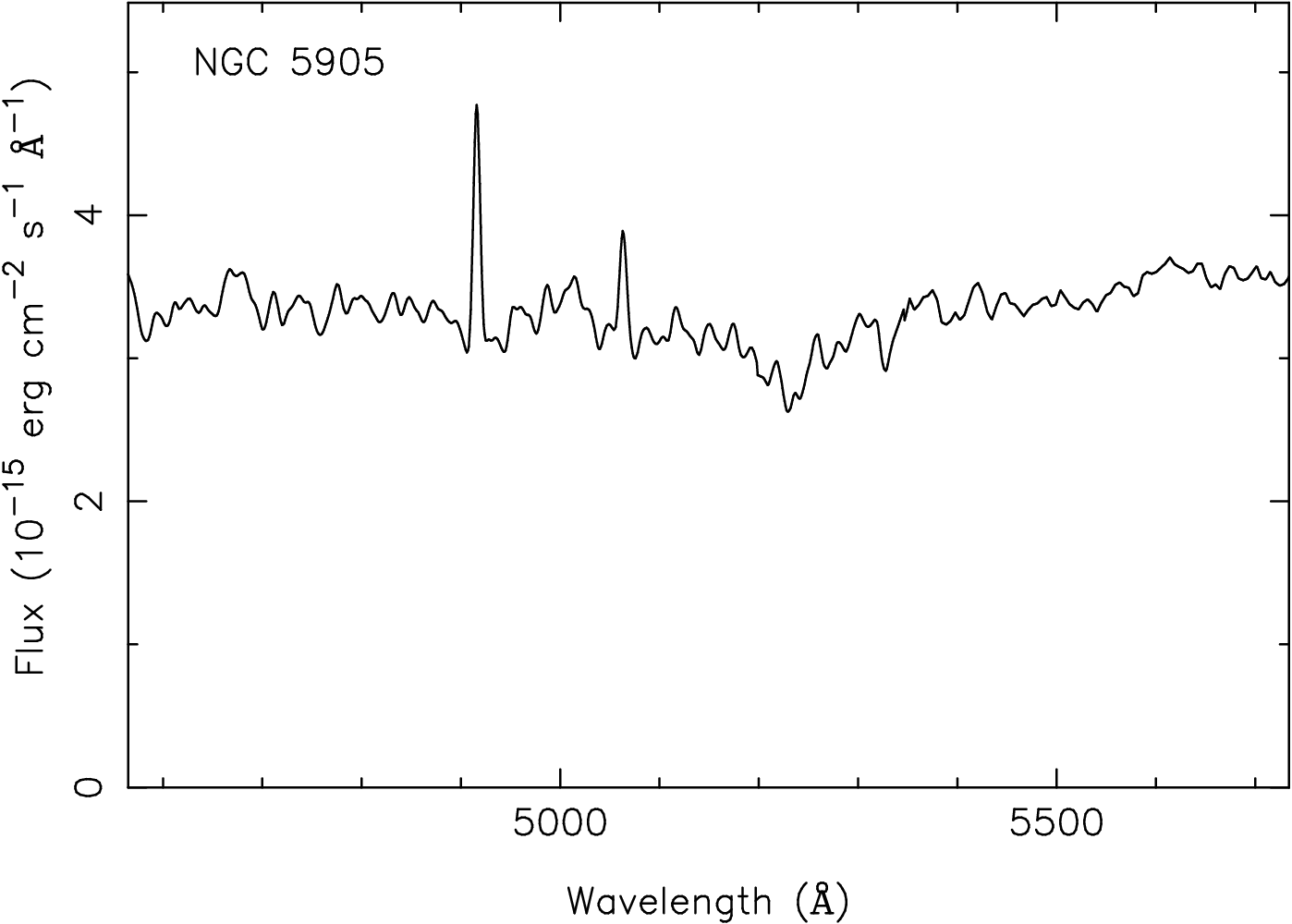}
\includegraphics[width=4.1 cm, angle=270]{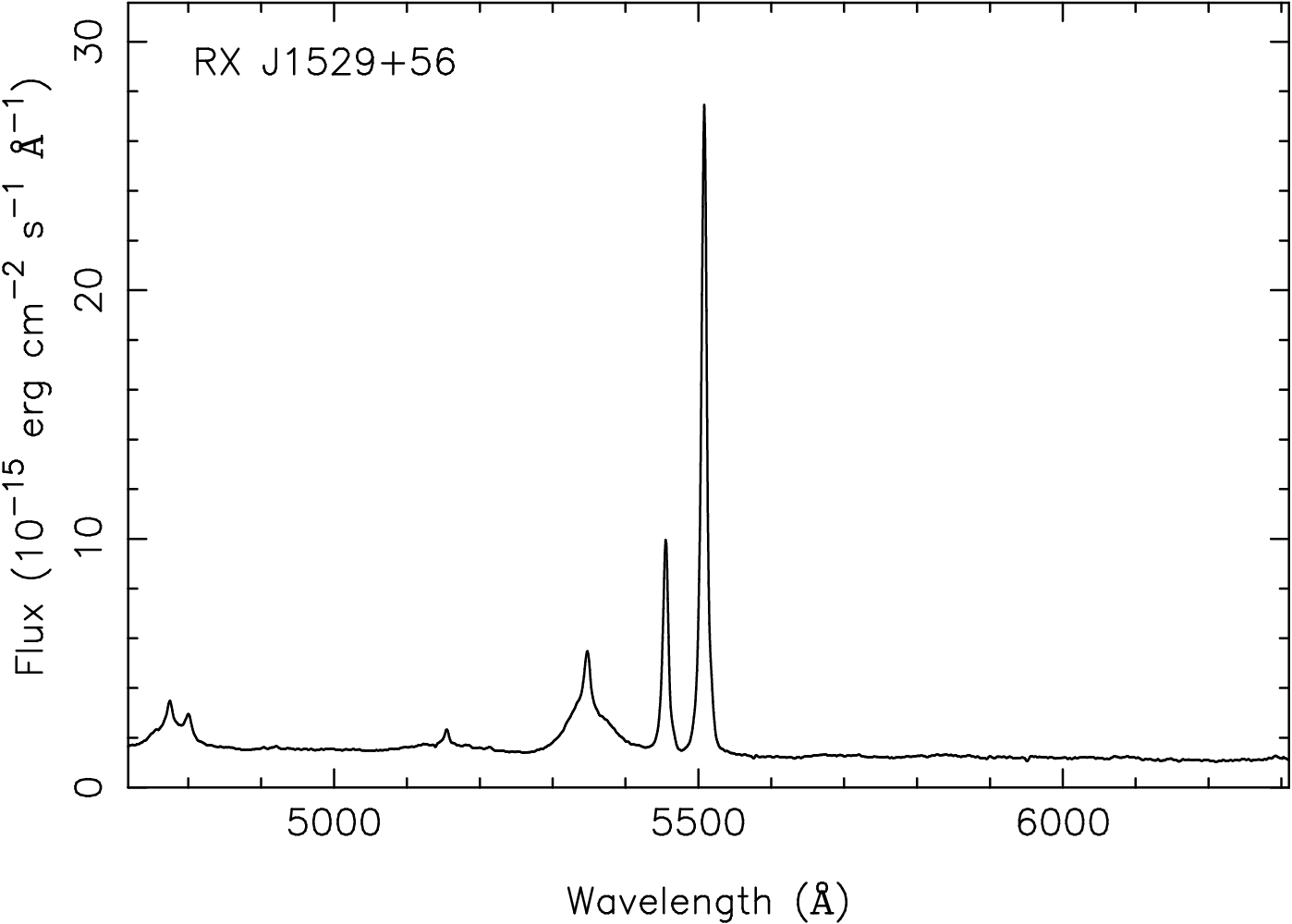}
}
\hbox{
\includegraphics[width=4.1 cm, angle=270]{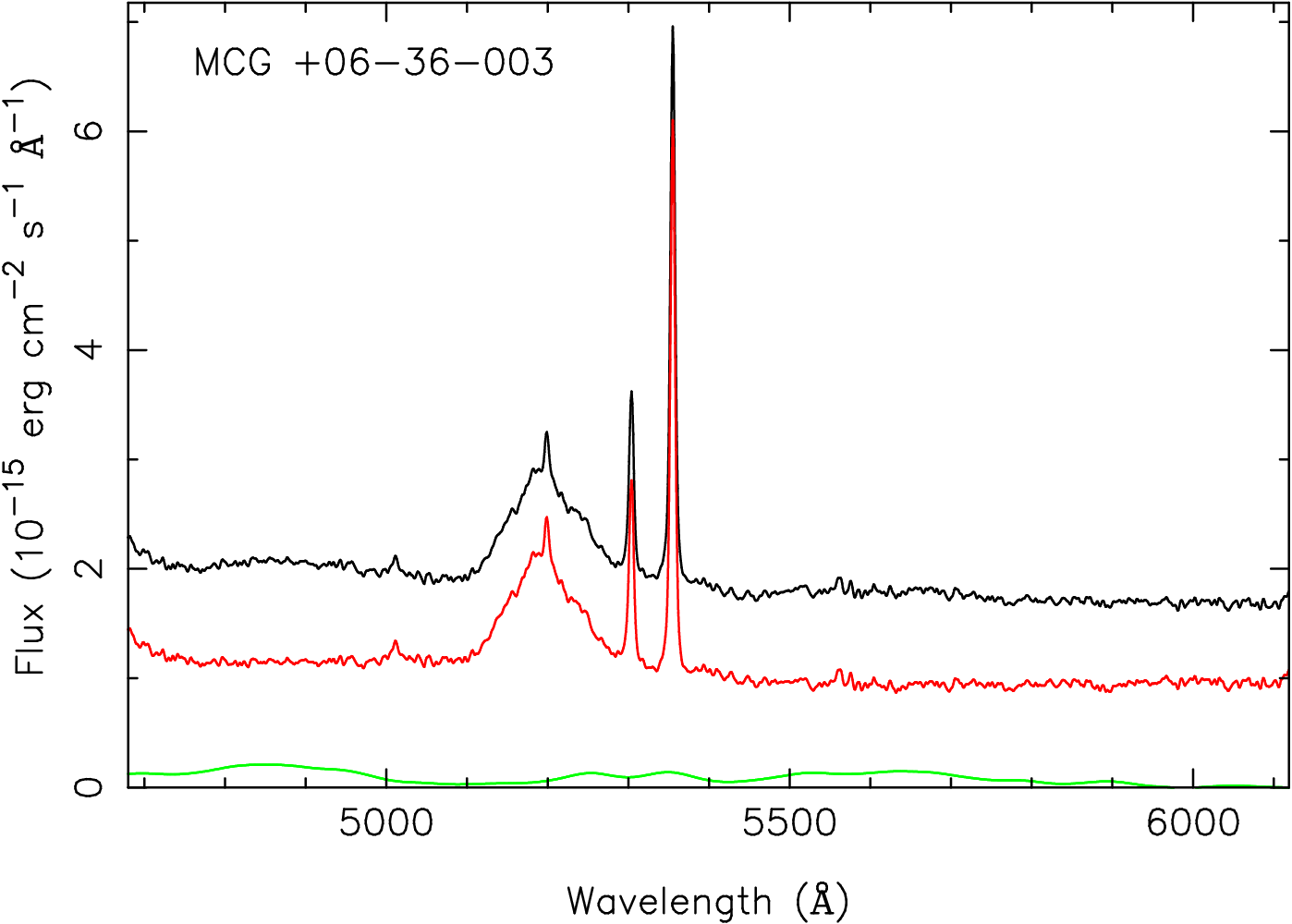}
\includegraphics[width=4.1 cm, angle=270]{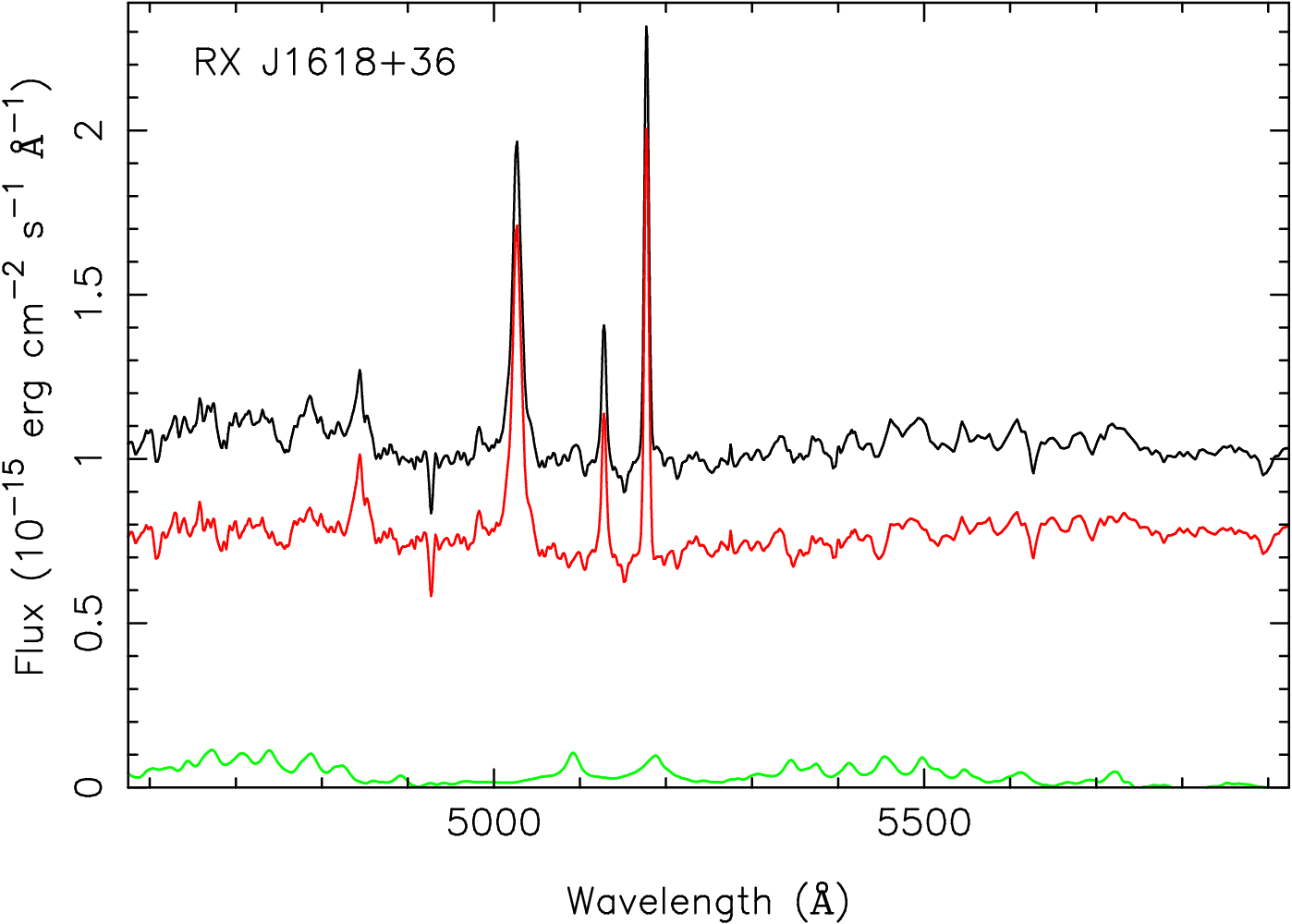}
\includegraphics[width=4.1 cm, angle=270]{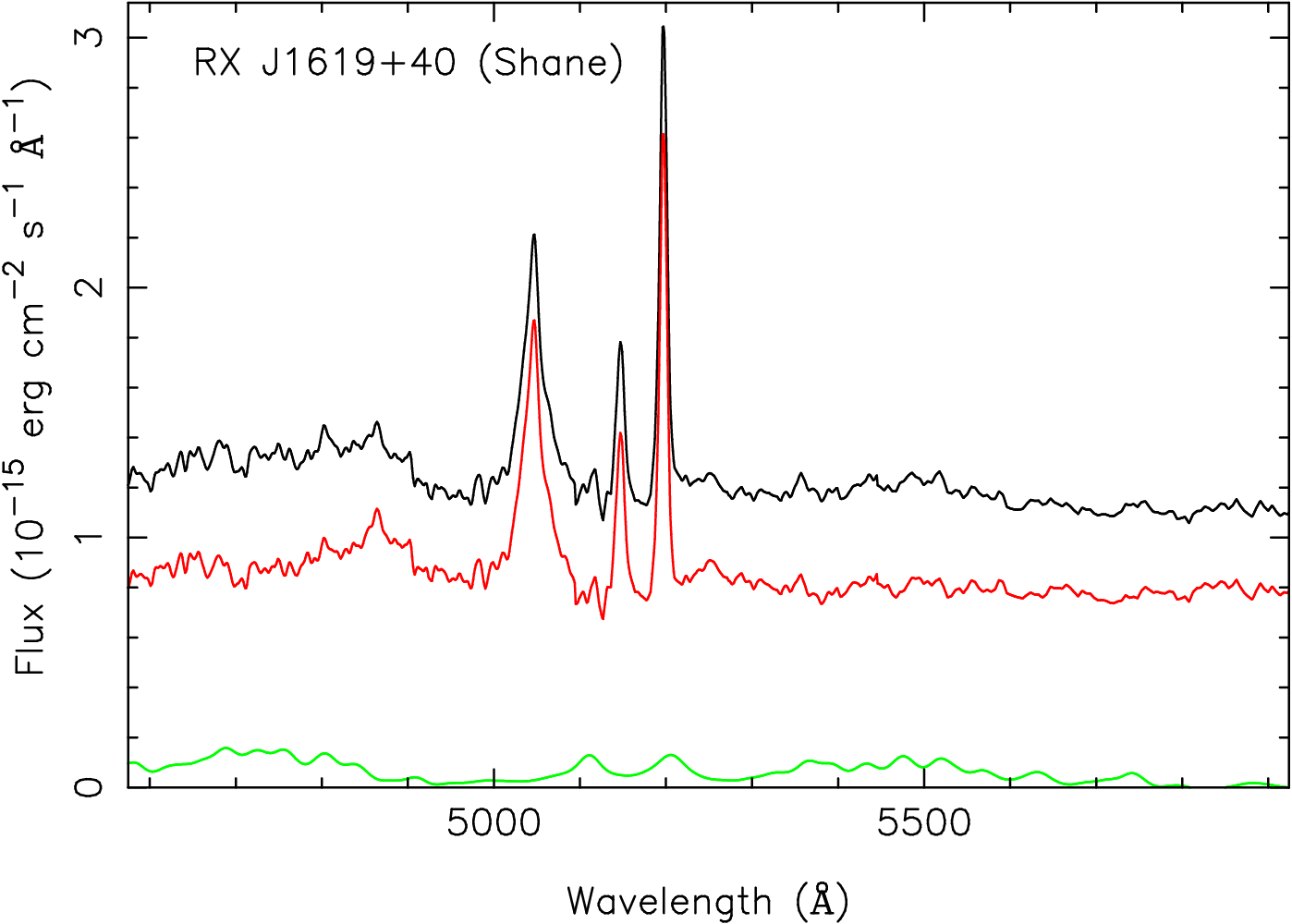}
}
\hbox{
\includegraphics[width=4.1 cm, angle=270]{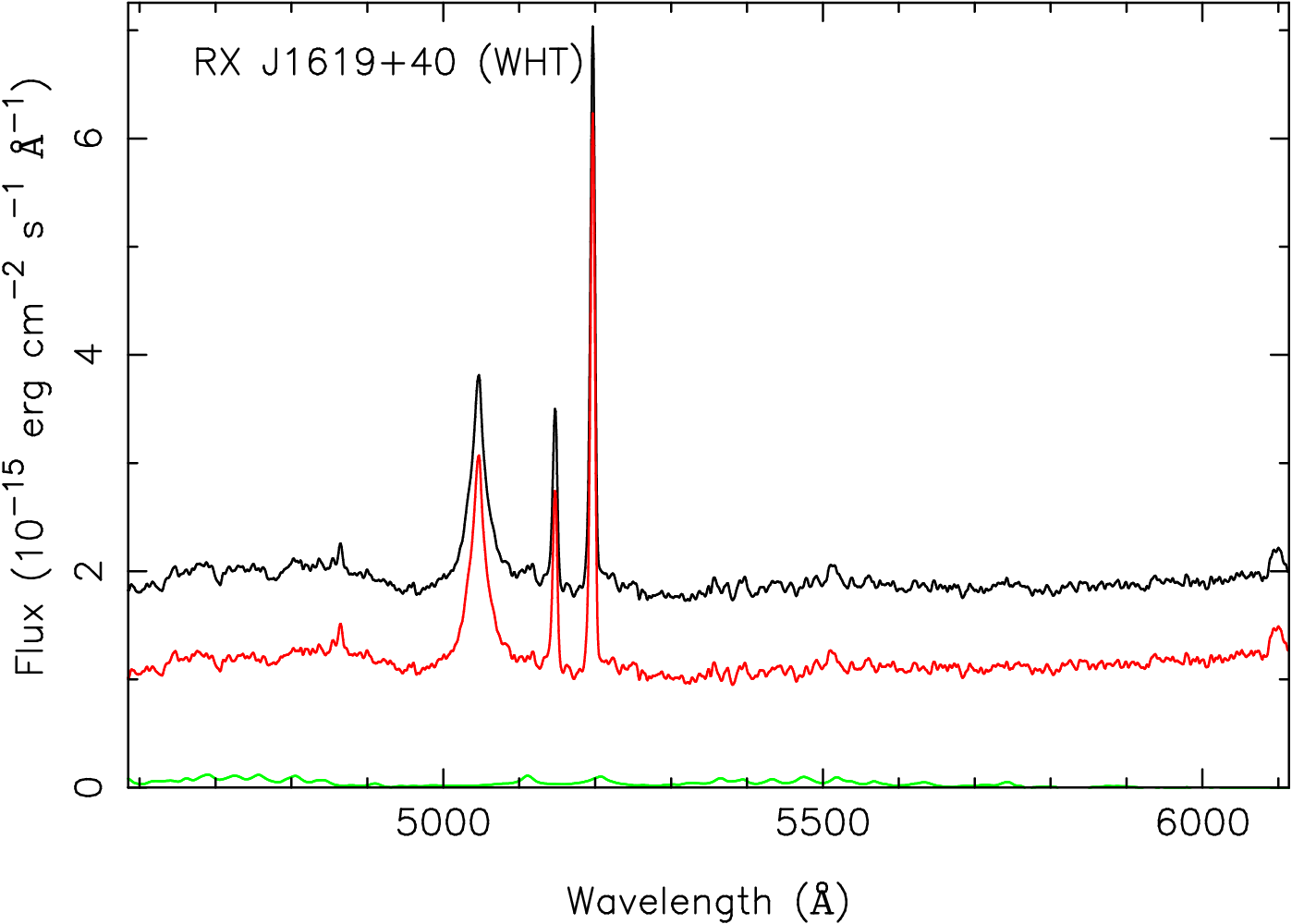}
\includegraphics[width=4.1 cm, angle=270]{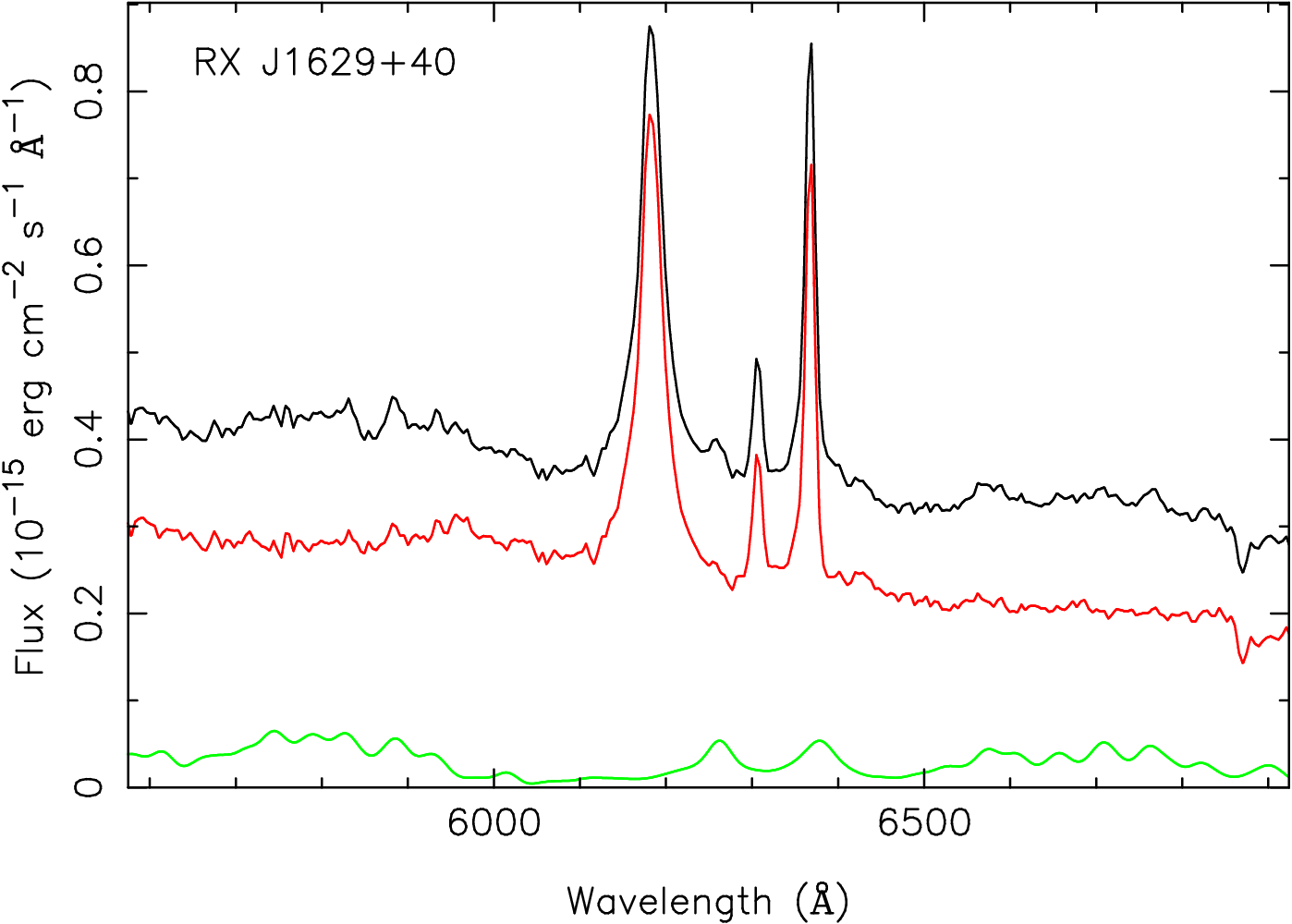}
\includegraphics[width=4.1 cm, angle=270]{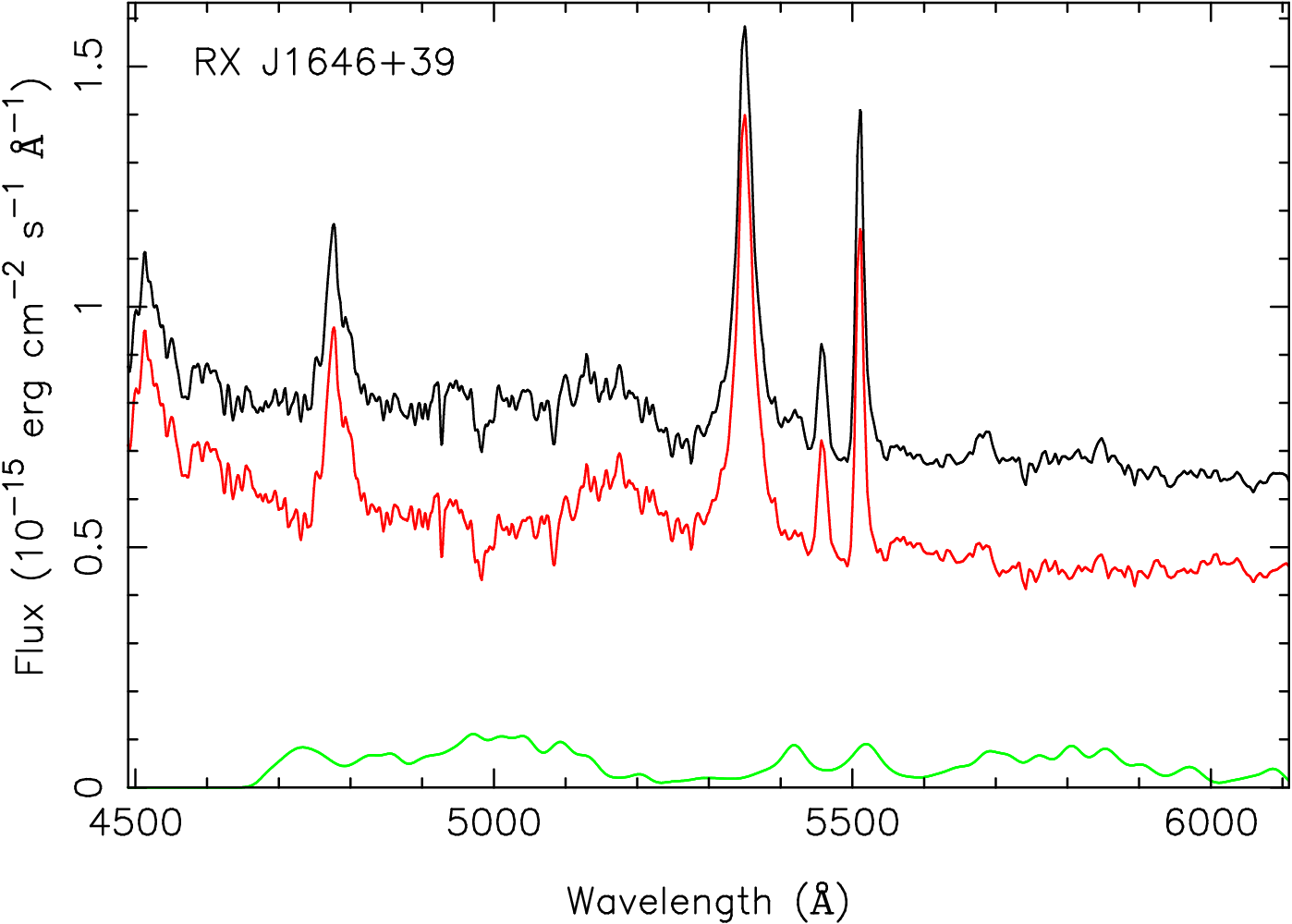}
}
\end{figure*}

\label{lastpage}
\end{document}